%% file: main.tex
\documentclass[
reprint,
amsmath,
amssymb,
aps,
]{revtex4-2}

\usepackage{graphicx}
\usepackage{dcolumn}
\usepackage[colorlinks=true, linkcolor=teal, citecolor=cyan]{hyperref}
\usepackage[all]{hypcap}
\usepackage{chngcntr}
\usepackage{ulem}
\usepackage[europeanresistors, straightvoltages]{circuitikz}
\usepackage{fourier}
\usepackage{multirow}
\usepackage{comment}

\newcommand{\op}[1]{\boldsymbol{#1}}
\newcommand{\ket}[1]{|#1\rangle}
\newcommand{\braket}[1]{\langle#1\rangle}

\begin{document}

\preprint{APS/123-QED}

\title{Enhancing dissipative cat qubit protection by squeezing}

\author{R. Rousseau\(^{1, 2}\)}
\author{D. Ruiz\(^{1, 3}\)}
\author{E. Albertinale\(^1\)}

\author{P. d'Avezac\(^1\)}
\author{D. Banys\(^1\)}
\author{U. Blandin\(^1\)}
\author{N. Bourdaud\(^1\)}
\author{G. Campanaro\(^1\)}
\author{G. Cardoso\(^1\)}
\author{N. Cottet\(^1\)}
\author{C. Cullip\(^1\)}
\author{S. Deléglise\(^2\)}
\author{L. Devanz\(^1\)}
\author{A. Devulder\(^1\)}
\author{A. Essig\(^1\)}
\author{P. Février\(^1\)}
\author{A. Gicquel\(^1\)}
\author{E. Gouzien\(^1\)}
\author{A. Gras\(^1\)}
\author{J. Guillaud\(^1\)}
\author{E. Gümü\c{s}\(^1\)}
\author{M. Hall\'en\(^1\)}
\author{A. Jacob\(^1\)}
\author{P. Magnard\(^1\)}
\author{A. Marquet\(^1\)}
\author{S. Miklass\(^1\)}
\author{T. Peronnin\(^1\)}
\author{S. Polis\(^1\)}
\author{F. Rautschke\(^1\)}
\author{U. R\'eglade\(^1\)}
\author{J. Roul\(^1\)}
\author{J. Stevens\(^1\)}
\author{J. Solard\(^1\)}
\author{A. Thomas\(^1\)}
\author{J.-L. Ville\(^1\)}
\author{P. Wan-Fat\(^1\)}

\author{R. Lescanne\(^1\)}
\author{Z. Leghtas\(^3\)}
\author{J. Cohen\(^1\)}
\author{S. Jezouin\(^1\)}
\author{A. Murani\(^1\)}
\affiliation{\(^1\)Alice \& Bob, 53 Bd du Général Martial Valin, 75015 Paris, France}
\affiliation{\(^2\)Laboratoire Kastler Brossel, Sorbonne Université, CNRS, ENS-Université,PSL, Collège de France, 4 place Jussieu, 75005 Paris, France}
\affiliation{\mbox{\(^3\)Laboratoire de Physique de l'Ecole normale supérieure, ENS-PSL, CNRS, Sorbonne Université, Université Paris Cité,} \mbox{Centre Automatique et Systèmes, Mines Paris, Université PSL, Inria, Paris, France}}

\date{\today}

\input{abstract}

\maketitle

\input{introduction}
\input{part1_moon_cats_first_experiemental_results}
\input{part2_error_rates_scaling}
\input{part3_zeno_error_rates}
\input{conclusion}
\input{acknowledgments}

\bibliographystyle{ieeetr}
\bibliography{MoonCatPaper}

\newpage
\appendix
\counterwithin{figure}{section}
\renewcommand{\thefigure}{\Alph{section}\arabic{figure}}
\input{1_device_fabrication_and_wiring.tex}
\input{2_circuit_hamiltonian.tex}
\input{3_calibrations_of_parametric_pumps.tex}
\input{4_system_calibration.tex}
\input{5_lifetimes_measurements.tex}
\input{6_zeno_gate_measurement.tex}
\input{7_scalings.tex}
\input{8_theory.tex}

\end{document}

%% file: abstract.tex
\begin{abstract}
    Dissipative cat-qubits are a promising architecture for quantum processors due to their built-in quantum error correction.
    By leveraging two-photon stabilization, they achieve an exponentially suppressed bit-flip error rate as the distance in phase-space between their basis states increases, incurring only a linear increase in phase-flip rate.
    This property substantially reduces the number of qubits required for fault-tolerant quantum computation.
    Here, we implement a squeezing deformation of the cat qubit basis states, further extending the bit-flip time while minimally affecting the phase-flip rate.
    We demonstrate a steep reduction in the bit-flip error rate with increasing mean photon number, characterized by a scaling exponent \(\gamma=4.3\), rising by a factor of \(74\) per added photon.
    Specifically, we measure bit-flip times of 22 seconds for a phase-flip time of \(1.3\; \mathrm{\mu s}\) in a squeezed cat qubit with an average photon number \(\bar{n}=4.1\), a 160-fold improvement in bit-flip time compared to a standard cat.
    Moreover, we demonstrate a two-fold reduction in \(Z\)-gate infidelity, with an estimated phase-flip probability of \(\epsilon_X = 0.085\) and a bit-flip probability of \(\epsilon_Z = 2.65 \cdot 10^{-9}\) which confirms the gate bias-preserving property.
    This simple yet effective technique enhances cat qubit performances without requiring design modification, moving multi-cat architectures closer to fault-tolerant quantum computation.
\end{abstract}

%% file: introduction.tex
\begin{figure}[th!]
    \centering
    \includegraphics[width=0.48\textwidth]{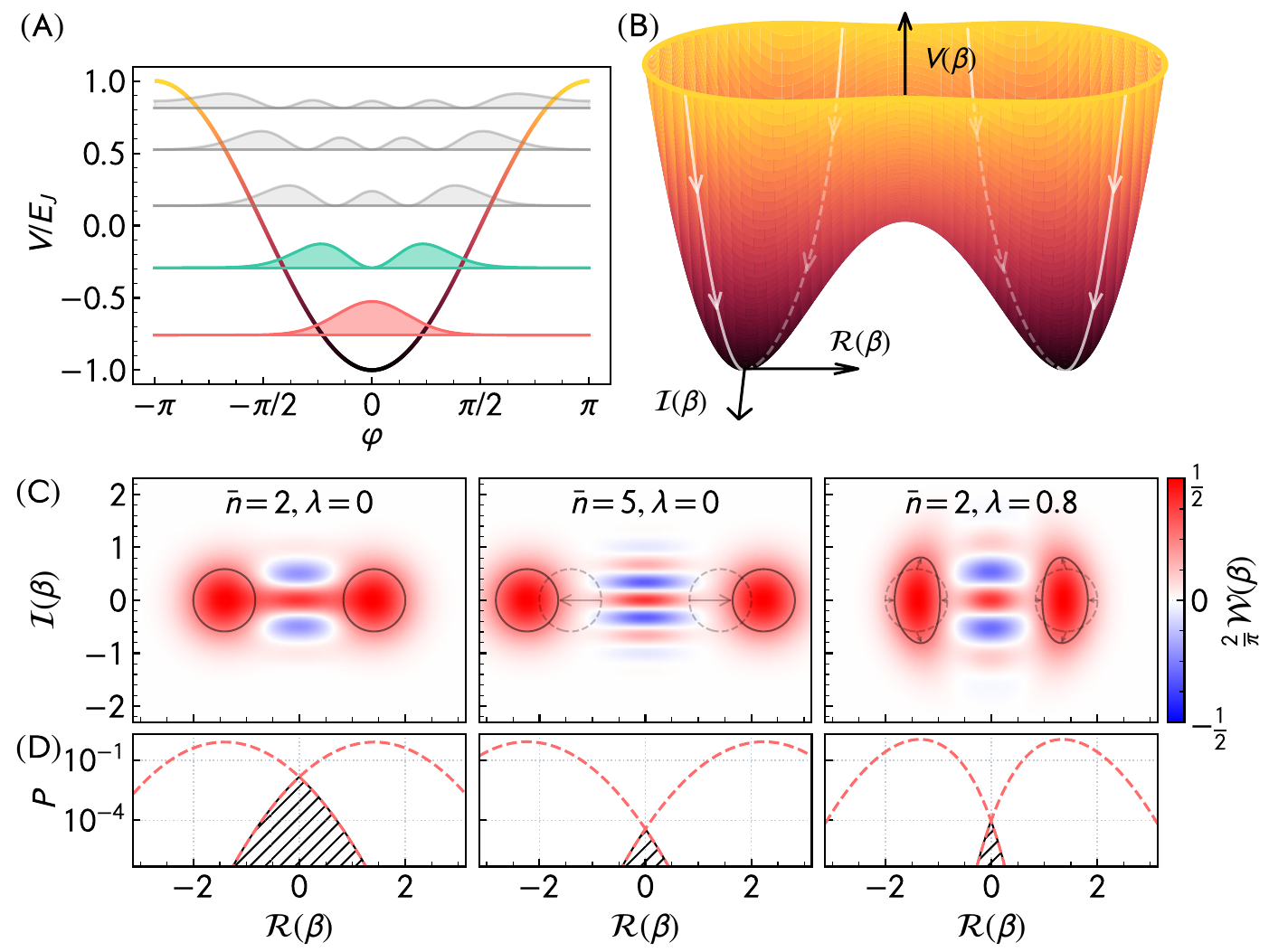}
    \caption{Principle of protection enhancement by squeezing.
        (A) Cosine potential (colored solid line) of a standard qubit (transmon) as a function of the reduced superconducting phase \(\varphi \).
        Horizontal lines indicate the corresponding energy eigenvalues, and filled areas represent the squared moduli of the wavefunctions.
        Highlighted in red and green, the two lowest-energy states define the qubit manifold and have overlapping support.
        (B) Pseudo-potential (colored surface) of a stabilized cat qubit overlaid with semiclassical trajectories (white lines).
        The two cat-qubit basis states lie in each wells, their separation ensures robust bit-flip protection.
        (C) Wigner functions \(\mathcal{W}(\beta)\) of cat states.
        The first two panels show coherent state-based standard cat states while the third panel illustrates a squeezed moon cat state.
        Solid black lines delineate the extent of the basis states at \(\mathcal{W} = 0.5/\pi \), and dashed lines indicate the extent of the left panel cat state.
        Arrows indicate modifications of the basis state.
        (D) Probability density of the cat basis states (red dashed lines) plotted in logarithmic scale, as a function of \(\mathcal{R}(\beta)\).
        The dashed region highlights the overlap between the two basis states.
    }\label{fig:fig1}
\end{figure}

More efficient quantum technologies can emerge through the careful choice of how information is encoded.
For example, rather than increasing the photon flux in LIGO, enhancing sensitivity to optical path using squeezed states of light has led to unprecedented gravitational wave detection accuracy~\cite{jia_squeezing_2024}.
Conversely, qubits can become more resilient to noise if they are encoded non-locally (Fig.~\ref{fig:fig1}A,B)~\cite{shor_fault-tolerant_1996}, as in Majorana~\cite{kitaev_unpaired_2001} or GKP states~\cite{campagne-ibarcq_quantum_2020}.
This opens the door to hardware-efficient quantum processor architectures, whereas standard approaches demand hundreds to thousands physical qubits per logical qubit to correct both bit-flip and phase-flip errors~\cite{gidney_how_2021}.
Among these solutions are dissipative cat qubits~\cite{mirrahimi_dynamically_2014}.
By dynamically stabilizing coherent states of a harmonic oscillator with opposite phase, an exponential protection against bit-flips is achieved~\cite{lescanne_exponential_2020}, leading to a significant reduction of qubit count overhead~\cite{gouzien_performance_2023}.
In this work, we extend this stabilization scheme to squeezed cat states.

A cat qubit lives in the manifold spanned by the basis states
\(\ket{\alpha}\) and \(\ket{-\alpha}\), coherent states of a harmonic oscillator, where \( \alpha \) is the cat amplitude.
To protect the basis states from local perturbations such as single-photon loss and dephasing, a stabilization mechanism is required.
It either consists in using the Hamiltonian gap of a parametrically driven non-linear oscillator~\cite{puri_bias-preserving_2020, grimm_kerr-cat_2020, frattini_squeezed_2022}, or a specifically engineered two-photon dissipation~\cite{mirrahimi_dynamically_2014, leghtas_confining_2015,touzard_coherent_2018, lescanne_exponential_2020, berdou_one_2023, reglade_quantum_2024,marquet_autoparametric_2024, putterman_preserving_2024}, which is our focus here.
As the mean photon number \(\bar{n} = |\alpha|^2\) grows, the coherent states become well separated in phase space, exhibiting exponentially small overlap (Fig.~\ref{fig:fig1}D).
Consequently, the bit-flip rate is exponentially suppressed, \(\Gamma_Z \sim \exp(-\gamma \bar{n})\), with a scaling exponents bounded to \(\gamma = 2\)~\cite{dubovitskii_bit-flip_2024, putterman_preserving_2024}, resulting in bit-flip times reaching few hundreds of seconds in recent experiments~\cite{berdou_one_2023, reglade_quantum_2024}.
However, this increased separation makes the superposition states more susceptible to decoherence: the phase-flip rate increases.
Crucially, this increase is only linear, \(\Gamma_X \sim \bar{n}\), yielding a significant bias between bit-flip and phase-flip errors.
The strategy for cat qubit based quantum error correction consists in leveraging this bias and correcting the remaining phase-flip error by redundantly encoding the information in several cat qubits~\cite{guillaud_repetition_2019,chamberland_building_2022, putterman_hardware-efficient_2024}.
This considerably limits the hardware overhead compared to relying on redundancy for both bit-flips and phase-flips~\cite{gouzien_performance_2023, ruiz_ldpc-cat_2024}.
However, to efficiently correct the phase-flips, the cat qubit phase-flip rate should remain well below a threshold and hence the photon number \(\bar{n}\) has to remain limited.
As a result, having a given bit-flip suppression with the least number of photon, or equivalently, a strong scaling exponent \(\gamma \) is critical.

Instead of increasing the coherent states amplitude, the overlap between the coherent states can be reduced by squeezing them (Fig.~\ref{fig:fig1}D).
Under the right noise conditions, this results in an enhancement of the bit-flip resilience without affecting the phase-flip rate and recent theoretical studies~\cite{hillmann_quantum_2023,xu_autonomous_2023,rojkov_stabilization_2024} on squeezed cat qubits have demonstrated that their scaling exponent could indeed exceed the bound \(\gamma = 2\).
Squeezed cat states have recently been generated using heralding techniques~\cite{jeannic_slowing_2018} and deterministic optimal control~\cite{pan_protecting_2023}, demonstrating some level of protection against decoherence~\cite{filip_amplification_2001,schlegel_quantum_2022}, but the active stabilization of a squeezed cat qubit manifold has not been demonstrated.
In this work, we propose and implement a scheme to dissipatively stabilize squeezed cat qubits, that does not require any design change compared to previous implementations of standard cat qubits~\cite{reglade_quantum_2024}.
By deforming the states, we are able to reach a scaling exponent \(\gamma = 4.3 \), largely exceeding the attainable bound of non-squeezed dissipative cats.
We measure a 160-fold improvement in bit-flip time at a fixed phase flip time.
Furthermore, notable improvements in gate performances were achieved.

%% file: part1_moon_cats_first_experiemental_results.tex
\section{Dissipation engineering for squeezed-like cat qubits}\label{sec:level2}

\begin{figure*}[ht]
    \centering
    \includegraphics[width=\textwidth]{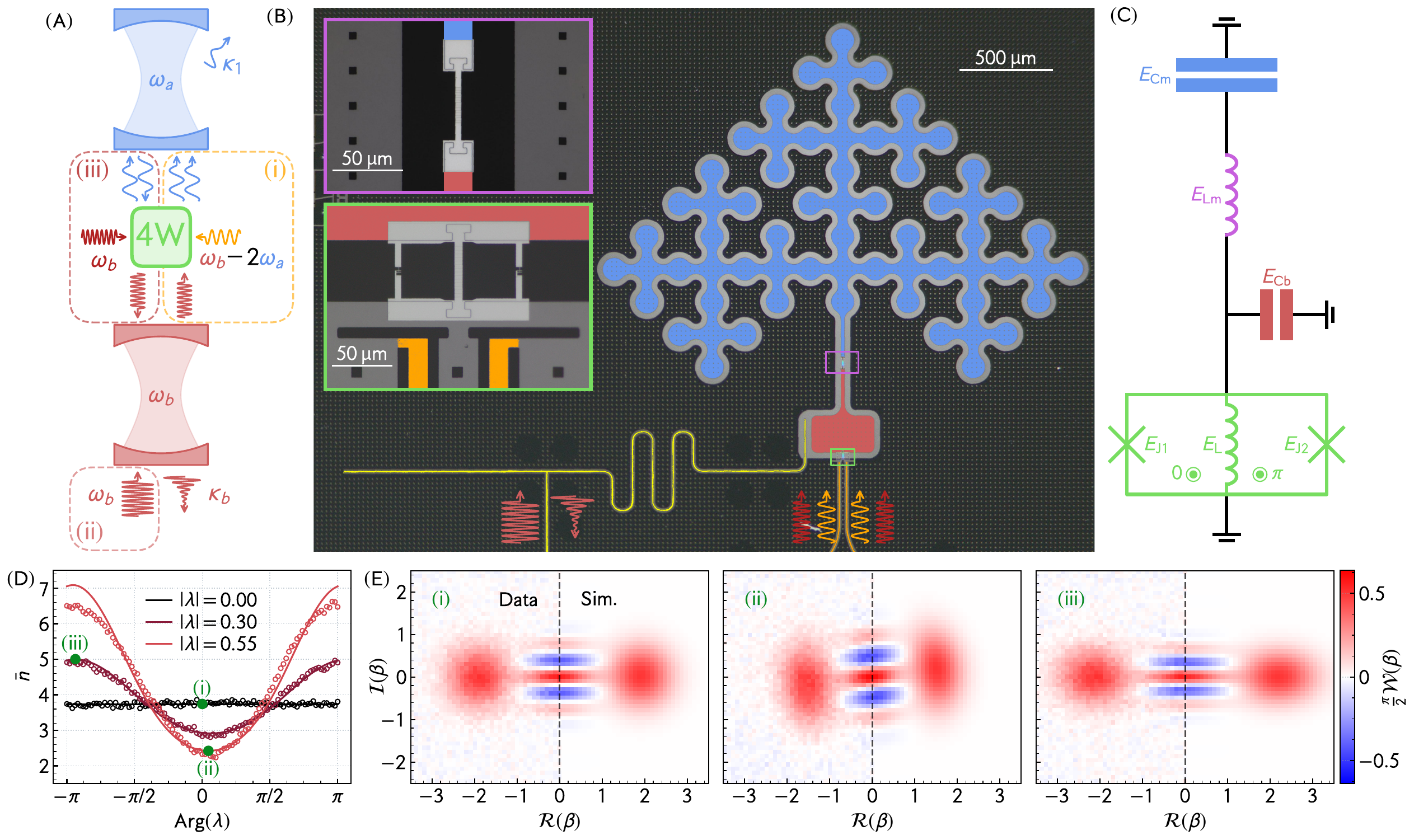}
    \caption{Moon cat qubit implementation and calibration.
        (A) Schematic of the dissipative squeezed cat qubit circuit.
        A high-Q memory mode (blue) is coupled to a low-Q buffer mode (red) via a four-wave mixing element (green).
        To engineer the Hamiltonian of Eq.~\ref{eq:full_hamiltonian}, we activate 3 processes: (i) a 2-photon exchange between the memory (blue arrows) and the buffer (red arrow) with a pump at \(\omega_b - 2\omega_a\)(orange arrow), (ii) a buffer drive (red arrow), which together stabilize standard cats and (iii) a longitudinal coupling with a separate pump at \(\omega_b\) (dark red arrow) which conditionally displaces the buffer (red arrow) depending on the memory photon number (blue arrows).
        (B) Colored optical micrograph of a twin sample.
        The fractal-shaped capacitor (blue) is connected to a planar capacitor (red) through a chain of 29 Josephson junctions (see purple inset).
        The red capacitor is shunted to ground via the Asymmetrically Threaded SQUID (ATS), shown in the green inset.
        The ATS consists of a SQUID shunted by a chain of 23 Josephson junctions, forming two loops that are threaded by an external magnetic flux generated by two flux-lines (orange overlay).
        A quarter-wavelength coplanar waveguide resonator (yellow) is capacitively coupled to the red capacitor and galvanically coupled to a transmission line, thereby filtering the memory mode to mitigate its direct coupling to the line.
        The same transmission line injects the buffer drive and enables output signal measurement.
        (C) Equivalent lumped-element circuit of the cat qubit.
        (D) Determination of the deformation parameter \(\lambda \).
        Measured (dots) and fitted (solid lines) mean photon number stabilized in the memory, plotted as a function of the phase \(\mathrm{Arg}(\lambda)\) for various \(|\lambda|\) values (color).
        (E) Measured Wigner functions of the memory at selected operating points (marked in panel D), illustrating the cat state deformation.
        In each subpanel, the left side of the dashed vertical line shows experimental data; the right side shows simulated Wigner functions using the fitted \(\lambda \) values from panel (D).
    }\label{fig:fig2}
\end{figure*}

To benefit from the advantage of squeezing cat qubits, it is required to stabilize their corresponding code space.
One possibility is to engineer a loss operator whose dark states are squeezed states~\cite{xu_autonomous_2023,hillmann_quantum_2023}.
Here, we engineer a simplified dissipator
\begin{equation}
    \label{eq:Moon_dissipator}
    \op{L_2}(\alpha, \lambda) = \sqrt{\kappa_2} \left(\op{a}^2 - \alpha^2 + \lambda (\op{a}^\dag \op{a} - \alpha^2)\right),
\end{equation}
where \(\op{a}\) is the annihilation operator of the memory mode hosting the cat qubit, \(\kappa_2\) is the two-photon dissipation rate, \(\alpha \) sets the amplitude of the stabilized basis states \( \bigl \{ \ket{\alpha}_\lambda, \ket{-\alpha}_\lambda \bigr \} \), and \(\lambda \) is the deformation parameter quantifying the deviation from coherent states (Appendix~\ref{ap:moon_cat_theory_sq_comparison}).
Unlike standard cat states (\(\lambda=0\), Fig.~\ref{fig:fig1}C first two panels), their Wigner functions are deformed proportionally to \(\lambda \) along a circular path, reflecting the rotational symmetry of the \(\op{a}^\dag \op{a}\) term in Eq.~(\ref{eq:Moon_dissipator}) (Fig.~\ref{fig:fig1}C third panel).
Owing to their visual resemblance to half-moons, we colloquially refer to this cat qubit variation as \textit{moon cat}.
In the limit of small \(\lambda \), moon cat states become equivalent to squeezed cat states, and numerical simulations confirm that for \(\lambda \le 1\), the moon cat qubit performs comparably to an ideal squeezed cat qubit (Appendix~\ref{ap:moon_cat_theory_sq_comparison}).

To engineer this dissipator (Eq.~\ref{eq:Moon_dissipator}), we introduce an ancillary buffer mode mediating the coupling between the memory and the environment.
Compared to standard cat stabilization~\cite{mirrahimi_dynamically_2014, leghtas_confining_2015,touzard_coherent_2018, lescanne_exponential_2020,berdou_one_2023, reglade_quantum_2024, marquet_autoparametric_2024,putterman_preserving_2024}, our engineered memory-buffer Hamiltonian contains an additional term proportional to \(\lambda \):
\begin{equation}
    \label{eq:full_hamiltonian}
    \op{H}/\hbar = g_2 \left(\op{a}^2 - \alpha^2 + \lambda (\op{a}^\dag \op{a} - \alpha^2)\right)\op{b}^\dag + h.c.,
\end{equation}
where \(\op{b}\) is the buffer annihilation operator and \(g_2\) is the two-to-one photon exchange rate between the memory and buffer modes.
By operating in the regime \(8|g_2\alpha|\ll\kappa_b\), where \(\kappa_b\) is the buffer single photon loss rate, the buffer can be adiabatically eliminated~\cite{reglade_quantum_2024}, yielding the dissipator (Eq.~\ref{eq:Moon_dissipator}) with \(\kappa_2 = 4|g_2|^2/\kappa_b \) (Appendix~\ref{ap:moon_cat_theory}).

We implement these dynamics on a superconducting circuit platform (Fig.~\ref{fig:fig2}A-C) featuring a lumped-element memory mode at \(\omega_a/2\pi = 1.1\) GHz with intrinsic loss rate \(\kappa_1/2\pi = 2.3\) kHz and mean thermal photon number \(n_{\rm th} = 0.93\), alongside a buffer mode at \(\omega_b/2\pi = 7.9\) GHz with loss rate \(\kappa_b/2\pi = 18.2\) MHz.
The choice of a low frequency memory is an attempt to reduce dielectric losses, at the expense of an elevated mean thermal photon population.
Both modes participate into a non-linear element known as the Asymmetrically Threaded SQUID (ATS) \cite{lescanne_exponential_2020}.
It is composed of two parallel Josephson junctions shunted by an inductance, threaded by two external magnetic fluxes (Fig.~\ref{fig:fig2}A-C) set to a sweet spot.
This element enables the engineering of the necessary odd order Hamiltonian interactions between the memory and the buffer, upon driving with an alternating flux (pump).

The memory-buffer Hamiltonian (Eq.~\ref{eq:full_hamiltonian}) comprises three types of terms to be engineered \(\op{a}^2\op{b}^\dag \), \(\op{b}\) and \(\op{a}^\dag \op{a}\op{b}^\dag \).
The first is induced by a pump at frequency \(\omega_b - 2\omega_a\).
We reach a rate \(g_2/2\pi = 1.3\) MHz resulting in an estimated \(\kappa_2/2\pi = 0.37\) MHz (Appendix~\ref{ap:calibrations_of_parametric_pumps}).
The second term is obtained from a simple resonant drive on the buffer mode with strength \( -g_2 \alpha^2 (1+\lambda)\).
The third term, the only element added to deform the cat states, is a longitudinal type of interaction with strength \( g_2\lambda \).
It is induced by a pump at frequency \(\omega_b\) on the ATS.
Crucially, adding this term did not require any design change compared to a standard cat qubit circuit.
The longitudinal coupling also enables the readout of the memory mean photon number (Appendix~\ref{ap:longitudinal_photon_number_measurement}) and photon number parity used for Wigner tomography~\cite{reglade_quantum_2024}.
Note that deforming the cat states with a fixed amplitude \(\alpha \) requires both the longitudinal coupling and the buffer drive to increase.

\begin{figure*}[ht]
    \centering
    \includegraphics[width=\textwidth]{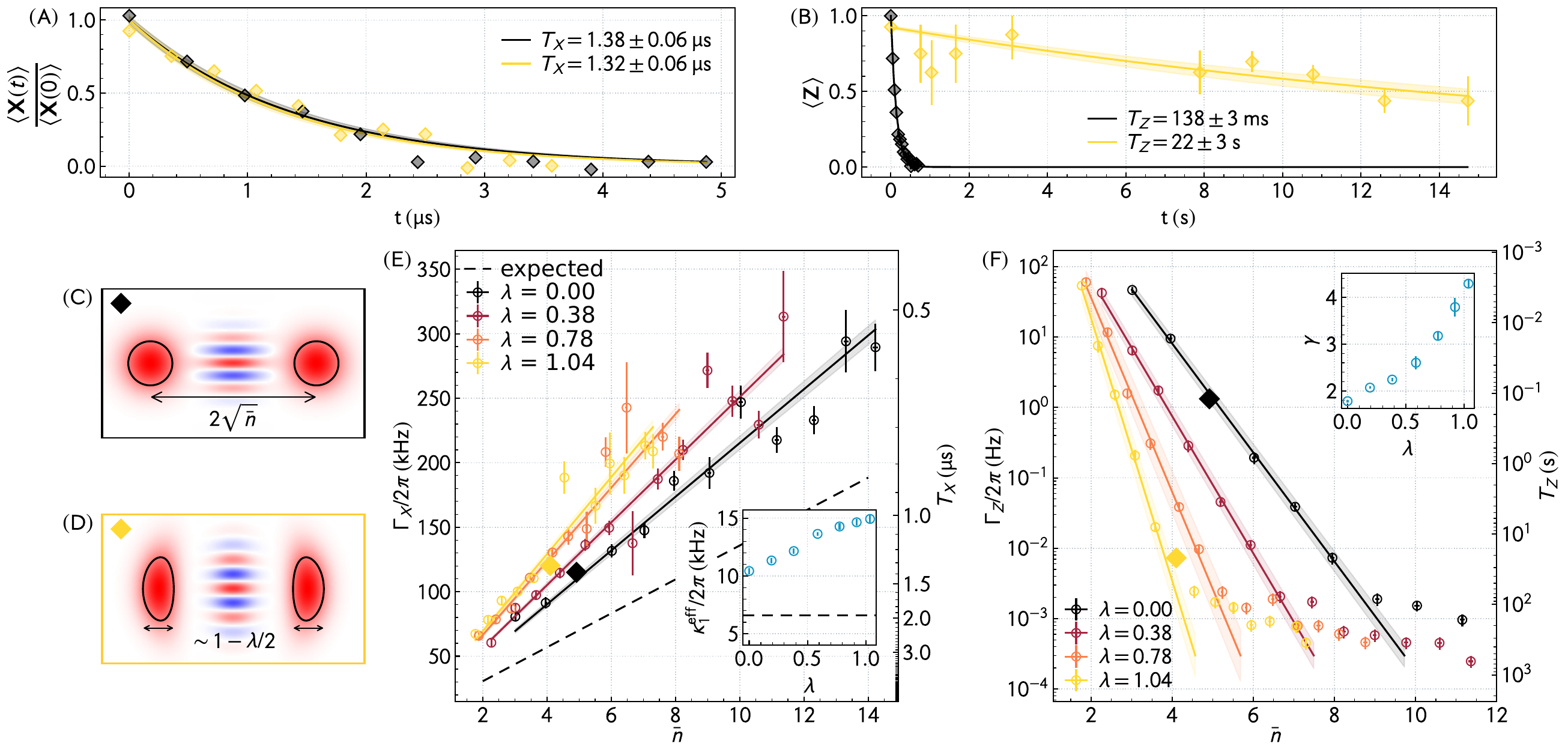}
    \caption{Enhancing bit-flip protection by squeezing.
        (A-B) Measured (diamonds) and fitted (solid lines) \(\braket{\op{X}} \) (resp.~\(\braket{\op{Z}} \)) evolution of a standard cat qubit \(\bigr \{\bar{n}=4.9, \lambda=0\bigl \} \) (black) and a moon cat qubit \(\bigr \{\bar{n}=4.1, \lambda=1.04\bigl \} \) (yellow).
        The phase-flip time \(T_X \) (resp.\ bit-flip time \(T_Z \)) is extracted from the fit.
        The moon cat qubit bit-flip time is 160 times larger than of the standard cat qubit, with the same phase-flip time.
        (C-D) Simulated Wigner functions of a cat state for the two operating points (same color scale as Fig.~\ref{fig:fig1}A-B).
        The solid lines delineate the extent of the basis states at \(\mathcal{W} = 0.5/\pi \).
        In (C), the black arrow indicates how the larger photon number increases the distance between the basis states.
        In (D), the arrow highlights the reduced width of the basis states upon squeezing.
        (E) Measured phase-flip rate (circles) showing a linear increase with mean photon number \(\bar{n}\).
        Solid lines are linear fits and half their slope defines the effective single-photon loss rate, \(\kappa_1^{\rm eff}\).
        The inset displays \(\kappa_1^{\rm eff}\) as a function of \(\lambda \).
        The dashed black lines represent the expected linear scaling and \(\kappa_1^{\rm eff}\) values based on the memory parameters.
        (F) Measured bit-flip rate (circles) plotted on a logarithmic scale.
        Solid lines are exponential fits, \(\Gamma_Z(\bar{n}) = A e^{-\gamma\bar{n}}\), applied to the low photon number region where the exponential scaling factor remains constant.
        For fixed \(\bar{n}\), the decay rate decreases exponentially with increased squeezing of the cat basis state.
        The inset shows the exponential scaling factor \(\gamma \) as a function of \(\lambda \).
        All error bars represent \(\pm 1\sigma \) (1 standard deviation) of uncertainty.
    }\label{fig:fig3}
\end{figure*}

The deformation parameter \(\lambda \) is a complex quantity whose phase determines the squeezing direction.
In particular, \(\lambda \) real and positive (resp.\ negative) yields squeezed (resp.\ anti-squeezed) cat states.
At fixed buffer drive amplitude, this leads to smaller (resp.\ higher) photon number.
We calibrate \(\lambda \) by measuring the memory's steady-state mean photon number across various longitudinal pump amplitudes and phases (Fig.\ref{fig:fig2}D).
A single parameter fit to these data then determines \(\lambda \).
Wigner function measurements at selected \(\lambda \) values (Fig.\ref{fig:fig2}E) confirm that lower (resp.\ higher) photon numbers correspond to squeezing (resp.\ anti-squeezing).
Although the model accurately captures the squeezing regime, it underestimates the photon number in anti-squeezed states, suggesting unmodeled nonlinearities.
Finally, using the fitted \(\lambda \), we reconstruct the cat states Wigner functions from the dissipator's kernel states (Appendix~\ref{ap:lambda_calibration}) and find good agreement with the measured Wigner tomographies especially, in the squeezing regime (Fig.~\ref{fig:fig2}E).

%% file: part2_error_rates_scaling.tex
\section{Idle decay rates}\label{sec:level3}

To highlight the benefits of squeezing on decay rates, we first compare a standard cat and a moon cat with parameters that yield the same phase-flip rate (Fig.~\ref{fig:fig3}C-D).
The decay rates are measured as follows.
We determine the bit-flip rate \(\Gamma_Z \) by fitting the exponential decay of \(\braket{\op{Z}}\) measured on a stabilized cat qubit initialized in \(\ket{\alpha}_\lambda \) or \(\ket{-\alpha}_\lambda \) after an idle time \(t\).
In the moon cat case, we first prepare coherent states with a memory displacement of amplitude \(\alpha \) and then turn on the dissipator to rapidly converge to the deformed states.
Measuring \(\op{Z}\) is done by a memory displacement followed by a photon number measurement~\cite{reglade_quantum_2024}.
Similarly, the phase-flip rate \(\Gamma_X \) is obtained by monitoring the decay of \(\braket{\op{X}}\) after initializing in the moon cat states \(\propto\ket{\alpha}_\lambda + \ket{-\alpha}_\lambda \) (Appendix~\ref{ap:phase_flip_measurement}).
As in standard cats, the initialization is done by turning on the dissipator on vacuum and \(\op{X}\) is merely the photon number parity.
Both cats exhibit the same phase-flip time \(T_X= 1/\Gamma_X = 1.3\;\mu\mathrm{s}\) (Fig.~\ref{fig:fig3}A) but the bit-flip time \(T_Z = 1/\Gamma_Z\) increases dramatically from \(138 \) ms to \(22\) s in the moon case -- a \(160\)-fold improvement (Fig.~\ref{fig:fig3}B).
Because \(T_Z\) can easily reach several seconds, our bit-flip measurement procedure is prohibitively long.
To address this, we sample \(t\) adaptively, using a Bayesian algorithm that maximizes the information flow (Appendix~\ref{ap:bayesian_adaptive_measurement}).

We measured the idling error rates across various cat photon numbers \(\bar{n}\) and deformation parameter \(\lambda \).
The bit-flip rate \(\Gamma_Z\) decreases exponentially with \(\bar{n}\), \(\Gamma_Z(\bar{n}) \sim \exp(-\gamma \bar{n})\) (Fig.~\ref{fig:fig3}F).
Notably, squeezed cat states show a steeper exponential suppression of bit-flips compared to standard cat states.
As \(\lambda \) increases from \(0\) to \(1.04\), \(\gamma \) rises from \(1.8\) to \(4.3\) (Fig.~\ref{fig:fig3}F inset), exceeding the theoretical limit of \(\gamma=2\) for standard cats~\cite{mirrahimi_dynamically_2014, gautier_combined_2022, dubovitskii_bit-flip_2024}.
Despite the slight prefactor increase in \(\Gamma_z(\bar{n})\) for \(\lambda \gtrsim 0.7\), the concurrent increase in scaling factor preserves the improvement in bit-flip time \(T_Z\) (Appendix~\ref{ap:scalings}).
Finally, as in past experiments~\cite{reglade_quantum_2024,putterman_preserving_2024}, we observe a saturation in bit-flip rate below \(10\) mHz that does not depend strongly on the deformation parameter.
Its origin is not yet understood, but it may stem from ionizing impacts that disrupts superconductivity~\cite{harrington_synchronous_2024, mcewen_resolving_2022}.

\begin{figure*}[ht]
    \centering
    \includegraphics[width=\textwidth]{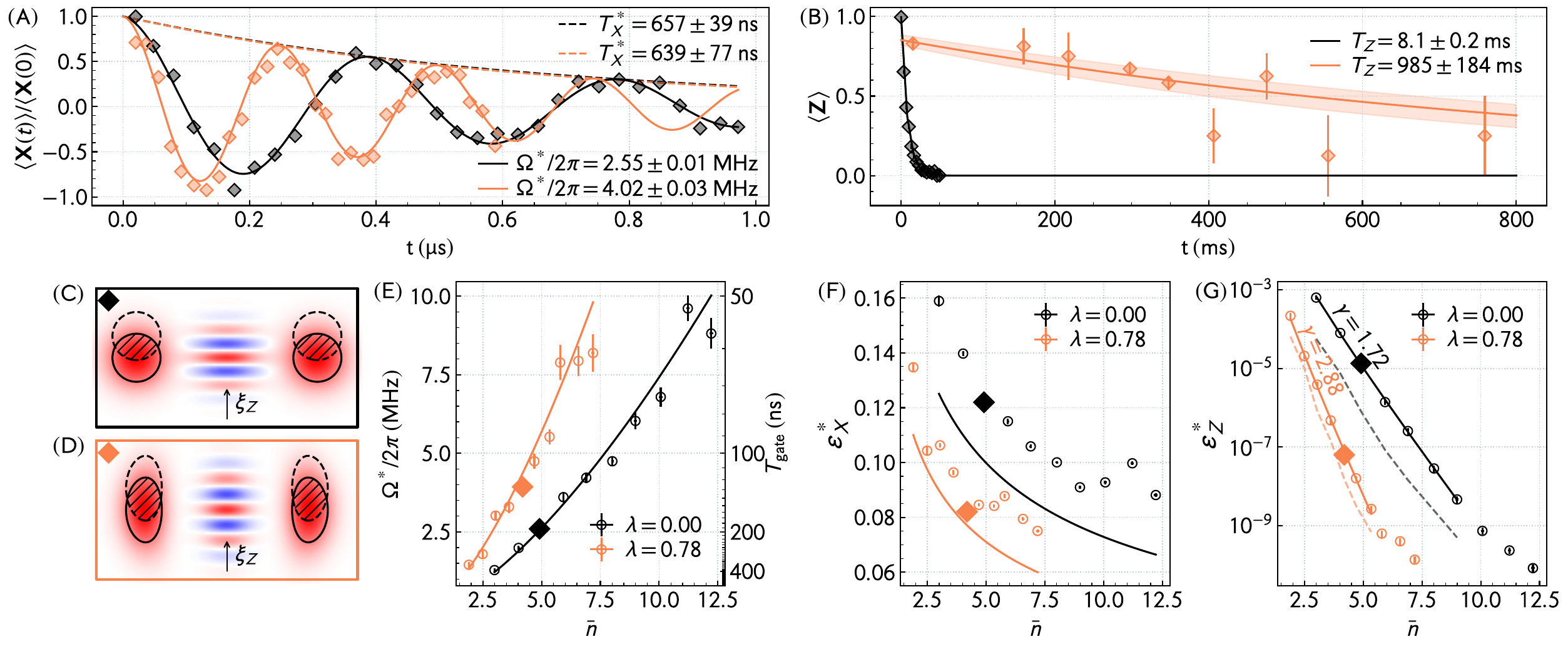}
    \caption{Enhancing \(Z\)-gate fidelity by squeezing.
        (A-B) Measured (diamonds) and fitted (solid lines) \(\braket{\op{X}} \) (resp.\ \(\braket{\op{Z}} \)) evolution during an optimal \(Z\)-gate  (see main text) of a standard cat qubit \(\bigr \{\bar{n}=4.9, \lambda=0\bigl \} \) (black) and a moon cat qubit \(\bigr \{\bar{n}=4.2, \lambda=0.78\bigl \} \) (orange).
        The gate Rabi rate \(\Omega \) and phase-flip probability \(\epsilon_X\) (resp.\ bit-flip probability \(\epsilon_Z\)) are extracted from the fit.
        With the same phase-flip rate, the moon cat qubit gate is faster with less errors.
        (C-D) Simulated Wigner functions of a cat state for the two operating points (same color scale as Fig.~\ref{fig:fig1}A-B).
        The solid lines delineate the extent of the basis states at \(\mathcal{W} = 0.5/\pi \), while the dashed lines correspond to orthogonally displaced basis states.
        The dashed region highlights the overlap between the two states, which is larger for the moon cat, suggesting reduced non-adiabatic errors.
        (E) Optimal gate Rabi rate \(\Omega \) as a function of the cat size for various squeezing strengths.
        (F) and (G) show the phase-flip and bit-flip probability, respectively, for these optimal gates.
        In (E) and (F), solid lines are simulations based on independently measured parameters, while in (G) they represent an exponential fit to the data.
        Dashed lines in (G) mark the baseline bit-flip error, computed from the idling bit-flip error rates (Fig.~\ref{fig:fig3}F).
        All error bars represent \(\pm 1\sigma \) (1 standard deviation) of uncertainty.
    }\label{fig:fig4}
\end{figure*}

Beyond the exponential bit-flip suppression, we confirm the expected linear increase of the phase-flip rate \(\Gamma_X\) with \(\bar{n}\) as shown in Fig.~\ref{fig:fig3}E.
However, the measured slopes exceed the expected values (dashed line) when considering both the bare single photon loss of the memory \(\kappa_1\) and its thermal population \(n_{\rm th}\).
The slope, denoted \(2\kappa_1^{\rm eff}\), should read \(2\kappa_1^{\rm eff} = 2 (\kappa_\uparrow +\kappa_\downarrow) = 2 \kappa_1 (1+2 n_{\rm th})\) since both photon loss \(\kappa_\downarrow \) and photon gain \(\kappa_\uparrow \) are parity flipping processes.
We observe a first increase when stabilizing standard cat states, an effect also noted in previous experiments~\cite{lescanne_exponential_2020}, which here amounts to a factor of \(1.6\).
This increase in \(\kappa_1^{\rm eff}\) may stem from memory heating or spurious phase-flipping processes induced by the two-photon pump~\cite{carde_flux-pump_2024,putterman_preserving_2024}.
As we deform the cats, \(\kappa_1^{\rm eff}\) increases further, by an additional factor of \(1.4\) at \(\lambda = 1.04\), likely for similar reasons involving the longitudinal pump (Fig.~\ref{fig:fig3}E inset).
Despite this, any loss in phase-flip time remains negligible compared to the orders of magnitude gains in bit-flip time, underlining the power of phase space engineering for cat qubits.

%% file: part3_zeno_error_rates.tex
\section{Zeno gates}\label{sec:level4}

Moon cats offer an advantage not only in reducing idling error rates but also in improving the Zeno \(Z\) gate.
This gate is realized by applying a drive at \(\omega_a\) orthogonal to the cat qubit axis, \(\op{H}_Z/\hbar = \xi_Z^*\op{a}+\xi_Z\op{a}^\dag \), where \(\xi_Z\) is the displacement rate.
Under strong stabilization, the basis states remain pinned while the fringes roll in phase space.
The \(Z\) rotation rate reads \(\Omega = 4\alpha \xi_Z\), with negligible corrections due to squeezing (Appendix~\ref{ap:zeno_theory}).

During the \(Z\) gate, qubit coherences are susceptible to both idle processes captured by \(\kappa_1^{\rm eff}\) and non-adiabatic errors, arising from the displacement of the state out of the cat qubit subspace.
While the former are only slightly affected by squeezing (Section\ \ref{sec:level3}), the latter are reduced thanks to two compounding effects.
First, applying a squeezing deformation increases the overlap between the displaced states and the basis states (Fig.~\ref{fig:fig4}C-D), thereby dividing non-adiabatic errors by a factor \(1+\lambda \)~\cite{chamberland_building_2022}.
Second, the cat qubit confinement rate is multiplied by the same factor, offering additional protection against non-adiabatic errors.
Consequently, to first-order, the total phase-flip rate during a \(Z\) rotation is well approximated by
\begin{equation*}
    \Gamma_X = 2\kappa_1^\mathrm{eff}\bar{n} + \frac{2\xi_Z^2}{4\kappa_2\bar{n}(1+\lambda)^2}
\end{equation*}
(Appendix~\ref{ap:zeno_theory}).
Because increasing the drive strength speeds up gates but also increases non-adiabatic errors, there exists an optimal rotation rate \(\Omega^*\) minimizing the total phase-flip errors \(\epsilon_X \)~\cite{chamberland_building_2022} (Appendix~\ref{ap:zeno_theory}).
At this optimum, the non-adiabatic error rate equals the idle error rate, yielding an optimal phase-flip rate \(\Gamma_X^* = 4\kappa_1^{\rm eff}\bar{n}\), which depends only on the cat photon number.

We determine \(\Omega^*\) by measuring \(\braket{\op{X}}\) after the qubit is initialized in a cat state as a function of drive amplitude and duration (Appendix~\ref{ap:zeno_gate_measurement}).
We compare a standard cat and a moon cat that have the same phase-flip time \(T^*_X = 0.6\;\mathrm{\mu s}\) under optimal rotation rate (Fig.~\ref{fig:fig3}C-D).
We find the moon cat optimal gate is \(1.6\) times faster, reducing the \(Z\) gate phase-flip probability \(\epsilon^*_X \) by a factor \(1.5\) (Fig.~\ref{fig:fig3}A).
Simultaneously, the bit-flip time \(T_Z\) improves by a factor of \(120 \), leading to an overall \(220\)-fold improvement in estimated bit-flip probability \(\epsilon^*_Z \) during the gate (Fig.~\ref{fig:fig3}B).

We determined the optimal \(Z-\)gate for various cat photon numbers and deformation parameters.
As expected from the previous discussion, increasing \(\bar{n}\) or \(\lambda \) raises the optimal rotation rate and reduces the gate phase-flip error (Fig.~\ref{fig:fig4}E-F), consistent with the reduced non-adiabatic errors.
The first order formulae reproduce well the behavior of \(\Omega^*\) without any fitting parameter.
However, we observe a discrepancy in the measured \(\epsilon^*_X \) which we attribute to the single mode approximation of the underlying model (Appendix~\ref{ap:two_photon_pump_calibration}).
Notably, this discrepancy diminishes with growing \(\lambda \), probably because squeezing increases confinement, reinforcing the approximations validity.

Crucially, the exponential bit-flip protection persists during gate operation, with moon cats outperforming standard cats (Fig.\ref{fig:fig4}G).
As quantified for the highlighted cat states above, this enhanced performance can be attributed to two key factors: a lower intrinsic idle bit-flip rate yields a \(35\)-fold improvement, and the shorter gate duration associated with moon cats contributes to an additional \(1.6\)-fold.
Additionally, the degradation of the bit-flip rate under the Zeno drive (comparison with dashed lines in Fig.\ref{fig:fig4}G) is less severe for moon cats, providing a further \(4\)-fold enhancement.
This reduced degradation is due to the stronger effective confinement of moon cats, which mitigates the leakage and resulting bit-flips induced by the strong Zeno drive.

%% file: conclusion.tex
\section{\label{sec:level5}Conclusion}

In conclusion, we have demonstrated that squeezing enhances the dissipative protection of cat qubits.
Without altering the circuit design and merely adding one additional parametric pump, we introduce a particular squeezing deformation which gives the moon cat qubit its shape.
This deformation, controlled by the parameter \(\lambda \), acts as an extra knob to tune the basis states overlap without increasing the photon number thereby suppressing bit-flip errors without significantly affecting the phase-flip channel.
In practice, we report an increase in the bit-flip exponential scaling factor up to \(\gamma =4.3\), surpassing the limit \(\gamma=2\) holding for coherent state-based dissipative cats qubits.
For a moon cat with a phase flip time \(T_X = 1.3 \;\mu \mathrm{s}\), we measure a bit-flip time \(T_Z = 22 \; \mathrm{s}\) compared to \(138 \; \mathrm{ms}\) for a standard cat with same phase-flip time, achieving a \(160\)-fold improvement.
Like the two-photon pump, the pump required for squeezing slightly degrades the bare performance of the system.
Nevertheless, the gain from squeezing vastly surpasses this loss.
The \(Z \)-gate benefits from the extended bit-flip time, along with a reduction of non-adiabatic phase-flip errors due to the extra confinement that squeezing provides.
We expect that two qubit gates, such as the CNOT gate, will similarly gain from these performances enhancements.
Our results show that the phase-space distribution of cat qubits can be tailored to significantly improve performances, bringing them closer to practical use in fault-tolerant quantum computing.

%% file: acknowledgments.tex
\begin{acknowledgments}
    The authors thank Steven Touzard for his insight on lowering the cat-qubit frequency and Pierre Guilmin, Gaspard Beugnot, Benjamin Huard, Philippe Campagne-Ibarcq, Emmanuel Flurin and Mazyar Mirrahimi for their fruitful discussions and feedback on the manuscript. We thank the SPEC at CEA Saclay for providing nano-fabrication facilities.
    This research is partially funded by the OpenSuperQ+ Horizon Europe project (grant agreement 101113946).
    This research is partially funded by the \textit{Usine à Chat} project from France 2030.

    \textbf{Data availability:} The data that support the findings of this work are available from the corresponding author upon a request.

    \textbf{Code availability:} The code used for data analysis and visualization is available from the corresponding author upon a request.

    \textbf{Author contributions:} R.R., D.R., J.C. and A.M. conceived the experiment. R.R. and R.L. designed the chip with F.R. providing support. F.R., S.M., M.H. and D.B. designed the microwave simulation software. P.d'A., N.B., A.G., A.T. and P.W.-F. designed the data acquisition software. U.B., P.F., A.G., P.M., S.P. and J.R. designed and built the experimental set-up. J.S. and G.C. fabricated the chip. E.A., G.C., N.C., C.C., L.D., A.D., A.E., E.G., S.J. P.M., T.P., J.S. and J.-L.V. provided experimental support. J.G. and E.G. provided theory support. R.R. measured the device and analyzed the data. R.R., D.R. and J.C. performed theoretical derivations and analysis. R.R., A.M., E.A., R.L. and Z.L. wrote the manuscript with input from all authors.

    \textbf{Corresponding author:} Correspondence should be addressed to A. Murani: anil.murani@alice-bob.com.

    \textbf{Ethics declarations - Competing interests} Authors affiliated with Alice \& Bob (A\&B) have financial interest in the company.
    ZL is a share-holder of A\&B.
    The remaining authors declare no competing interests.
\end{acknowledgments}

%% file: 1_device_fabrication_and_wiring.tex
\section{Device description, fabrication and wiring}\label{ap:fab}

\subsection{Fabrication and packaging}
The device is fabricated by patterning the circuit on a Tantalum layer sputtered on a Sapphire substrate.
Josephson junctions are then evaporated in Aluminum using the Dolan-bridge technique.
In the following we detail the steps of the circuit fabrication process.

\subsubsection{Wafer preparation}
The wafer used for device consists in a 430 \(\mu\mathrm{m}\)-thick sputtered with a 200 nm-thick \(\alpha \)-phase tantalum layer.
At the beginning of the fabrication process the wafer is cleaned in an acetone bath and rinsed with isopropanol (IPA).

\subsubsection{Tantalum patterning}
Patterning of the circuit is realized using optical laser lithography and reactive ion etching.
The wafer is dried for 1 minute on a hotplate at 120\textdegree{}C, let cool down and coated with a 500 nm-thick layer of S1805 resist, which is then patterned with a Heidelberg-\(\mu\)MLA (390-nm laser diode).
The resist is then developed for 1 minute in MF319 and rinsed with water.
Subsequently we use a CF4 gas at 20 sccm and 0.002 mbar to perform reactive ion etching on the patterned wafer.
The sample is then cleaned with ultrasounds in an acetone bath at 45\textdegree{}C and then undergoes an oxygen ashing step (3 minutes, 0.035 mbar, 200 W) to strip the resist residues.

\subsubsection{Josephson junctions}
Josephson junction are patterned with electron beam lithography and evaporated using the Dolan-bridge technique.
To this aim, the wafer is coated with a  bi-layer of MAA EL13 (650 nm, baked 3 min at 185\textdegree{}C) and PMMA A3 (120 nm, baked 30 min at 185\textdegree{}C).
Finally, a thin layer (20 nm) of conductive resist (Electra92) is spinned on top to improve charge-evacuation during lithography.
Junctions are patterned using an electron beam at 20 kV, then the conductive coating is stripped using de-ionized water and the bi-layer is developed for 2 minutes in a 3:1 solution of IPA and de-ionized water at 6 \textdegree{}C.
Finally, a gentle oxygen descum (10s, 0.75mbar, 75W) is performed on the developed wafer, in order to clean the resist mask and remove the resist residues remaining after development.

Junction evaporation is performed in a Plassys electron-beam evaporator.
The wafer is loaded and pumped to a base pressure of 4.5E-9 mbar, then it undergoes an argon milling (15s, 35mA, 500V) at the two evaporation angles \(\pm \)30\textdegree.
The chamber is then re-pumped and the first 35 nm layer of aluminum is evaporated at +30\textdegree{}.
Oxydation is done in a pure-oxygen environment for 10 minutes at a pressure of 20 mbar.
The second 70 nm-thick aluminum layer is evaporated at an angle of -30\textdegree{}and a final oxydation capping layer is grown by exposing the sample to a 100 mbar oxygen atmosphere for 20 minutes.
After the evaporation, the resist is lifted-off for 90 minutes in acetone at 65\textdegree{}C, then sonicated for 2 minutes.

\subsubsection{Device packaging}
To package the chip for measurement, the wafer is diced using a PMMA protective coating, which is then stripped in acetone at 65\textdegree{}C and ultrasounds.
Finally, the chip is mounted in a sample holder and wire-bonded to a printed-circuit-board.

\subsection{Room-temperature setup}\label{ap:room_temp_setup}
\begin{figure*}
    \centering
    \includegraphics[height=0.96\textheight]{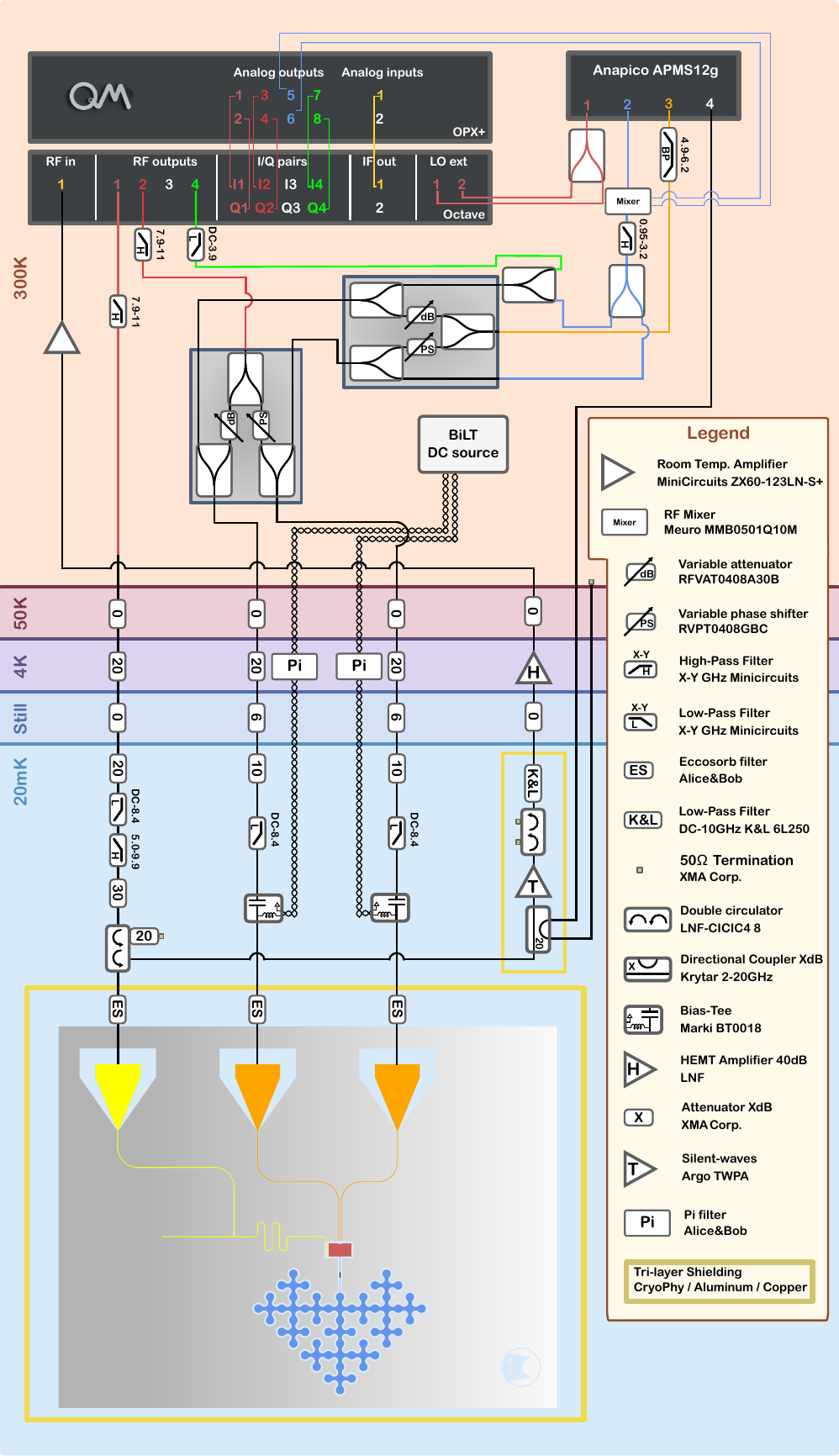}
    \caption{Wiring diagram schematics of the experiment.}\label{fig:fridge_map}
\end{figure*}

The room-temperature (RT) setup used to control and readout the system is showed in Fig.~\ref{fig:fridge_map}.
The DC signals used to bias the ATS are generated using a BiLT voltage source.
The RF signals to drive the buffer, memory, longitudinal interaction, and reset are generated through frequency up-conversion of IQ signals outputted from the analog channels of a Quantum Machine (QM).

As described in Section~\ref{ap:compensation_module}, the two-photon and longitudinal pumps need to be compensated in order to cancel respectively the Stark shift induced on buffer and memory, and the displacement of the buffer.
To this aim, we use compensation modules on the two-photon and longitudinal RT pump lines (Appendix~\ref{ap:compensation_module}).

\subsection{Low-temperature setup}

The mixing chamber setup is composed of attenuation and filtering of the flux bias lines.
A bias-T is used to combine the filtered DC flux bias and the filtered RF pumps.
The filtered buffer signal is connected through a double circulator to the Purcell filter port.
The output signal is amplified by an Argo Traveling Wave Parametric Amplifier (TWPA) and HEMT amplifier.

%% file: 2_circuit_hamiltonian.tex
\section{Circuit Hamiltonian}\label{ap:circuit_hamiltonian}

\subsection{Circuit model}

\begin{figure}
    \centering
    \begin{circuitikz}[scale=0.9]
        \draw
        (0, 0) node[ground]{}
        (0, 0) to[C, l=\(E_{\rm Cb} \)]
        (0, 3) to[short]
        (3, 3) to[short] (3, 2.5)
        (3, 0) to[short]
        (3, 0.5) to[L, l=\(E_{\rm L}\), v=\(\varphi \)] (3, 2.5)
        (3, 0) node[ground]{}
        (3, 0.5) to[short] (1.5, 0.5)
        (1.5, 2.5) to[short] (4.5, 2.5)
        (4.5, 0.5) to[short] (3, 0.5)
        (3, 3) to[L, l=\(E_{\rm Lm}\)]
        (6.5, 3) to[short] (6.5, 2.5)
        (6.5, 0.5) to[C, l=\(E_{\rm Cm}\), v=\(\varphi_{\rm m}\)] (6.5, 2.5)
        (6.5, 0.5) to[short]
        (6.5, 0) node[ground]{};

        \draw
        (1.5, 0.5) to[barrier, l=\(E_{\rm J1} \mkern-18mu\)] (1.5, 2.5)
        (4.5, 0.5) to[barrier, l=\(E_{\rm J2} \mkern-18mu\)] (4.5, 2.5);

        \draw[->]   (1.8, 0.8) circle(1.5mm) node[right] {\( \; \varphi_{\rm L}\)};
        \fill   (1.8, 0.8) circle(0.5mm);

        \draw[->]   (4.2, 0.8) circle(1.5mm) node[left] {\(\varphi_{\rm R}\; \)};
        \fill   (4.2, 0.8) circle(0.5mm);
    \end{circuitikz}
    \caption{Lumped element equivalent of the superconducting circuit.
        The circuit consist of a buffer (resp.\ memory) capacitance of energy \(E_{\rm Cb}\) (resp. \(E_{\rm Cm}\)).
        The superconducting phase difference across the memory capacitance is denoted \(\varphi_{\rm m}\), and it is connected to the ATS trough a junction chain modeled as a linear inductance with energy \(E_{\rm Lm}\).
        The buffer capacitance is shunted to ground by the ATS element.
        The central junction chain is modeled as a linear inductance with energy \(E_{\rm L}\) and phase difference \(\varphi \), flanked by two Josephson junctions with energy \(E_{\rm J1/2}\), forming two loops threaded by externals magnetic fluxes \(\varphi_{\rm L/R}\).
    }\label{fig:circuit}
\end{figure}
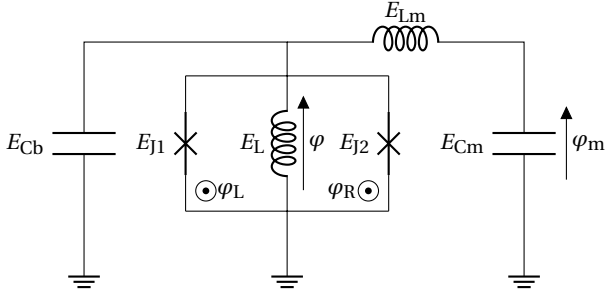

To derive the circuit Hamiltonian, we first simplify the superconducting circuit to a lumped element model (Fig~\ref{fig:circuit}).
The potential energy of the Asymmetrically Threaded SQUID (ATS) is given by,
\begin{multline*}
    U(\varphi, \varphi_\Delta, \varphi_\Sigma) = \frac{E_{\rm L}}{2}\varphi^2 - 2 E_{\rm J} \cos(\varphi_\Sigma)\cos(\varphi_\Delta + \varphi) \\
    + 2\Delta E_{\rm J} \sin(\varphi_\Sigma)\sin(\varphi_\Delta + \varphi),
\end{multline*}
where \(\varphi \) is the superconducting phase difference across the ATS, \(\varphi_{\Sigma/\Delta} = (\varphi_{\rm L} \pm \varphi_{\rm R})/2\), \(E_{\rm L}\) is the ATS inductance energy, \(E_{\rm J} = (E_{\rm J1} + E_{\rm J2})/2\) is the average junction energy and \(\Delta E_{\rm J} = (E_{\rm J1} - E_{\rm J2})/2\) represents the junction asymmetry.
At an arbitrary flux-bias, for small perturbations around the potential minima, the ATS potential can be approximated as \(U(\varphi, \varphi_\Delta, \varphi_\Sigma) \sim  \frac{1}{2}E_{\rm L}^{\rm eff}\varphi^2\), simplifying the circuit by replacing the ATS with an effective inductance.
The equations of motion for the simplified circuit are:
\begin{equation*}
    \begin{pmatrix}
        \ddot{\varphi} \\
        \ddot{\varphi}_{\rm m}
    \end{pmatrix}
    =
    -
    \begin{pmatrix}
        \omega_{\rm a, 0}^2  & -\omega_{\rm a, 0}^2                                        \\
        -\omega_{\rm b, 0}^2 & \omega_{\rm b, 0}^2\left(1+\frac{L_{\rm eff}}{L_{m}}\right)
    \end{pmatrix}
    \begin{pmatrix}
        \varphi \\
        \varphi_{\rm m}
    \end{pmatrix},
\end{equation*}
where \(\omega_{\rm a/b, 0}^2 = 8 E_{\rm Lm} E_{\rm Ca/b}\).
By diagonalizing the dynamics matrix, we obtain the modes frequencies of the system,
\begin{multline*}
    \omega_{\rm a/b}^2 = \frac{\omega_{\rm a,0}^2}{2} + \frac{\omega_{\rm b,0}^2}{2} \left(1+\frac{L_{\rm eff}}{L_{m}}\right) \\
    \pm \sqrt{\frac{\omega_{\rm a,0}^4}{4} + \frac{\omega_{\rm b,0}^4}{4}{\left(1+\frac{L_{\rm eff}}{L_{m}}\right)}^2 + \frac{\omega_{\rm a,0}^2 \omega_{\rm b,0}^2}{2} \left(1-\frac{L_{\rm eff}}{L_{m}}\right)}.
\end{multline*}

\subsection{Flux-map fit}
\begin{figure}
    \centering
    \includegraphics[width=0.48\textwidth]{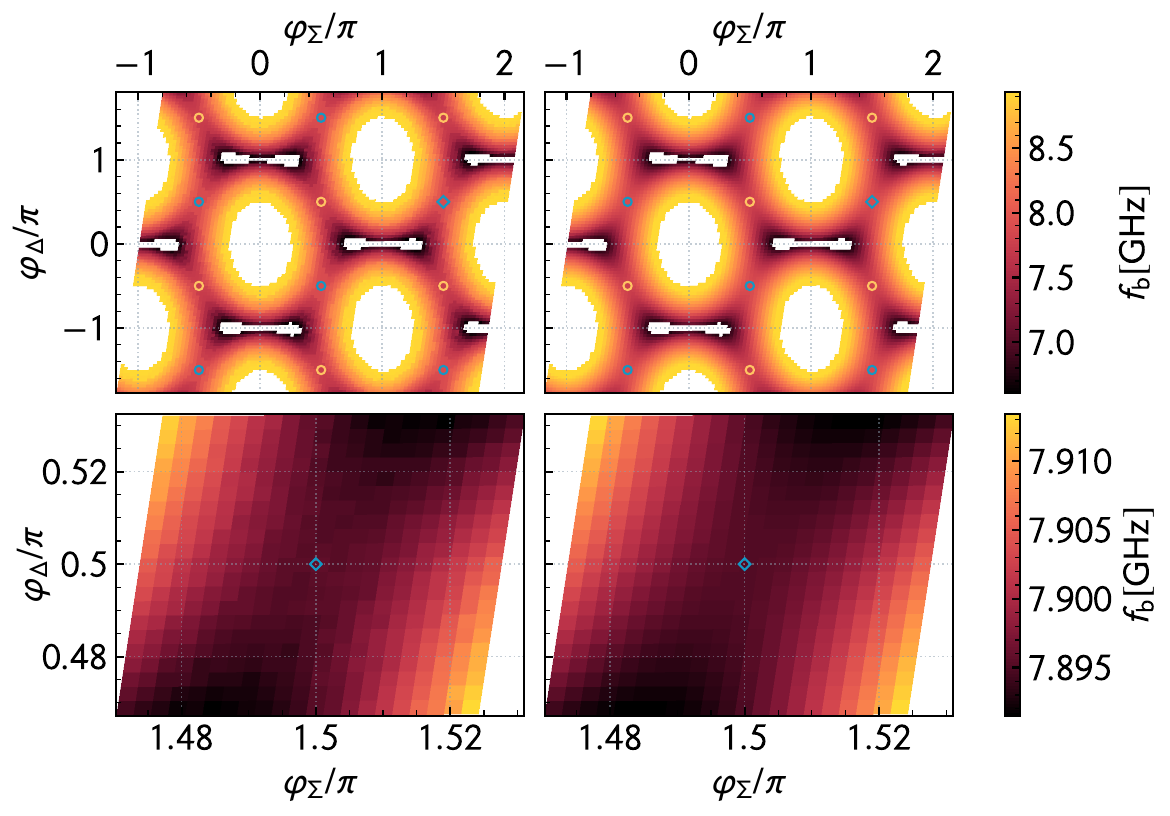}
    \caption{Measured (left) and simulated (right) buffer frequency as a function of the ATS flux bias.
        Circles are placed at the saddle points, with colors (orange or red) indicating the two non-equivalent saddle points resulting from the side junction asymmetry.
        The red diamond marks the experimental working point at \((\varphi_\Sigma, \varphi_\Delta) = (3\pi/2, \pi/2)\).
        The bottom panels provide zoomed-in views of this saddle point.
    }\label{fig:flux_map}
\end{figure}

To thread the ATS loops with magnetic flux, a DC current is sent through two coplanar waveguide (CPW) near the ATS (Fig.~\ref{fig:fig2}B in main text).
The ATS flux bias can be expressed as a function of the DC bias by the following formula:
\begin{equation*}
    \begin{pmatrix}
        \varphi_\Sigma \\
        \varphi_\Delta
    \end{pmatrix}
    = \frac{1}{2}
    T \cdot M
    \begin{pmatrix}
        -I_{\rm L} \\
        I_{\rm R}
    \end{pmatrix}
    + \begin{pmatrix}
        \varphi_{\Sigma, 0} \\
        \varphi_{\Delta, 0}
    \end{pmatrix},
\end{equation*}
with transformation matrix
\(T = \begin{pmatrix}
    1 & 1  \\
    1 & -1
\end{pmatrix}\) and mutual inductance matrix
\(
M = \begin{pmatrix}
    L_L    & M_{LR} \\
    M_{RL} & L_R
\end{pmatrix}\).
The negative sign on \(I_{\rm L}\) account for the flux lines design, while the offset vector compensates for any trapped static magnetic fluxes.
Using this formulas, we can fit the buffer frequency as a function of the flux bias (Fig.~\ref{fig:flux_map}), and the extracted parameters are summarized in Table~\ref{tab:flux_map_parameters}.
For the remainder of the text, the static part of the magnetic flux is set to the frequencies saddle point \((\varphi_\Sigma, \varphi_\Delta) = (3\pi/2, \pi/2)\).
The modes parameters at this working point are also listed in Table~\ref{tab:flux_map_parameters}.
\begin{table}
    \begin{center}
        \setlength{\tabcolsep}{8pt}
        \renewcommand{\arraystretch}{1.1}
        \label{tab:flux_map_parameters}
        \begin{tabular}{|c|c|c|c|}
            \hline
            \(E_{\rm Ca}/h\)         & \(9.6 \;\mathrm{MHz}\)  & \(E_{\rm Cb}/h\)               & \(110 \;\mathrm{MHz}\)  \\
            \hline
            \(E_{\rm Lm}/h\)         & \(35.3 \;\mathrm{GHz}\) & \(E_{\rm L}/h\)                & \(33.0 \;\mathrm{GHz}\) \\
            \hline
            \(E_{\rm J}/h\)          & \(18.1 \;\mathrm{GHz}\) & \(\Delta E_{\rm J}/E_{\rm J}\) & \(-2.95 \% \)           \\
            \hline
            \hline
            \(\omega_{\rm a}/2\pi \) & \(1.08 \;\mathrm{GHz}\) & \(\omega_{\rm b}/2\pi \)       & \(7.90 \;\mathrm{GHz}\) \\
            \hline
            \(\kappa_{\rm a}/2\pi \) & \(2.3 \;\mathrm{kHz}\)  & \(\kappa_{\rm b}/2\pi \)       & \(18.2 \;\mathrm{MHz}\) \\
            \hline
            \(\varphi_{\rm a}\)      & \(0.098\)               & \(\varphi_{\rm b}\)            & \(0.234\)               \\
            \hline
        \end{tabular}
        \caption{Superconducting circuit parameters and mode parameters at the experiment working point.
            The capacitance energies are simulated using microwave simulations, while the other circuit parameters are extracted from the flux map fit (Fig.~\ref{fig:flux_map}).
            The modes frequencies and loss rates were measured at the experiment working point, and the zero-point fluctuations were deduced from the modes parameters.}
    \end{center}
\end{table}

\subsection{Pumping of the ATS}\label{ap:ats_pumping}

In the following sections, we combine DC flux bias with AC flux pumps, yielding \(\varphi_\Sigma(t) = \frac{3\pi}{2} + \delta(t)\) and \(\varphi_\Delta(t) =  -\frac{\pi}{2} + \sigma(t)\).
By simplifying the ATS potential, the system Hamiltonian can be expressed as:
\begin{equation*}
    \begin{aligned}
        \op{H} =       & \hbar \omega_a \op{a}^\dag \op{a} + \hbar \omega_b \op{b}^\dag \op{b}  + \frac{E_L}{2}{(\op{\varphi} - \delta(t))}^2 - \frac{E_L}{2}\op{\varphi}^2 \\
                       & + 2E_{\rm J}\sin(\sigma(t))\sin(\op{\varphi}) - 2\Delta E_{\rm J}\left(\cos(\sigma(t))\cos(\op{\varphi}) + \frac{\op{\varphi}}{2}\right),          \\
        \op{\varphi} = & \varphi_a(\op{a}+\op{a}^\dag) + \varphi_b(\op{b}+\op{b}^\dag).
    \end{aligned}
\end{equation*}
Here \(\varphi_{a}\) and \(\varphi_{b}\) represents the zero-point fluctuations of the memory and buffer modes in the central inductance of the ATS.
The time-dependant part of the differential flux, \(\delta(t) \), is placed into the ATS array superconducting phase difference~\cite{you_circuit_2019}.
Neglecting the asymmetry of the side junctions and assuming small time-dependant flux variations, \(|\sigma(t)| \ll 1\), the ATS potential can be approximated as \(\op{U}(t) = -E_L\delta(t)\op{\varphi} + 2E_{\rm J}\sigma(t) \sin(\op{\varphi})\).
Finally, since the zero-point fluctuations of the memory and buffer modes in the ATS central inductance are small, \(\varphi_a, \varphi_b \ll 1\), we can further simplify the potential to:
\begin{equation}
    \label{eq:reduced_ats_potential}
    \op{U}(t) = (2E_J\sigma(t) - E_L\delta(t))\op{\varphi} - \frac{E_{\rm J}}{3} \sigma(t) \op{\varphi}^3.
\end{equation}

In the following sections, we consider a coherent modulation of the ATS fluxes with angular frequency \(\omega_p \) and respective amplitudes \(\sigma \) and \(\delta \), resulting in \(\sigma(t) = \sigma \mathcal{R}(e^{i\omega_p t})\) and \(\delta(t) = \mathcal{R}(\delta e^{i\omega_p t})\).
The first term in Eq.~\ref{eq:reduced_ats_potential}, should be interpreted as a linear drive on the modes with a complex amplitude of:
\begin{equation}
    \label{eq:xi1_expr}
    \xi_{1, a/b} = \varphi_{a/b}(2E_{\rm J}\sigma - E_{\rm L}\delta)/\hbar,
\end{equation}
which is the pre-dominant effect of the pump on the system.
This is the sole effect when the \(\sigma = 0\), a condition used to create a memory mode drive by applying a pump with angular frequency \(\omega_a\) in the differential mode of the ATS.
In a general case, this displacement cancels when the two pumps are in phase and \(\sigma/\delta = 2E_{\rm J}/E_{\rm L}\).
In the frame of the displaced modes, \(\op{a} \rightarrow \op{\tilde{a}} + \zeta_a e^{-i \omega_p t}\) and \(\op{b} \rightarrow \op{\tilde{b}} + \zeta_b e^{-i \omega_p t}\), where:
\begin{equation*}
    \zeta_{a/b} \underset{t \gg 1/\kappa_{a/b}}{\longrightarrow} \frac{-i\xi_{1, a/b}}{2i(\omega_{a/b} - \omega_p) + \kappa_{a/b}},
\end{equation*}
after applying a rotating-wave approximation (RWA), the potential reads \(U = \hbar \Delta_a \op{\tilde{a}}^\dag \op{\tilde{a}} + \hbar \Delta_b \op{\tilde{b}}^\dag \op{\tilde{b}} + \frac{E_{\rm J}}{3}\sigma(t)\op{\varphi}^3\).
The frequency Stark-shift induced by the pump on the modes is given by:
\begin{equation}
    \label{eq:stark_shift_expr}
    \Delta_{a/b} = \frac{2}{3}\varphi_{a/b}^2E_{\rm J}\sigma (\varphi_a\mathcal{R}(\zeta_a) + \varphi_b\mathcal{R}(\zeta_b)),
\end{equation}
which vanishes when the linear drive term cancels out.
The Hamiltonian of the systems finally reads,
\begin{equation}
    \label{eq:pumped_hamiltonian}
    \op{H} = \hbar \omega_{a, 1} \op{a}^\dag \op{a} + \hbar \omega_{b, 1} \op{b}^\dag \op{b} + \frac{E_{\rm J}}{3}\sigma(t)\op{\varphi}^3,
\end{equation}
where the tilde notation has been omitted, and the Stark-shift has been absorbed into the mode frequencies \(\omega_{a/b, 1} = \omega_{a/b} + \Delta_{a/b}\).

We now consider a pump with angular frequency \(\omega_p = \omega_{b, 1} - 2\omega_{a, 1} \) and move to the frame of each modes.
After performing a RWA on the Hamiltonian from Eq.~\ref{eq:pumped_hamiltonian} we find:
\begin{equation}
    \op{H}_{\omega_p = \omega_{b, 1} - 2\omega_{a, 1}} = \hbar g_2 \left( \op{a}^2\op{b}^\dag + \op{a}^{\dag2}\op{b} \right),
\end{equation}
where \(g_2 = \frac{1}{2} E_{\rm J} \sigma \varphi_{a}^2 \varphi_b / \hbar \).
In this case, the pump mediates a two-to-one photon exchange between the memory and buffer modes.
Now considering a pump with angular frequency \(\omega_p = \omega_{b, 1}\), and applying the same procedure, we obtain:
\begin{equation}
    \op{H}_{\omega_p = \omega_{b, 1}} = \hbar g_{\rm l} \op{a}^\dag \op{a} (\op{b} + \op{b}^\dag) + \hbar g_{\rm sp}(\op{b}^{\dag2}\op{b} + \op{b}^\dag\op{b}^2).
\end{equation}
Here, the first term, with strength \(g_{\rm l} = E_{\rm J} \sigma \varphi_{a}^2 \varphi_b / \hbar \), represents the longitudinal coupling, while the second term, with strength \(g_{\rm sp} = \frac{1}{2}E_{\rm J} \sigma \varphi_{b}^3 / \hbar \), is a spurious term.

Finally, considering the junction asymmetry term we previously neglected, and assuming small pump amplitude and phase, the potential simplifies to:
\begin{equation*}
    \op{U}(t) = 2\Delta E_{\rm J} \sigma(t)^2 (1- \frac{1}{2}\op{\varphi}^2).
\end{equation*}
For a pump with angular frequency \(\omega_p = (\omega_{b, 1} - \omega_{a, 1})/2\), and after performing a rotating wave approximation, we obtain a beam-splitter Hamiltonian between the buffer and memory mode: \(\op{H}_{\omega_p = (\omega_{b, 1} - \omega_{a, 1})/2} = \hbar g_{\rm reset}(\op{a}\op{b}^\dag + \op{a}^\dag\op{b})\), where \(g_{\rm reset} = \Delta E_{\rm J} \sigma^2 \varphi_a \varphi_b / 8\hbar \).
In the regime where \(g_{\rm reset} \ll \kappa_b\), it induces a dissipation rate for the memory mode, \(\kappa_1^{\rm reset} = 4g_{\rm reset}^2/\kappa_b\)~\cite{reglade_quantum_2024}.
In the experiment, this mechanism was used to reset and cool the memory mode.

%% file: 3_calibrations_of_parametric_pumps.tex
\section{Calibration of parametric pumps}\label{ap:calibrations_of_parametric_pumps}

\begin{table}
    \begin{center}
        \setlength{\tabcolsep}{8pt}
        \renewcommand{\arraystretch}{1.15}
        \label{tab:system_parameters}
        \begin{tabular}{|c|c|c|c|}
            \hline
            \(g_2/2\pi \)                  & \(1.3 \;\mathrm{MHz}\)  & \(\kappa_2/2\pi \)                 & \(0.37 \;\mathrm{MHz}\)        \\
            \hline
            \(g_{\rm l}^{\rm meas}/2\pi \) & \(0.76 \;\mathrm{MHz}\) & \(g_{\rm l}/2\pi \)                & \(0\) -- \(1.3\;\mathrm{MHz}\) \\
            \hline
            \hline
            \(\kappa_{\rm a}/2\pi \)       & \(2.3 \;\mathrm{kHz}\)  & \(\kappa_{\rm a}^{\rm 2ph}/2\pi \) & \(3.65 \;\mathrm{kHz}\)        \\
            \hline
            \(\kappa_{\phi}/2\pi \)        & \(10.5 \;\mathrm{kHz}\) & \(\kappa_{\phi}^{\rm 2ph}/2\pi \)  & \(18.8 \;\mathrm{kHz}\)        \\
            \hline
            \hline
            \(K_4/2\pi \)                  & \(14.7 \;\mathrm{kHz}\) & \(K_6/2\pi \)                      & \(6.8 \;\mathrm{Hz}\)          \\
            \hline
        \end{tabular}
        \caption{System parameters calibrated in the following sections.
            The measurement longitudinal amplitude strength is noted \(g_{\rm l}^{\rm meas}\), while the one used for Moon cat stabilization is noted \(g_{\rm l}\).
            The memory loss rate (resp.\ dephasing rate) when the two-photon pump is played is noted \(\kappa_{\rm a}^{\rm 2ph}\) (resp. \(\kappa_{\phi}^{\rm 2ph}\)).
        }
    \end{center}
\end{table}

\subsection{Longitudinal pump for photon number measurement}\label{ap:longitudinal_photon_number_measurement}

\begin{figure}
    \centering
    \includegraphics[width=0.48\textwidth]{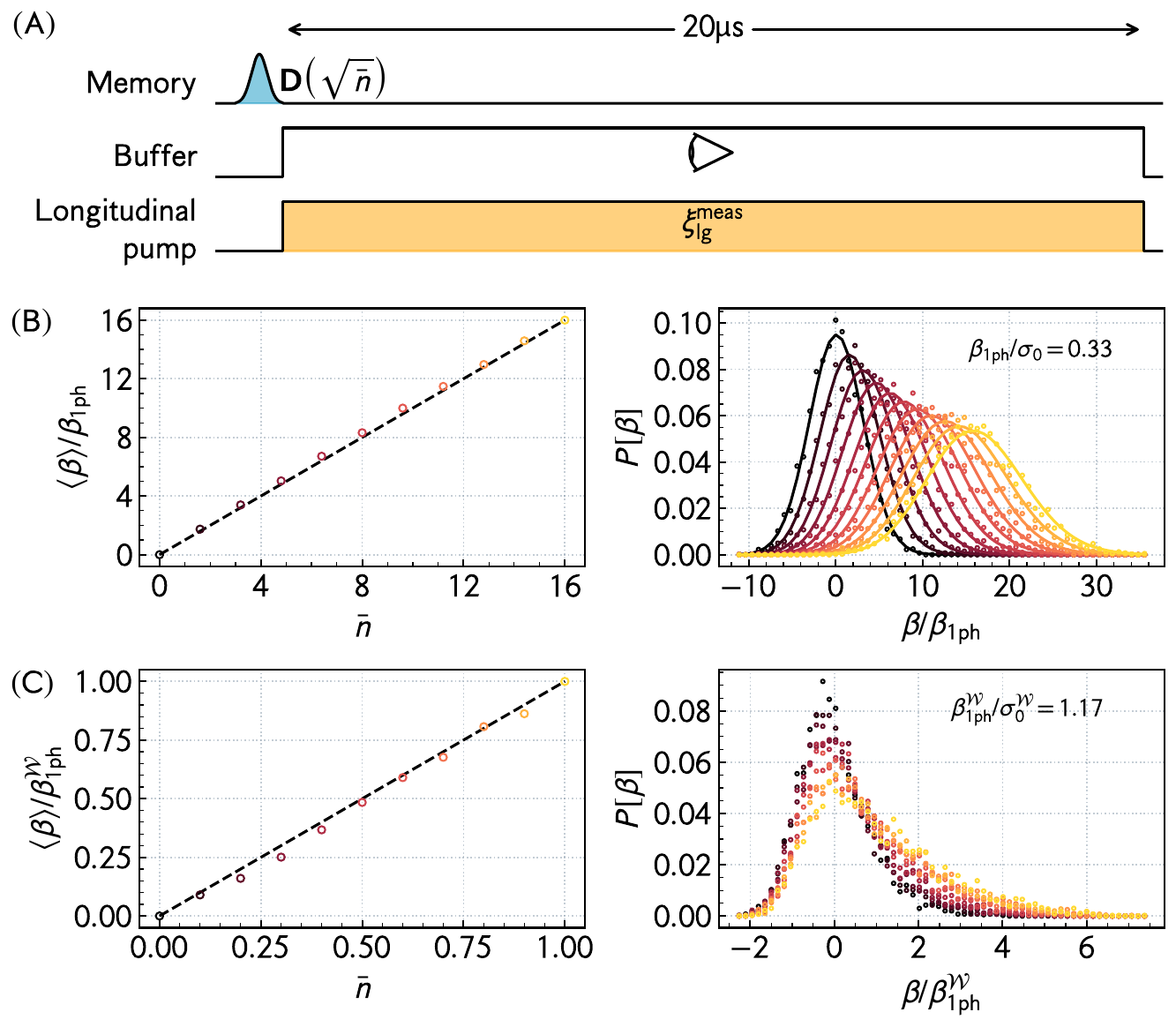}
    \caption{Longitudinal memory photon number readout calibration.
        (A) Pulse sequence for the photon number measurement calibration.
        A gaussian pulse displaces the memory state into a coherent state.
        The longitudinal pump is played while measuring the output signal of the buffer mode.
        (B) The buffer output signal \(\beta \), rescaled by the average signal for one photon in the memory \(\beta_{\rm 1ph}\), as a function of the mean number of photons in the memory.
        The left panel shows the linear relationship between the mean signal value and \(\bar{n}\).
        The right panel displays the measurement histograms (dots) for each \(\bar{n}\) values.
        The solid lines represents the fit to the histograms.
        The readout signal-to-noise ratio (SNR) is estimated by the ratio of the average signal for one photon to the standard deviation of the 0 photon histogram \(\sigma_0\).
        (C) Same panels as in (B), but for a different regime of the longitudinal pump, optimized to measure the Wigner function of the memory states, denoted by the subscript \(\mathcal{W}\).
        The histograms in the right panel deviate from the model due to the influence of the spurious term, which distorts the histograms.
        However, the relationship between the mean signal value and \(\bar{n}\) remains linear, and the SNR is increased by a factor of \(3.5\).
    }\label{fig:longtitudinal_calibration}
\end{figure}

In this section, we describe the calibration of the photon number measurement using the longitudinal pump.
When the longitudinal pump is applied the system's dynamics in the frame of the modes is governed by:
\begin{equation*}
    \begin{aligned}
        \op{H_{\rm l}} = & \hbar \xi_1 (\op{b} + \op{b}^\dag) + \hbar g_{\rm l} \op{a}^\dag \op{a} (\op{b} + \op{b}^\dag) + \hbar g_{\rm sp}(\op{b}^{\dag2}\op{b} + \op{b}^\dag\op{b}^2), \\
        \op{L_{\rm a}} = & \sqrt{\kappa_{\rm a}} \op{a} \; \; \text{and} \; \; \op{L_{\rm b}} = \sqrt{\kappa_{\rm b}} \op{b}.
    \end{aligned}
\end{equation*}
The first term represents a linear displacement on the buffer mode.
The second term is the longitudinal interaction to be engineered, displacing the buffer modes proportionally to the number of photons in the memory.
The third term is a spurious term, acting as a self-amplification mechanism on the buffer.
If we were to neglect this spurious term as well as the linear drive, the steady-state buffer operator in the Heisenberg picture would be \(\op{b_{\rm ss}} = -2i g_{\rm l} \op{a}^\dag\op{a}/\kappa_{\rm b}\).
The spurious term on this state can be seen as a drive with amplitude \(g_{\rm sp} \langle \op{b_{\rm ss}}^\dag \op{b_{\rm ss}} \rangle \) in a mean field approximation.
To justify if we can neglect it, the displacement strength must be much smaller than the effective longitudinal interaction strength, \( \frac{g_{\rm sp} \langle \op{b_{\rm ss}}^\dag \op{b_{\rm ss}} \rangle} {g_{\rm l} \langle \op{a}^\dag \op{a} \rangle}  \ll 1\).
This condition limits the maximum number of photons that can be linearly mapped to the buffer mode to \(\bar{n} \ll 2 \frac{\kappa_{\rm b}^2}{4 g_l^2} \frac{\varphi_{\rm a}^2}{\varphi_{\rm b}^2}\), which evaluates to approximately 50 photons with our system parameters.

In the following we will place ourselves in the linear regime, where the spurious term can be neglected.
In this regime, we relate the measurement of the buffer mode to the memory photon number measurement.
Owing to its fast dynamics, we assume that the buffer reaches its steady state immediately after activating the longitudinal pump.
Furthermore, since the memory lifetime is much longer than the measurement duration, we neglect the losses of the memory.
In this regime, the buffer operator is given by \(\op{b_{\rm ss}} = -2i (\xi_1 \op{1} + g_{\rm l} \op{a}^\dag\op{a})/\kappa_{\rm b}\).
Without loss of generality, we assume \(\langle \op{b_{\rm ss}} \rangle \) to be real.
This stochastic differential equation for this quadrature signal \(y(t)\) can be written as~\cite{wiseman_quantum_2009}:
\begin{equation*}
    \begin{aligned}
        dy_t & = \sqrt{\eta} \sqrt{\frac{\kappa_b}{2}} \mathrm{Tr}(\op{b_{\rm ss}} \rho_t + \rho_t \op{b_{\rm ss}}^\dag) dt + dW_t                            \\
             & = \sqrt{\eta} \sqrt{\frac{\kappa_{\rm l}}{2}} 2 \mathrm{Tr}(\op{a}^\dag\op{a} \rho_t) dt + \sqrt{8 \eta}\frac{\xi_1}{\kappa_{\rm b}} dt+ dW_t,
    \end{aligned}
\end{equation*}
where \(\rho_t\) is the state density matrix at time \(t \), \(\eta \) is the measurement quantum efficiency, \(dW_t \) is the Wiener process satisfying Ito rules \(dW_t^2 = dt\), and \(\kappa_{\rm l} = \frac{4g_{\rm l}^2}{\kappa_b}\) is the effective coupling rate for the photon number measurement channel.
The measurement signal is the integrated signal over a time \(T \), multiplied by an amplification gain factor \(G\):
\begin{equation*}
    \begin{aligned}
        S = & \frac{\sqrt{G}}{T} \int_{0}^{T} dy_t                                                                                                                                   \\
        =   & \sqrt{8\eta G}\frac{\xi_1}{\kappa_{\rm b}} + \frac{\sqrt{G}}{T} \int_{0}^{T} \left(\sqrt{2\eta \kappa_{\rm l}} \mathrm{Tr}(\op{a}^\dag\op{a} \rho_t) dt + dW_t\right).
    \end{aligned}
\end{equation*}
The linear displacement on the buffer produces an offset in the final signal, which is experimentally calibrated by measuring the signal for an empty memory.
We will now remove this offset to simplify the expressions.

For the following calculations, we model the weak measurement as a strong measurement of the memory Fock states for the initial time.
This approach effectively assumes that the projection time \({(\eta \kappa_l)}^{-1}\) is negligible compared to the measurement time \(T\).
For a coherent state in the memory with amplitude \(\alpha \), the projection probabilities onto the Fock state are given by \(P(n) = |\braket{n|\alpha}|^2 = \frac{|\alpha|^{2n}}{n!} e^{-|\alpha|^2}\).
The measurement signal \(S_n\) is then integrated and follows a normal distribution centered at \(\upsilon n = \sqrt{2\eta G \kappa_l} n\) with a variance \(\sigma_0^2 = G/T\).
The total signal probability distribution is the sum of the signals for each Fock state, weighted by the Poisson distribution of the Fock states:
\begin{equation*}
    f_{S}(s) = \sum_n f_{S_n}(s)P(n) = \frac{e^{-|\alpha|^2}}{\sqrt{2\pi\sigma_0^2}}\sum_n e^{-\frac{{(s-\upsilon n)}^2}{2\sigma_0^2}}\frac{|\alpha|^{2n}}{n!}.
\end{equation*}
It follows that \(\braket{S} = |\alpha|^2\upsilon \) and \(\sigma_S^2 = \sigma_0^2 + \upsilon^2|\alpha|^2\).
This expression can be generalized to any memory state by replacing the Poisson distribution with the memory state distribution.
Using a displaced thermal state, we are able to fit the measurement histogram (Fig.~\ref{fig:longtitudinal_calibration}B).

To maximize the signal-to-noise ratio in the very low memory photon numbers regime, in particular for the use case of parity measurements, we select a different working point for the longitudinal pump compared to the regime used for large photon number measurement.
Specifically, we increase the linear drive strength on the buffer mode, allowing the buffer state to reach amplitudes where the spurious terms can no longer be neglected.
This approach leverages the spurious term as a signal amplifier, enhancing the signal-to-noise ratio at the cost of distorting the readout histograms compared to the previous theoretical model (Fig.~\ref{fig:longtitudinal_calibration}C).

\subsection{Two-photon pump}\label{ap:two_photon_pump_calibration}

\begin{figure}
    \centering
    \includegraphics[width=0.48\textwidth]{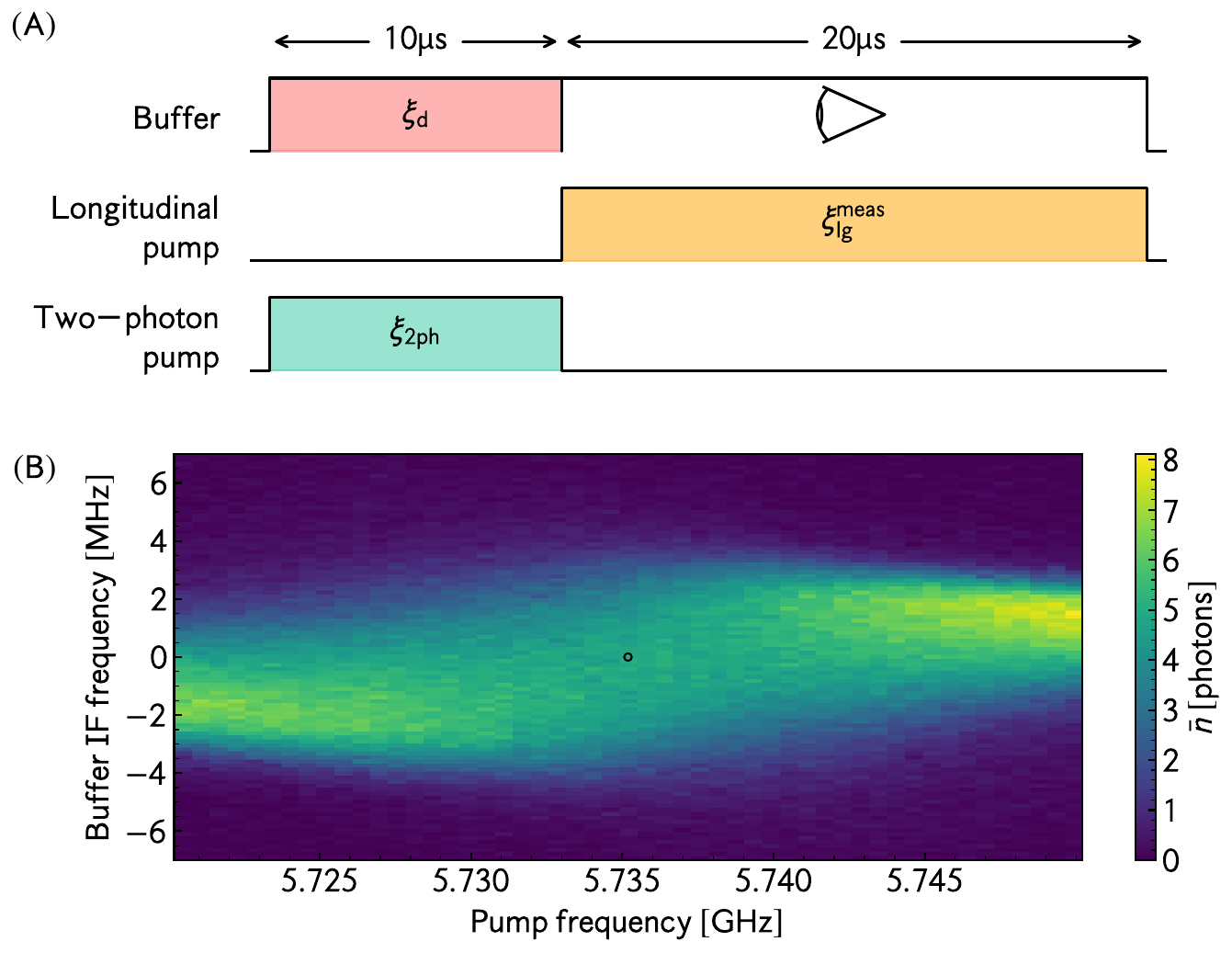}
    \caption{Calibration of the two-photon pump and buffer frequencies.
        The steady-state mean photon number (color) in the memory mode is measured as a function of the two-photon pump frequency and  the buffer intermediary frequency (IF).
        The frequencies are calibrated by locating the saddle point in the experimental results (black circle).
    }\label{fig:diamond}
\end{figure}

In this section we detail the calibration process for the two-photon pump.

First, we calibrate the two-photon pump and buffer drive frequencies \(\omega_{\rm p}\), \(\omega_{\rm d}\). We recall that our buffer drive is generated with an arbitrary waveform generator at intermediate frequencies (IF), up-converted from an local oscillator reference tone (LO), such that the buffer is driven at frequency \(\omega_{\rm d} = \omega_{\rm d}^{\rm LO} + \omega_{\rm d}^{\rm IF}\) (Appendix~\ref{ap:room_temp_setup}).
We measure the steady-state mean photon number \(\bar{n}\) in the memory mode using the longitudinal interaction (see main text), as a function of \(\omega_{\rm p}\) and \(\omega_{\rm d}^{\rm IF}\), while  \(\omega_{\rm d}^{\rm LO}\) is tied to \(\omega_p\) by the relation \(\omega_{\rm d}^{\rm LO} = \omega_{\rm p} + 2 \omega_{\rm a}\), where \(\omega_{\rm a}\) is the memory mode frequency, dressed by the presence of the pumps~\cite{berdou_one_2023}.
When the two-photon pump is detuned, the strength of the effective two-photon interaction term \(g_2\) is reduced, causing an increase of \(\bar{n}\).
Conversely, when the buffer pump is detuned, the effective buffer drive strength decreases, leading to a drop in \(\bar{n}\).
These two effects create a saddle point in \(\bar{n}\) which can be used to calibrate \(\omega_{\rm p}\) and \(\omega_{\rm d}\) (Fig.~\ref{fig:diamond}).

\begin{figure}
    \centering
    \includegraphics[width=0.48\textwidth]{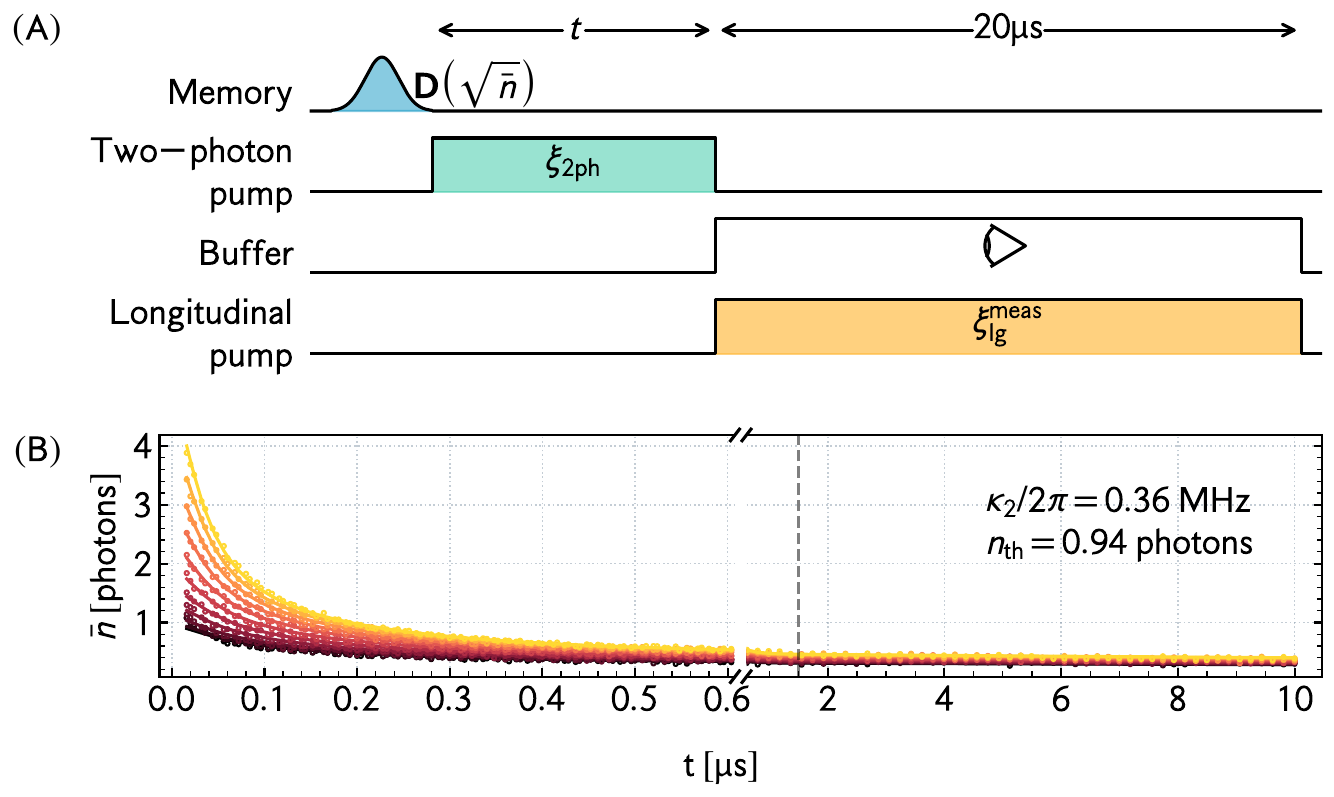}
    \caption{Memory state deflation measurement.
        (A) Pulse sequence of the experiment: a gaussian pulse displaces the initial memory thermal state, with a mean photon number \(n_{\rm th}\), by \(\sqrt{\bar{n}}\).
        The two-photon pump is then played for a variable duration \(t\), and finally, the memory photon number is measured by activating the longitudinal pump and recording the buffer output field.
        (B) Measured memory photon number (circles) as a function of the two-photon pump duration and initial displacement (color).
        The solid lines represent the results of a quantum fit to the data.
        The dashed black lines indicate the selected deflation time used for mapping the parity of an arbitrary state onto the photon number in the 0/1 Fock manifold.
    }\label{fig:kappa_2_cal}
\end{figure}

Next we displace the memory state and apply the two-photon pump for varying durations, and then measure \(\bar{n}\) (Fig.~\ref{fig:kappa_2_cal}).
For this measurement, the memory is prepared in a thermal state using a passive reset.
The trajectories are fitted using a quantum model with the following Hamiltonian and Lindblad loss operators:
\begin{equation*}
    \begin{split}
         & \op{H} = \hbar g_2^*\op{a}^2\op{b}^\dag + \hbar g_2\op{a}^{2\dag}\op{b}, \\
         & \op{L_{\rm b}} = \sqrt{\kappa_{\rm b}}\op{b},                            \\
         & \op{L_{\downarrow}} = \sqrt{\kappa_1 (n_\mathrm{th} + 1)}\op{a},         \\
         & \op{L_{\uparrow}} = \sqrt{\kappa_1 n_\mathrm{th}}\op{a}^\dag,
    \end{split}
\end{equation*}
where the initial states are displaced thermal states with a mean thermal photon number \(n_\mathrm{th}\).
The fitted value of \(n_\mathrm{th}\) matches the one estimated via the measurement of the Wigner function of the thermal state (Fig.~\ref{fig:photon_number_calibration}E).
Additionally, this measurement allows us to calibrate the two-photon pump duration required to map the memory state parity to the 0/1 manifold, which is essential for the parity measurement.

\begin{figure}
    \centering
    \includegraphics[width=0.48\textwidth]{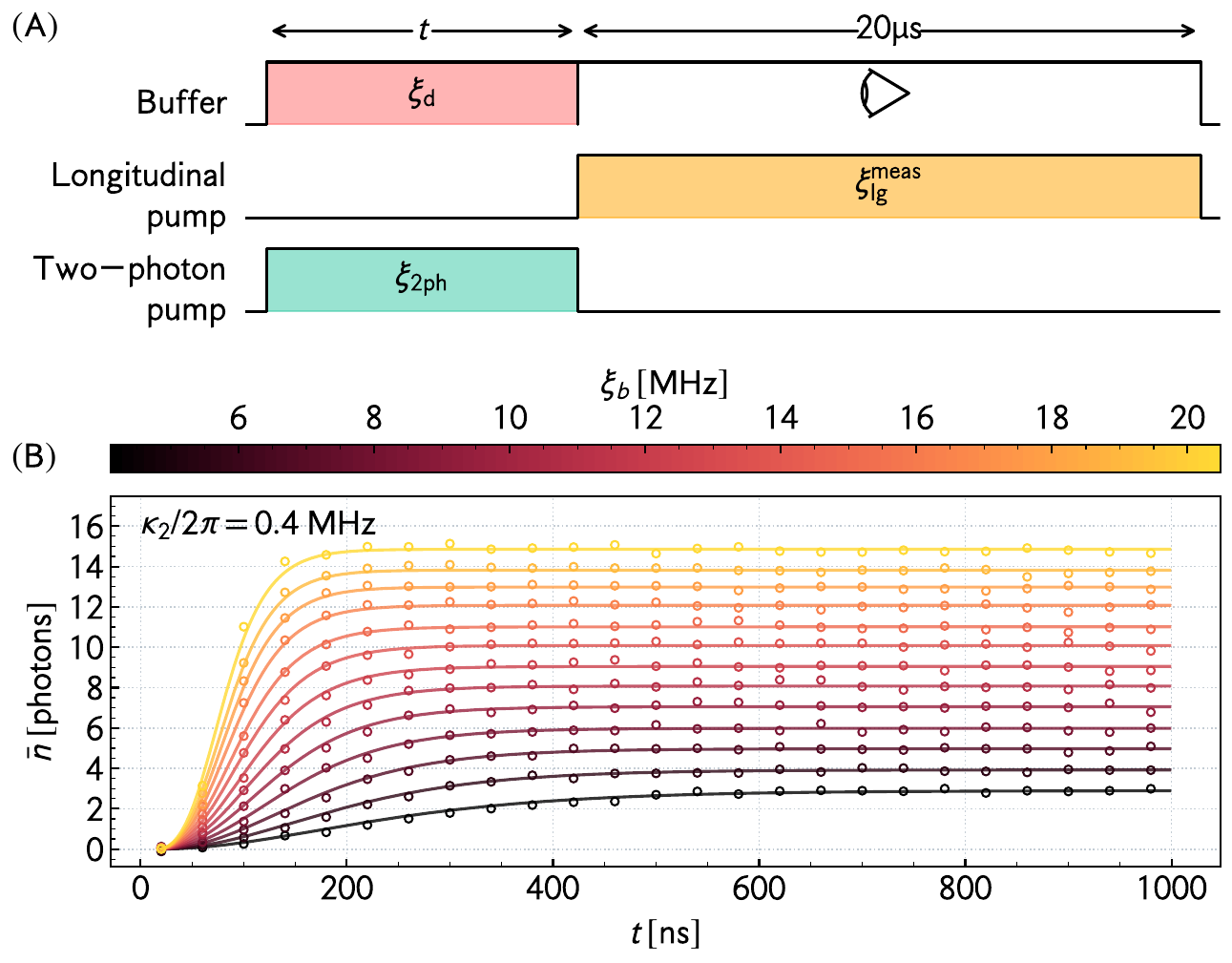}
    \caption{Cat state inflation measurement.
        (A) Pulse sequence of the experiment: the two-photon pump and buffer drive of amplitude \(\xi_b\) are played simultaneously for a variable duration \(t \), stabilizing cat subspaces of different sizes.
        The mean memory photon number is then measured by applying the longitudinal pump while recording the buffer output field.
        (B) Measured memory photon number (circles) as a function of the two-photon pump duration and buffer drive amplitude (color).
        The solid lines represents the results of a quantum fit to the data.
    }\label{fig:cat_inflation}
\end{figure}

Finally, applying the two-photon pump and a buffer drive with amplitude \(\xi_b\) stabilizes a cat state with a mean photon number \(\bar{n} = |\alpha|^2 = |\xi_b/g_2|\), at a rate \(\kappa_{\rm conf} = 4\kappa_2|\alpha|^2\), where \(\kappa_2 = \frac{4g_2^2}{\kappa_{\rm b}}\)~\cite{reglade_quantum_2024}.
We measured the cat size as a function of the cat manifold stabilization duration \(t_{\rm inflate}\) and \(\xi_b\) (Fig.~\ref{fig:cat_inflation}).
The data were fitted using a quantum model with the Lindblad loss operators
\begin{equation*}
    \begin{split}
        \op{L_2}            & = \sqrt{\kappa_2}(\op{a}^2 - \alpha^2),      \\
        \op{L_{\downarrow}} & = \sqrt{\kappa_1 (n_\mathrm{th} + 1)}\op{a}, \\
        \op{L_{\uparrow}}   & = \sqrt{\kappa_1 n_\mathrm{th}}\op{a}^\dag.
    \end{split}
\end{equation*}
The fitted two-photon loss rate \(\kappa_2/2\pi = 0.4\mathrm{\;MHz}\) qualitatively matches the one calibrated from the memory deflation measurement.
This measurement allows to calibrate \(t_{\rm inflate}\) for all cat sizes used in the experiment.

\subsection{Compensation module}\label{ap:compensation_module}

\begin{figure}
    \centering
    \begin{circuitikz}
        \draw(0, 0) to[short, i=\(i\)] (1, 0);

        \draw(1, 0) to[short] (1.5, .5) to[short] (1.7, .5) to[short] (1.7, .9) to[short] (2.5, .9) to[short] (2.5, .1) to[short] (1.7, .1) to[short] (1.7, .5);
        \node at (2.1, .5) [anchor=center] {\(V_{\rm att}\)};
        \draw(2.5, .5) to[short] (2.6, .5);
        \draw[dotted] (2.6, .5) to[short] (3, .5);
        \draw(3, .5) to[short] (3.1, .5) to[short, i=\(i_{\rm L}\)] (3.1, 1.5);

        \draw(1, 0) to[short] (1.5, -.5) to[short] (1.7, -.5) to[short] (1.7, -.9) to[short] (2.5, -.9) to[short] (2.5, -.1) to[short] (1.7, -.1) to[short] (1.7, -.5);
        \node at (2.1, -.5) [anchor=center] {\(V_\phi \)};
        \draw(2.5, -.5) to[short] (2.6, -.5);
        \draw[dotted] (2.6, -.5) to[short] (3, -.5);
        \draw(3, -.5) to[short] (3.1, -.5) to[short, i_=\(i_{\rm R}\)] (3.1, -1.5);

        \draw(3.3, 1.5) to[barrier] (4.8, 1.5) to[short] (4.8, -1.5) to[barrier] (3.3, -1.5) to[short] (3.3, 1.5);
        \draw(3.3, 0) to[L] (4.8, 0);
        \draw[->]   (3.9, 0.75) circle (1.5mm) node[right] {\( \; \; \varphi_{\rm L}\)};
        \fill   (3.9, 0.75) circle (0.5mm);
        \draw[->]   (3.9, -0.75) circle (1.5mm) node[right] {\( \; \; \varphi_{\rm R}\)};
        \fill   (3.9, -0.75) circle (0.5mm);
    \end{circuitikz}
    \caption{Schematic of the compensation module.
        The initial signal is split into two lines: the first passes through a variable attenuator, and the second through a variable phase shifter.
        These two lines then respectively pass by each of the ATS loops.
    }\label{fig:compensation_module}
\end{figure}
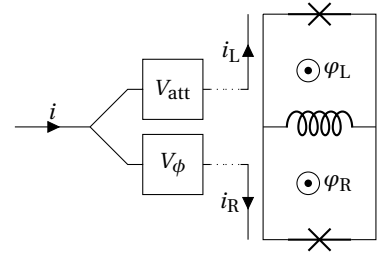

\begin{figure}
    \centering
    \includegraphics[width=0.48\textwidth]{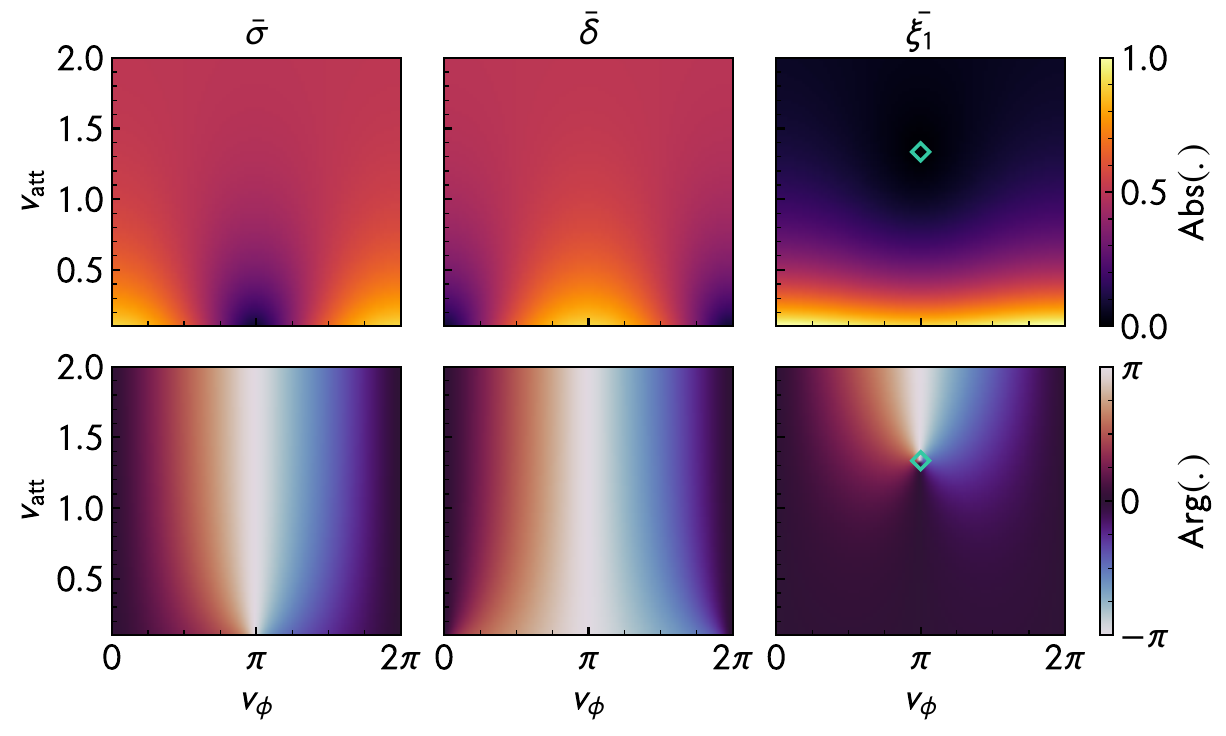}
    \caption{Simulated compensation maps.
        The reduced complex amplitudes of fluxes \(\bar{\sigma}\) and \(\bar{\delta}\) correspond to the fluxes divided by the absolute value of their maximum in the map, while the reduced complex amplitude of the linear drive is defined as \(\bar{\xi_1} = \xi_1 \hbar/(\varphi_{a/b} E_{\rm L})\).
        The absolute value (top) and phase (bottom) of the reduced sigma flux (left), delta flux (center), and linear drive term (right) are plotted (color) as a function of the attenuation and phase shift of the compensation module.
        A cancellation of \(\xi_1\) is observed as the phase undergoes a \(2\pi \) winding around it, marked by the green diamond.
    }\label{fig:compensation_module_theory}
\end{figure}

To control the time-dependent flux modulation of the ATS, we use a compensation module (Fig.~\ref{fig:compensation_module}).
In this section we present a simplified model for this module.
The input current \(\underline{i}(t) = \mathcal{R}(i(t)) = \mathcal{R}(ie^{i\omega_p t})\), with angular frequency \(\omega_p\), is split into two lines.
The first line passes through a variable phase attenuator, with voltage \(V_{\rm att}\), and the second through a variable phase shifter, with voltage \(V_\phi \).
The two currents can be expressed as:
\begin{equation*}
    \begin{aligned}
        i_{\rm L}(t) & = \frac{1}{2}10^{-v_{\rm att}}e^{i\omega_p l_{\rm L}/c} i(t), \\
        i_{\rm R}(t) & = \frac{1}{2}e^{iv_\phi}e^{i\omega_p l_{\rm R}/c} i(t),
    \end{aligned}
\end{equation*}
where \(v_{\rm att}\) and \(v_\phi \) are the renormalized attenuation and phase shift in dB and radians, respectively, \(l_{\rm L, R}\) are the length of the lines, and \(c\) is the speed of light in the cables.
Assuming each line only threads one ATS loops, with a mutual inductance \(M \), the common and differential fluxes can be expressed as:
\begin{gather*}
    \begin{pmatrix}
        \sigma(t) \\
        \delta(t)
    \end{pmatrix}
    = \frac{M}{4}
    \begin{pmatrix}
        1 & 1  \\
        1 & -1
    \end{pmatrix}
    \begin{pmatrix}
        -10^{-v_{\rm att}} \\
        e^{i(v_\phi - \omega_p \delta_{\rm l}/c)}
    \end{pmatrix}
    i(t),
\end{gather*}
where \(\delta_{\rm l} = l_{\rm L} - l_{\rm R}\) is the difference in cable length, and the phase of \(i\) has been chosen such that \(i_{\rm L}(t) \) is real to simplify expressions.
The resulting time dependent fluxes and \(\xi_1\) (Eq.~\ref{eq:xi1_expr}) are plotted in Fig.~\ref{fig:compensation_module_theory}.
The asymmetry in cable lengths, \(\delta_{\rm l} \), makes the phase calibration of the compensation module frequency-dependent.

\subsection{Longitudinal pump for Moon cat stabilization}

\begin{figure}
    \centering
    \includegraphics[width=0.48\textwidth]{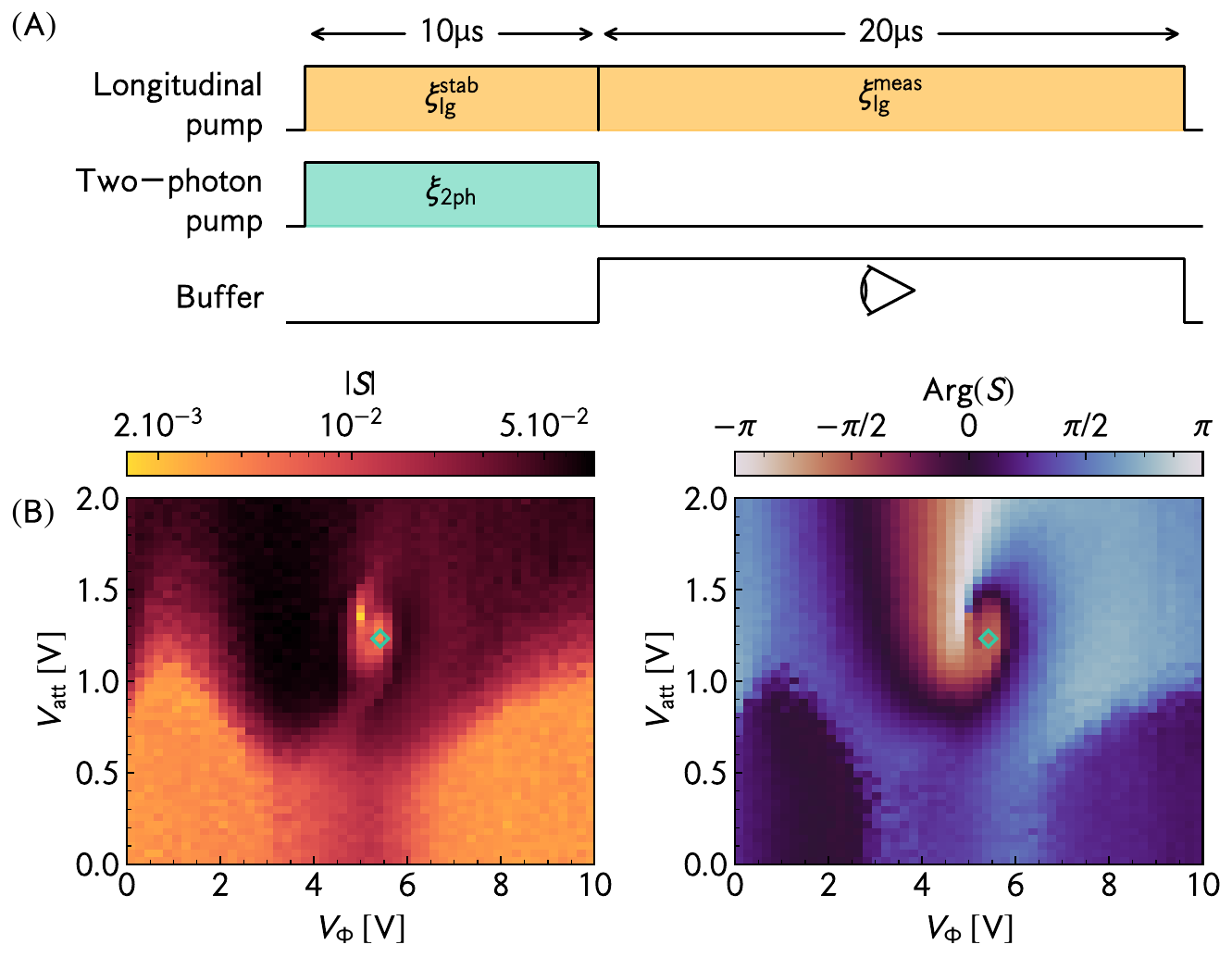}
    \caption{Longitudinal pump compensation map.
        (A) Pulse sequence of the experiment: the two-photon and longitudinal pumps are played for \(10 \; \mathrm{\mu s}\) stabilizing a cat state in the memory.
        The cat size is proportional to the strength of the buffer drive applied by the longitudinal pump.
        The buffer output signal is measured while the longitudinal pump is played, and the measured signal is proportional to the number of photons in the memory.
        (B) The absolute value and phase of the measured signal are plotted (color) as a function of the compensation module voltages.
        The green diamond indicates the compensation point where the linear drive of the longitudinal pump is cancelled.
        The large regions exhibiting a small signal (bottom left and right) corresponds to the longitudinal pump inducing a very large Stark-shift on the modes, which suppresses the two-photon conversion strength.
    }\label{fig:longitudinal_compensation}
\end{figure}

In this section we detail the calibration of the longitudinal pump for Moon cats stabilization.
As explained in Section~\ref{ap:ats_pumping}, applying a pump to the ATS generates a drive term on the modes (Eq.~\ref{eq:xi1_expr}) noted \(\xi_1\).
Since the longitudinal pump is resonant with the buffer mode, the drive term can be much larger than the capacitive buffer drive used for cat stabilization.
To achieve proper control of the cat state, we chose to cancel \(\xi_1\), which can be done by fine-tuning the attenuation and phase of the compensation module (Fig.~\ref{fig:compensation_module_theory}).
A simple approach to calibrate \(\xi_1\) would be to apply the longitudinal pump and measure the buffer output field \(S_{\xi_1} \propto \xi_1\), looking for a zero of the field.
Unfortunately, on-chip crosstalk \(S_\chi \), between the pump lines and the buffer output line is non negligible.
The output field then writes \(S = S_{\xi_1} + S_\chi \), meaning this method of calibration does not fully cancel \(\xi_1\), and a better strategy was needed.

By applying both the two-photon pump and the longitudinal pump, we stabilize cat states in the memory, with size \(\bar{n}(V_{\rm att}, V_\phi) \propto \xi_1\).
We then measure the number of photons in the memory using the longitudinal pump (Fig.~\ref{fig:longitudinal_compensation}A).
Since the calibration of the photon number measurement depends on the compensation point, the measured quantity \(S(V_{\rm att}, V_\phi) = \upsilon(V_{\rm att}, V_\phi) \bar{n}(V_{\rm att}, V_\phi)\) is proportional to the number of photons, with a complex scaling factor \(\upsilon \) that depends on the voltages applied to the compensation module (Appendix~\ref{ap:longitudinal_photon_number_measurement}).
By observing the measurement signal as a function of the compensation module voltages, two zeros of \(|S|\) can be seen: one with a \(2\pi \)-winding around it and one without (Fig.~\ref{fig:longitudinal_compensation}B).
Since \(\bar{n}(V_{\rm att}, V_\phi): \mathbb{R}^2\rightarrow\mathbb{R}\), the compensation module setting cancelling \(\xi_1\) corresponds to the zero without the \(2\pi \)-winding, with the other being a zero of the complex-valued function \(\upsilon(V_{\rm att}, V_\phi)\).
By cancelling the \(\xi_1\) term in the evolution Hamiltonian, the longitudinal pump induced Stark-shift on the modes is also cancelled (Eq.~\ref{eq:stark_shift_expr}).

\subsection{Memory drive coupling strength calibration}

\begin{figure}
    \centering
    \includegraphics[width=0.48\textwidth]{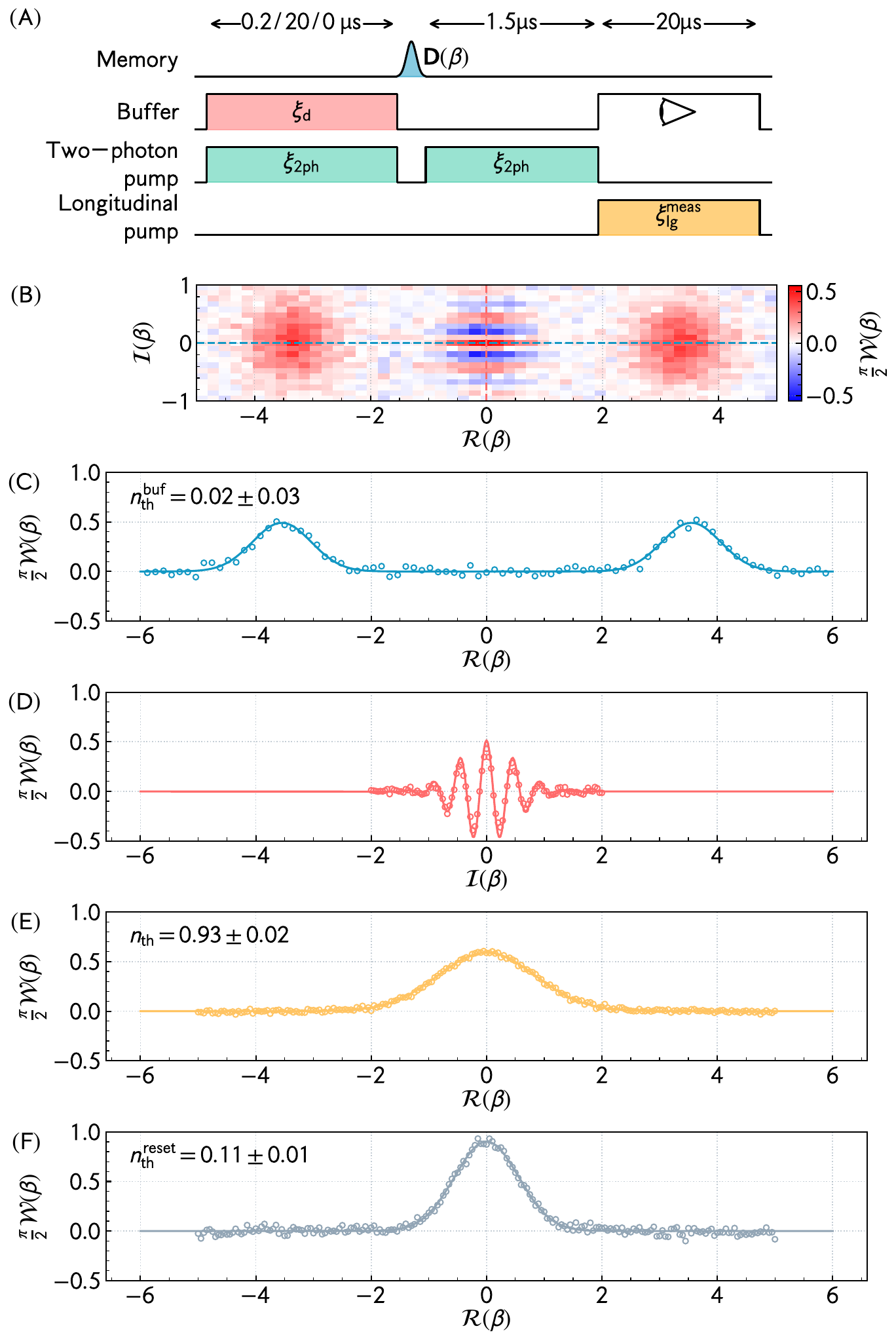}
    \caption{Memory drive strength calibration.
        (A) Pulse sequence for cat state stabilization followed by Wigner tomography.
        The two-photon pump and buffer drives are applied to stabilize a cat state in memory.
        A Gaussian-shaped displacement pulse is applied to the memory, followed by a projection into the Fock qubit manifold to map the parity to the photon number.
        The parity is then measured using a longitudinal photon number measurement.
        (B) Wigner tomography of a cat state with 11.6 photons stabilized for 200 ns.
        The blue (resp.\ red) dashed lines indicate the positions of the cuts shown in plots C, E and F (res.\ D).
        (C-F) Measured cuts (circles) and fitted curves (lines) of the memory Wigner function.
        (C) Real-axis cut of the Wigner function for a decohered cat state stabilized for \(20 \mathrm{\;\mu s}\).
        (D) Imaginary-axis cut of the Wigner function for a coherent cat state stabilized for \(200\;\mathrm{ns}\).
        (E) Real-axis cut of the Wigner function after a \(500 \mathrm{\;\mu s}\) passive reset.
        (F) Real-axis cut of the Wigner function after a \(10 \mathrm{\;\mu s}\) active reset.
    }\label{fig:photon_number_calibration}
\end{figure}

In this section, we detail the calibration of the memory drive coupling strength.
The memory is driven upon applying a differential flux of the ATS at the memory mode frequency, generating a drive term (Eq.~\ref{eq:xi1_expr}).
To calibrate the drive strength, we measure cuts of the Wigner function of the memory state.
The Wigner function of a state \(\rho \) can be expressed as the displaced parity of the state: \(\mathcal{W}(\beta) = \frac{2}{\pi}\mathrm{Tr}[\op{D}(-\beta)\rho \op{D}(\beta) \op{P}]\), where \(\op{P}\) is the parity operator.
Since the two-photon loss operator commutes with parity operator \(\op{P}\), deflating a memory state to the 0/1 Fock manifold maps the parity to the number of photons in the memory, which can be measured using the longitudinal interaction~\cite{reglade_quantum_2024}.
By using the analytical formulas of the Wigner function of a cat state~\cite{haroche_quantum_2006}, we can extract the memory drive strength from the cuts of the Wigner function (Fig.~\ref{fig:photon_number_calibration}C-D).
Measuring the memory Wigner function after a reset allows us to determine the thermal population of the memory.
After a passive reset, the memory is in a thermal state, and the Wigner function is a Gaussian centered at the origin with a standard deviation \(\sigma = \frac{1}{2}\sqrt{1+2n_{\rm th}}\).
Since the memory mode frequency is \(\omega_a/2\pi = 1.08\;\mathrm{GHz}\), its thermal population is non zero and determined to be \(n_{\rm th} = 0.93\) (Fig.~\ref{fig:photon_number_calibration}E), the same as the one measured in Section~\ref{ap:two_photon_pump_calibration}.
For all other measurements in this paper, we used an active reset, cooling the memory thermal population to \(n_{\rm th}^{\rm reset} = 0.11\) (Fig.~\ref{fig:photon_number_calibration}F).
Additionally, when stabilizing a cat state, the state is thermalized to the buffer temperature.
The Gaussian width of the decohered cat state logical states depends on the buffer thermal population, and we measured it to be \(n_{\rm th}^{\rm buf} = 0.02 \pm 0.03\) photons (Fig.~\ref{fig:photon_number_calibration}C).

%% file: 4_system_calibration.tex
\section{System calibration}\label{ap:system_calibration}

\subsection{Memory self-Kerr measurement}

\begin{figure}
    \centering
    \includegraphics[width=0.48\textwidth]{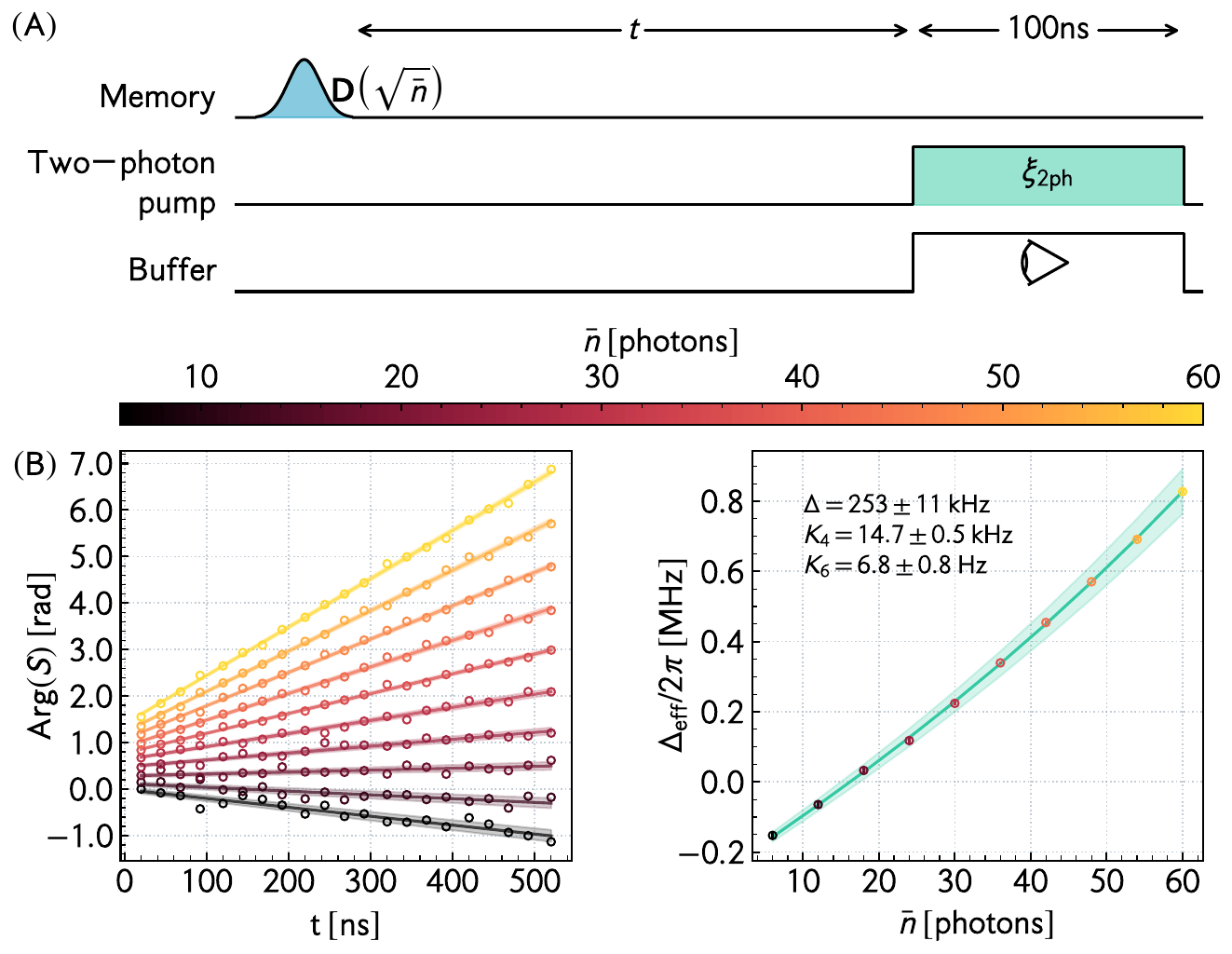}
    \caption{Memory self-Kerr measurement.
        (A) Pulse sequence of the measurement: The memory state is first displaced to a coherent state with amplitude \(\sqrt{\bar{n}}\), then the memory is allowed to evolve for a variable duration \(t\).
        Finally, we apply the two-photon pump and measure the output field of the buffer mode.
        (B) In the left panel, the phase of the signal (circles) is plotted for different coherent state amplitude (color) as a function of the evolution duration.
        The linear fit (lines) allows the extraction of the effective detuning.
        In the right panel, the extracted detunings are plotted as a function of the initial state amplitude.
        The fit (line) allows to extract the Kerr coefficients.
    }\label{fig:kerr_measurement}
\end{figure}

In this section, we detail the calibration technique of the memory self-kerr.
The memory is first displaced to a coherent state of amplitude \(\alpha \), and then allowed to evolve for a duration \(t\).
Finally, we apply the two-photon pump and measure the output field of the buffer mode (Fig.~\ref{fig:kerr_measurement}A), measuring the operator \(\op{a}^2\); this approach is valid for measurement durations short compared to \(1/\kappa_2\).
In the Heisenberg picture, the evolution of \(\op{a}^2\) under the Hamiltonian \(\op{H}/\hbar = \Delta \op{a}^\dag\op{a} - \frac{K_4}{2}\op{a}^{\dag 2}\op{a}^2 - \frac{K_6}{6}\op{a}^{\dag 3}\op{a}^3\) follows:
\begin{equation*}
    \frac{d\op{a}^2}{dt} = -i(\Delta - K_4 - 2K_4\op{a}^\dag \op{a} - K_6\op{a}^{\dag}\op{a}^2 - K_6\op{a}^{\dag2}\op{a}^3)\op{a}^2.
\end{equation*}
For small evolution time compared to \({(|\alpha|^2 K_4)}^{-1}\), \({(|\alpha|^4 K_6)}^{-1}\), \(T_1 \), and \(T_\phi \), the memory state remains in a coherent state with amplitude \(\alpha \).
The memory state undergoes an evolution with an effective detuning \(\Delta_{\mathrm{eff}} = 2\Delta - K_4 - 2K_4|\alpha|^2 - K_6|\alpha|^{3/2} - K_6|\alpha|^{5/2}\).
Fitting the phase of the output signal with a linear curve provides the effective detuning as the slope.
This detuning can the be fitted for different initial states amplitudes, yielding \(\Delta, K_4\) and \(K_6\) (Fig.~\ref{fig:kerr_measurement}B), referenced in Table~\ref{tab:system_parameters}.

\subsection{Memory dephasing measurement}

\begin{figure}
    \centering
    \includegraphics[width=0.48\textwidth]{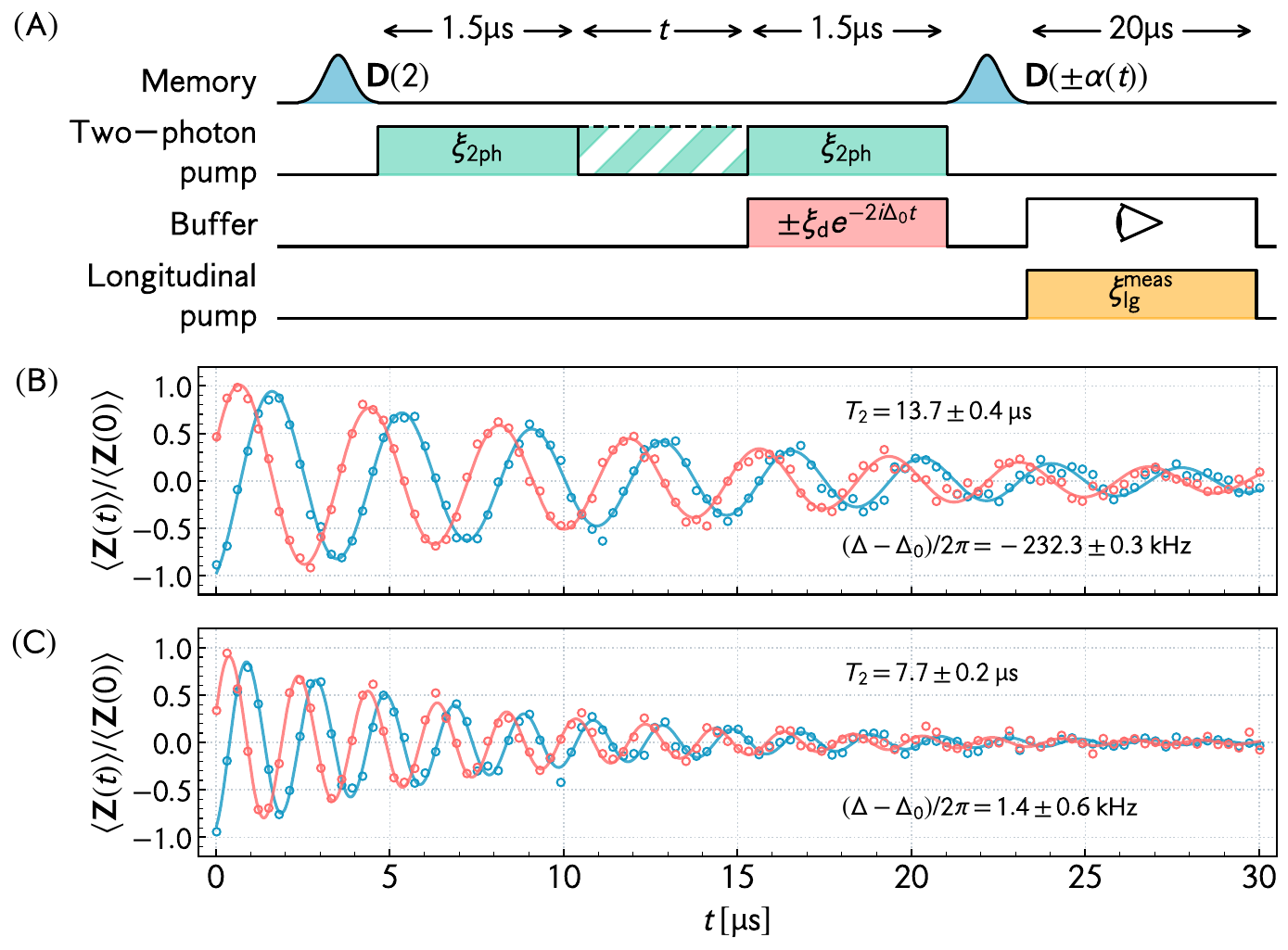}
    \caption{Memory \(T_2\) measurements.
        (A) Pulse sequence of the experiment: The memory is first displaced to a coherent state with 4 photons, and the state is then deflated to the 0/1 manifold.
        The state is allowed to evolve for a variable duration \(t\) with the two-photon pump either off (B) or on (C).
        The state is then reinflated to a cat state with phase \(\Delta_0 t\) or \(\Delta_0 t + \pi/2\), mapping the \(\op{X}\) or \(\op{Y}\) operator of the Fock qubit to the \(\op{Z}\) operator of the cat qubit.
        Finally, we measure the \(\op{Z}\) operator of the cat qubit.
        (B) Measured (circles) cat qubit \(\op{Z}\) operator as a function of the evolution time.
        Colors corresponds to the two phases of inflation at each time.
        The fit (lines) allows the extraction of the memory \(T_2\) and detuning \(\Delta \).
    }\label{fig:dephasing}
\end{figure}

In this section, we detail the measurement of the dephasing time of the memory mode.
The experiment consists of measuring the 0/1 manifold Fock qubit \(\op{X}\) and \(\op{Y}\) operators as a function of time~\cite{marquet_autoparametric_2024}.
To efficiently measure these operators, we map them to the \(\op{Z}\) operator of a cat qubit by inflating a cat state (Fig.~\ref{fig:dephasing}A).
The logical \(\op{Z}\) operator of the cat qubit is then measured by applying displacing the cat state and measuring the mean photon number.
An artificial detuning, \(\Delta_0 \), is introduced by rotating the buffer pulse that inflates the cat as a funtion of time, \(\xi_d e^{-2i\Delta_0 t}\), thereby rotating the angle of the inflated cat state.
The hamiltonian and loss operator governing the memory state evolution are:
\begin{align*}
    \op{H}              & = \hbar\Delta_0 \op{a}^\dag \op{a},          \\
    \op{L}_{\downarrow} & = \sqrt{\kappa_1 (n_\mathrm{th} + 1)}\op{a}, \\
    \op{L}_{\uparrow}   & = \sqrt{\kappa_1 n_\mathrm{th}}\op{a}^\dag,  \\
    \op{L}_\phi         & = \sqrt{2\kappa_\phi} \op{a}^\dag \op{a}.
\end{align*}
The \(\op{X}\) and \(\op{Y}\) operators evolve as two damped oscillations in quadrature, with a frequency \(\Delta - \Delta_0\) and a characteristic decay time \(T_2 = 1/(\kappa_1/2 + \kappa_\phi)\).
We extract the memory mode coherence time \(T_2\) from a fit to the data. Upon activating the two-photon pump, we observe an increase in the dephasing rate (see Table~\ref{tab:system_parameters}).

\subsection{Lambda calibration}\label{ap:lambda_calibration}

\begin{figure}
    \centering
    \includegraphics[width=0.48\textwidth]{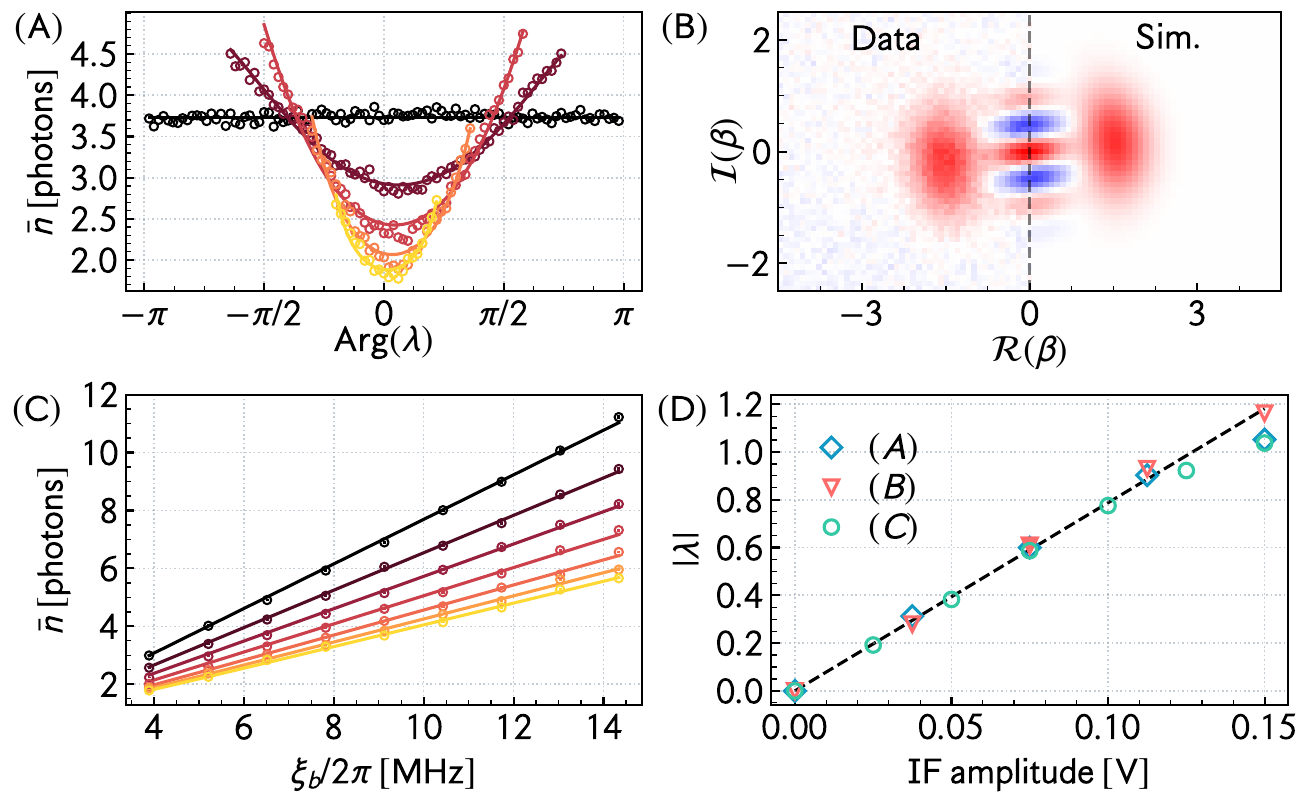}
    \caption{Calibration methods of \(|\lambda|\).
        (A) Fit of the memory steady-state mean photon number as function of \(\mathrm{Arg}(\lambda) \) for different amplitude of the longitudinal pump (color).
        The fit is restricted to the region around the minimum of the curves to avoid the influence of spurious terms.
        (B) Fit of a high resolution Wigner function of a Moon cat.
        (C) Fit of the mean photon number in the cat state as a function of the buffer drive amplitude for different longitudinal pump amplitudes (color).
        (D) \(|\lambda|\) as a function of the pump IF amplitude.
        The fit results of the other three panels are shown with 3 different markers.
        The filled triangular marker indicates the result from the fit in panel (B).
        The black dashed lines indicates the linear scaling of the \(|\lambda|\) with the pump IF amplitude.
    }\label{fig:lambda_amplitude_calibration}
\end{figure}

We used three different methods to calibrate the amplitude of \(\lambda \).
The first method, depicted in Fig.~\ref{fig:fig2}A of the main text, showed that experimental data did not match the simulation for anti-squeezing like deformations.
For such deformations, the Fock state state distribution is broader than the Poissonian distribution of coherent states, leading to greater sensibility to spurious terms, such as memory self-Kerr and cross-Kerr with the buffer mode.
To calibrate \(|\lambda|\) in the squeezing-like deformation regime, we fitted the curves near to the minimum photon number (Fig.~\ref{fig:lambda_amplitude_calibration}A).
The second method involved fitting high resolution Wigner functions of moon cats (Fig.~\ref{fig:lambda_amplitude_calibration}) using the Moon cat state analytical formula (Eq.~\ref{eq:moon_cat_analytical}).
The third method used the photon numbers measured during the scaling experiments.
We fitted the Moon cat sizes as a function of the buffer drive amplitude for different longitudinal pump amplitudes (Fig.~\ref{fig:lambda_amplitude_calibration}C).
The results from these three methods are shown in Fig.~\ref{fig:lambda_amplitude_calibration}D, and all methods show very good agreement.
However, in the high amplitude regime, the results do not match as well because the needed compensation calibration is highly precise in this region, and small drifts in the compensation module could explain the difference.
Additionally, the loss of the linear relationship between \(|\lambda|\) and the pulse amplitude could be due to the loss of linearity in the compensation module elements at high power.
The values used for all plots in the main text are those extracted from the fit of the cat size as a function of the buffer drive amplitude.

%% file: 5_lifetimes_measurements.tex
\section{Lifetimes measurements}\label{ap:lifetimes_measurements}

\subsection{Bayesian adaptive bit-flip measurement}\label{ap:bayesian_adaptive_measurement}

The measurement of the bit-flip rate \(\Gamma_X\) follows a three-steps procedure.
First, a resonant drive displaces the memory to either \(\ket{0}_L=\ket{\alpha}\) or \(\ket{1}_L=\ket{-\alpha}\) while the cat stabilization is off.
The latter is then turned on for a variable idling duration \(t\), stabilizing the cat manifold.
During this step, bit-flips may occur with a probability that depends on the average photon number \(\bar{n}\).
Finally, to measure the projection along the \(Z\)-axis, the stabilization is switched off, and a counter-displacement \({\cal D}(\mp\alpha)\) is applied to the memory.
A photon number measurement follows, yielding \(0\) if no bit-flip occurred, and \(4 |\alpha|^2\) otherwise.

Cat qubits can typically exhibit bit-flip times on the order of several hundred seconds.
This can lead to prohibitively long measurement durations if the sampling times are not selected carefully.

\subsubsection{Adaptive Bayesian method}

To minimize the measurement time while ensuring informative measurements, we applied a Bayesian adaptive design approach, adapted from~\cite{rainforth_modern_2024}.
The dynamics of the resulting bit-flip experiment are modeled by an exponentially decaying function:
\begin{equation*}
    \langle Z \rangle_t = C_\infty + (C_0-C_\infty)\exp(-\Gamma_Z t)
\end{equation*}
where the parameters to be estimated are \(\theta = \{\Gamma_Z, C_0, C_\infty \} \).
These parameters represents the decay rate \(\Gamma_Z \) of \(\braket{\op{Z}} \), the initial value \(C_0\) and final value \( C_\infty \) (offset).
The offset is included in the model only for bit-flip measurements performed during a Zeno gate, since the memory drive is the sole term that breaks parity symmetry.
Each run of the experiment involves selecting a measurement time \(t \), preparing the initial state \(\ket{0}_L \) or \(\ket{1}_L \), and measuring \(\op{Z} \), with the procedure repeated \(N \) times for each state.
Every measurement yields a result \(y_i \in \{0, 1\} \), where \(1 \) corresponds to the state being measured in \(\ket{0}_L \) and \(0 \) corresponds to \(\ket{1}_L \).

Bayesian Adaptive design provides a principled method for selecting the measurement time \(t \), aiming to maximize the \textit{information gain}.
The selected measurement time is the one that leads to the greatest increase in information.
We begin by modeling the number of bit-flips using a Poisson distribution.
The probability of the observing a total measurement outcome \(y = \sum_i y_i \) is given by:
\begin{equation*}
    \begin{split}
         & P(y=k|\theta, t, 0/1) = \binom{N}{k}{p_{0/1}}^k{(1-p_{0/1})}^{N-k},             \\
         & \text{where} ~~ p_{0/1} = (1 + C_\infty + (C_\infty \pm C_0)e^{-\Gamma_Z t})/2.
    \end{split}
\end{equation*}
The information gain in \(\Gamma_Z \) is defined as the reduction in Shannon entropy between the prior and posterior marginal distributions:
\begin{multline*}
    \mathrm{InfoGain}_{\Gamma_Z}(y, t) := \mathrm{H}[p(\Gamma_Z)] - \mathrm{H}[p(\Gamma_Z|y, t)] \\ =\mathbb{E}_{p(\Gamma_Z|y, t)}[\log{p(\Gamma_Z|y, t)}] - \mathbb{E}_{p(\Gamma_Z)}[\log{p(\Gamma_Z)}].
\end{multline*}
However, in our case, larger values of \(t \) lead to longer experiment durations.
To mitigate this, we aim to maximize the \textit{information flow}, defined as \(\mathrm{InfoFlow}(y, t) = \mathrm{InfoGain}(y, t)/t \), which is measured in bits per second.
Since the measurement outcome is a random variable, we focus on the \textit{expected information flow} (EIF) for \(\Gamma_Z \).
This is computed using the marginal distribution of possible outcomes \(p(y|t) = \mathbb{E}_{p(\theta)}[p(y|\theta, t)] \), as follows:
\begin{equation*}
    \begin{split}
        \mathrm{EIF}_{\Gamma_Z}(t) & = \mathbb{E}_{p(y|\theta, t)}[\mathrm{InfoFlow}_{\Gamma_Z}(y, t)]                                        \\
                                   & = \frac{1}{t}\mathbb{E}_{p(\Gamma_Z)p(y|\Gamma_Z, t)}\left[\log{\frac{p(y|\Gamma_Z, t)}{p(y|t)}}\right].
    \end{split}
\end{equation*}
After selecting the measurement time \(t \) that maximizes the EIF for the \(\Gamma_Z \) distribution, a Bayesian update is applied to the distribution \(p(\theta) \) after each measurement,
\begin{equation*}
    p(\theta) \rightarrow p(\theta|y, t) = \frac{p(\theta) p(y|\theta, t)}{p(y|t)}.
\end{equation*}

The algorithm was implemented using JAX~\cite{bradbury_jax_2018}, starting from a uniform prior distribution.
To accommodate the wide range of possible values, we opted to work with \(\log{\Gamma_Z} \), allowing us to manage variations across several orders of magnitude.
Measurements were performed iteratively until the desired uncertainty in \(\Gamma_Z \) estimation was achieved.

\subsubsection{Comparison to other methods}

\begin{figure}
    \centering
    \includegraphics[width=0.48\textwidth]{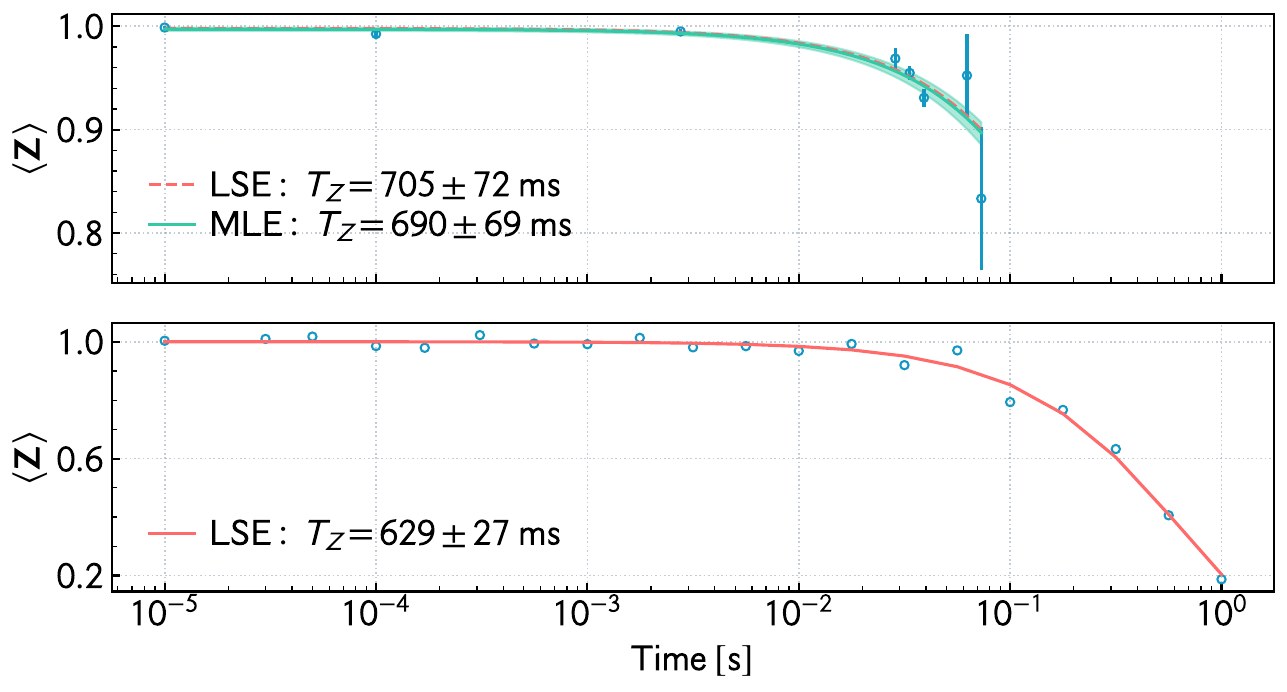}
    \caption{Bayesian adaptive (top panel) and fixed-time (bottom panel) bit-flip characteristic time measurements.
        In the Bayesian approach, measurement times are carefully selected to maximize the \textit{information flow} related to the bit-flip time.
        Each data point is computed from a different number of experimental shots, resulting in non-uniform error bars.
        The bit-flip time can be extracted either from the posterior distribution of \(\Gamma_Z \) corresponding to a Maximum Likelihood Estimation (MLE) or directly estimated using a Least Square Estimation (LSE) algorithm.
        Both methods yield results that are consistent with the LSE estimation obtained from the fixed-time measurements.
    }\label{fig:bayesian_bf_fixed}
\end{figure}

Once the adaptive bit-flip measurements are complete, the data can be analyzed in multiple ways.
One approach is to use the posterior distribution of \(\Gamma_Z \) estimated during the experiment, which corresponds to a Maximum Likelihood Estimator (MLE).
However, this method depends on the validity of the experimental model, and it may fail if the model does not accurately represent the experiment.
For instance, in the case of large cat sizes, the saturation of bit-flip times is not well understood and could indicate a breakdown in the model.
Alternatively, a Least Square Estimator (LSE) can be applied to the data, simplifying the model to an exponential decay.
It’s important to note that due to the adaptive nature of the algorithm, each data point might have been measured with a different number of experimental shots, resulting in variable uncertainty across the data.

Consider a data point measured with \(N \) shots.
Let \(x \) represent the number of shots where \(\op{Z}\) is measured as \(1\), following a Bernoulli distribution of parameter \(p \).
The unbiased estimator for \(p \) is given by \(\hat{p} = x/N \).
The variance of this estimator is \(\sigma_{\hat{p}}^2 = p(1-p)/N \).
A common approach, based on the law of large numbers, is to estimate the variance of the estimator as \(\hat{\sigma}_{\hat{p}}^2 = \hat{p}(1-\hat{p})/N \).
However, this approach can lead to counter-intuitive results when \(N \) is small.
For example, the variance estimate can be zero if \(x = 0 \) or \(x = N \), which is guaranteed to happen in the \(N = 1\) limit.
To address this issue, we employed a Bayesian approach to estimate the variance of the estimator.
We begin by assuming that the prior probability of the parameter \(p \) follows a Beta distribution, \(\mathrm{Beta}(\alpha, \beta) \), with the probability distribution function:
\begin{equation*}
    f(p;\alpha, \beta) = \frac{p^{\alpha-1}{(1-p)}^{\beta-1}}{\mathrm{B}(\alpha, \beta)},
\end{equation*}
where \(\mathrm{B}(\alpha, \beta) = \Gamma(\alpha)\Gamma(\beta)/\Gamma(\alpha + \beta)\) is the normalization constant and \(\Gamma \) denotes the gamma function.
The mean and variance of this distribution are given by: \(\mu = \alpha/(\alpha + \beta) \), \(\sigma^2 = \alpha\beta/({(\alpha + \beta)}^2(\alpha + \beta + 1)) \).
A key property of the Beta distribution is that the posterior probability of \(p \), after observing \(x \) successes and \(N-x \) failures, also follows a Beta distribution: \(\mathrm{Beta}(\alpha + x, \beta + N - x) \).
This makes the Beta distribution the conjugate prior of the Bernoulli distribution.
Assuming a uniform prior \((\alpha = \beta = 1)\), the Bayesian estimator for \(p \) is given by \(\hat{p}_{\rm bay} = (x + 1)/(N + 2) \), which is biased.
Un-biasing this estimator, we derive a better estimator for the variance of \(\hat{p}\),
\begin{equation*}
    \hat{\sigma}_{\hat{p}}^2 = \frac{(1+x)(1+N-x)}{N^2(N+3)}.
\end{equation*}
Using this uncertainty, we compute the LSE of the bit-flip time and its associated uncertainty (Fig.~\ref{fig:bayesian_bf_fixed}), confirming the results obtained via the MLE method.
When comparing the results of the Bayesian adaptive method to those from the fixed-time approach, where measurement times are predetermined, the estimated bit-flip times fall within the one-sigma interval uncertainty interval in both cases (Fig.~\ref{fig:bayesian_bf_fixed}).

\begin{figure}
    \centering
    \includegraphics[width=0.48\textwidth]{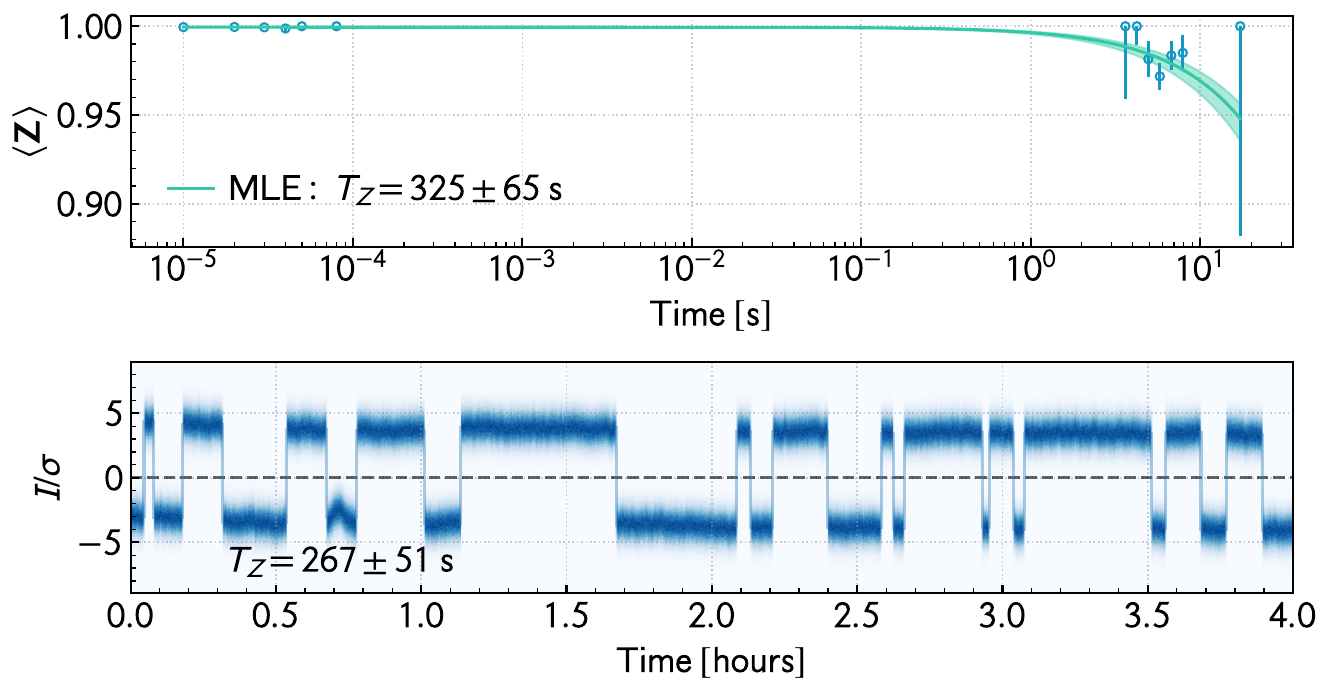}
    \caption{Bayesian adaptive (top panel) and trajectory (bottom panel) bit-flip characteristic time measurements.
        In the Bayesian adaptive method, sampling is concentrated at shorter times relative to the bit-flip time.
        Meanwhile, the trajectory of the \(\op{Z}\) observable is monitored over a span of 4 hours.
        The bit-flip time extracted from both methods are in agreement.
    }\label{fig:bayesian_bf_trajectory}
\end{figure}

We compared the adaptive method to trajectory-based measurements to asses its validity for bit-flips times longer than a few tens of seconds (Fig.~\ref{fig:bayesian_bf_trajectory}).
Trajectories are obtained by applying a continuous drive to the memory, orthogonal to the cat qubit axis, and measuring the output field of the buffer~\cite{reglade_quantum_2024}.
The phase of the drive relative to the cat axis must be carefully calibrated to avoid transferring population between the cat basis states.
However, even with perfect phase calibration, excessive drive amplitude can still induce bit-flips.
The estimated bit-flip times obtained using the adaptive method are consistent with those from the trajectory method, demonstrating the robustness of the adaptive approach.
Notably, the adaptive method focuses on measuring the bit-flip probability at timescales shorter than the bit-flip time, which is particularly relevant for quantum computing applications.
In such contexts, the target is to minimize bit-flip errors during the execution of algorithms, as these errors are typically uncorrected~\cite{gouzien_performance_2023}.

\subsubsection{Measurement time}

\begin{figure}
    \centering
    \includegraphics[width=0.384\textwidth]{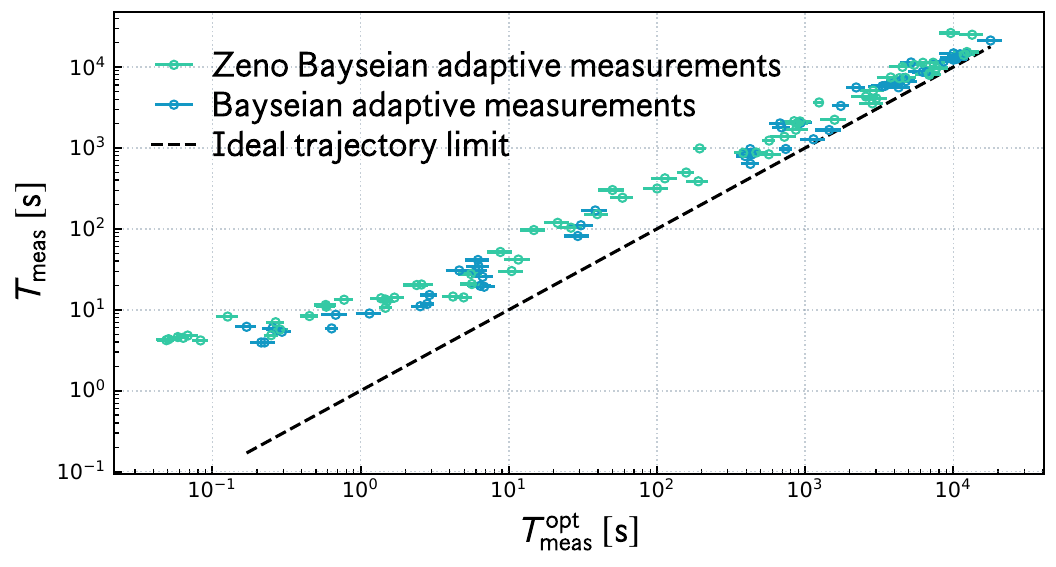}
    \caption{Comparison of the measurement duration of the adaptive method with the trajectory limit \(T_m{\rm meas}^{\rm opt}\).
    The measurement duration is plotted as a function of the optimal measurement duration \(T_{\rm meas}^{\rm opt}\) for the idle error rate scaling (blue circles) and Zeno gate error scaling (green circle).
    The black dashed line represents the ideal trajectory measurement limit.
    As the bit-flip time increases relative to the overhead durations, the adaptive method approaches this optimal limit more closely.
    }\label{fig:bayesian_bit_flip_measurement_time}
\end{figure}

To compare the measurement time of the adaptive method, we evaluate the duration required for a trajectory measurement to achieve a given uncertainty in the bit-flip time.
We consider a trajectory of \(N \) perfect measurements \( \{ Z_n \} _n \), each of duration \(t_m\), and estimate the jump probability \(p_j\) of the \(\op{Z} \) observable.
Without loss of generality, we assume \(Z_0 = 1\).
The expected value of the \(n\)-th measurement is given by \(\mathbb{E}[Z_n] = {(1-2p_j)}^n \).
Defining the jump probability rate as \(p_0 = p_j/t_m \) and the total time \(t = n t_m \), in the continuous measurement limit, the mean of the instantaneous observable \(Z(t)\) is given by \(\mathbb{E}[Z(t)] = \exp(-2 p_0 t) \).
Notably, the bit-flip rate is \(\Gamma_Z = 2 p_0 \).
To estimate \(p_j\), we define random variables \( \{ J_n \} _n \), where \(J_n = 1 \) if a jump occurs at time \(n t_m \), and \(J_n = 0 \) otherwise.
The jump probability estimator is \(\hat{p}_j = \sum_n J_n/n \), which is unbiased with variance \(\sigma_{\hat{p}_j}^2 = p_j(1-p_j)/N \).
The coefficient of variation \(\nu = \sigma_{ \hat{p}_j}/\hat{p}_j = \sqrt{(1-p_j)/p_j/N}\) tends to \(p_0 t \) in the continuous measurement limit.
Finally, in this ideal measurement setting, the time required to reach a given coefficient of variation \(\nu \) for the bit-flip time is:
\begin{equation*}
    T_{\rm meas}^{\rm opt} = 2 T_Z/\nu^2,
\end{equation*}
where \(T_Z = 1/\Gamma_Z \) is the bit-flip time.
As the bit-flip time increases, the measurement duration of the adaptive method approaches this optimal limit (Fig.~\ref{fig:bayesian_bit_flip_measurement_time}).
For shorter bit-flip times, however, the measurement duration in the adaptive method is dominated by overheads, such as preparation, measurement, and communication delays to the instruments.

\subsection{Phase-flip measurement}\label{ap:phase_flip_measurement}

\begin{figure}
    \centering
    \includegraphics[width=0.48\textwidth]{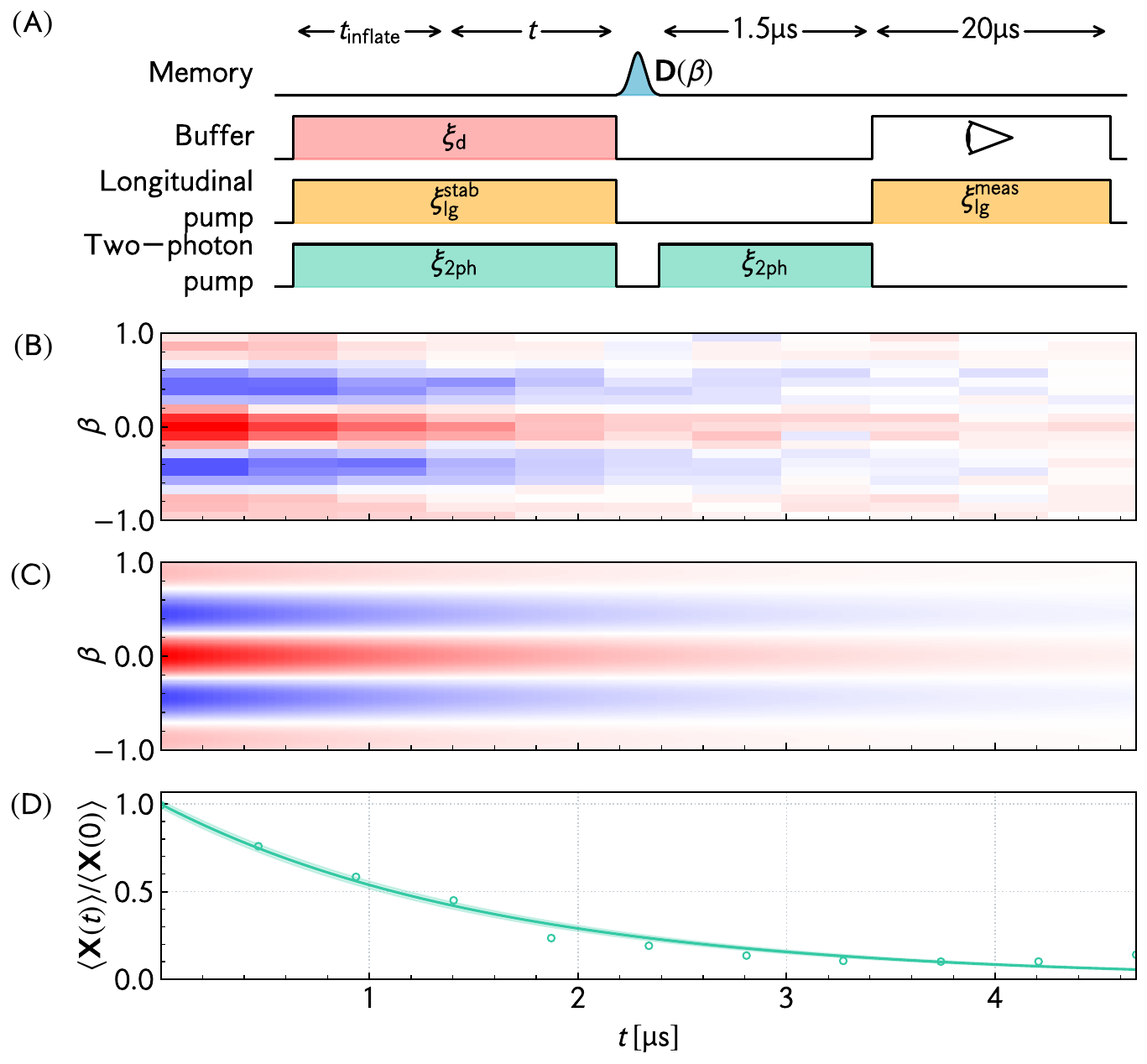}
    \caption{Phase-flip measurement.
        (A) Pulse sequence for the phase-flip measurement.
        A cat state is first prepared by stabilizing the cat manifold, after which it is stabilized for an additional variable duration \(t\).
        Finally, the fringes of the cat Wigner function are measured.
        (B) Measured decay of the fringes for a Moon cat with parameters \(\lambda = 0.59\) and \(\bar{n} = 3.2\) photons.
        (C) Fit of the experimental data from panel (B) using analytical expression for the fringes.
        (D) Extracted cat parity decay as a function of the stabilization duration \(t\).
    }\label{fig:cat_fringes_decay}
\end{figure}

In this section we detail the phase-flip lifetime measurement procedure.
The cat state is prepared in the memory by stabilizing the cat manifold for a duration \(t_{\rm stab}\), which depends on the size of the cat (Fig.~\ref{fig:cat_inflation}).
A cut of the fringes from the cat's Wigner function is then measured as a function of the additional stabilization time (Fig.~\ref{fig:cat_fringes_decay}).
The analytical expression for the cat's phase-flip time can be derived by considering the evolution of the parity operator \(\op{P} = e^{i\pi\op{a}^\dag \op{a}}\) in the Heisenberg picture.
Since the parity operator commutes with the Moon cat Lindblad loss operator (Eq.~\ref{eq:Moon_dissipator}) and the dephasing loss operators, we only need to account for the single-photon loss and gain Lindblad operators.
Using the Baker-Campbell-Hausdorff formula, we find:
\begin{align*}
    \frac{d\op{P}}{dt} & = \kappa_1(1+n_{\rm th})\mathcal{D}^\dag[\op{a}]\op{P} + \kappa_1 n_{\rm th}\mathcal{D}^\dag[\op{a}^\dag]\op{P} \\
                       & = -2\kappa_1(\op{a}^\dag \op{a} (1+2n_{\rm th}) + n_{\rm th})\op{P}.                                            \\
\end{align*}
For cat state with sufficiently large mean photon number, this value becomes independent of the cat's parity.
The phase-flip rate is given by:
\begin{equation}
    \label{eq:Gamma_X}
    \Gamma_X(\bar{n}) = 2 \kappa_1 \bar{n}(1+2n_{\rm th}) + 2\kappa_1 n_{\rm th}.
\end{equation}
The lifetime is extracted by fitting the decay of the fringes using the following analytical expression for the fringes:
\begin{equation*}
    \mathcal{W}(i b, t) = A e^{-t/T_{\rm X}} \cos(b \nu) e^{-\frac{b^2}{2\sigma^2}}.
\end{equation*}
For standard cat states, the wavelength of the fringes in phase space is exactly \(\nu = 1/4\alpha \), where \(\alpha \) represents the cat size, and the width of the fringe extent is \(\sigma = 1/2\).
For Moon cat states with a small deformation parameter \(\lambda \), numerical simulations indicates that \(\nu \) remains unchanged, while \(\sigma \simeq (1+\lambda)/2 \) increases.

%% file: 6_zeno_gate_measurement.tex
\section{Zeno gate measurement}\label{ap:zeno_gate_measurement}

\begin{figure}
    \centering
    \includegraphics[width=0.48\textwidth]{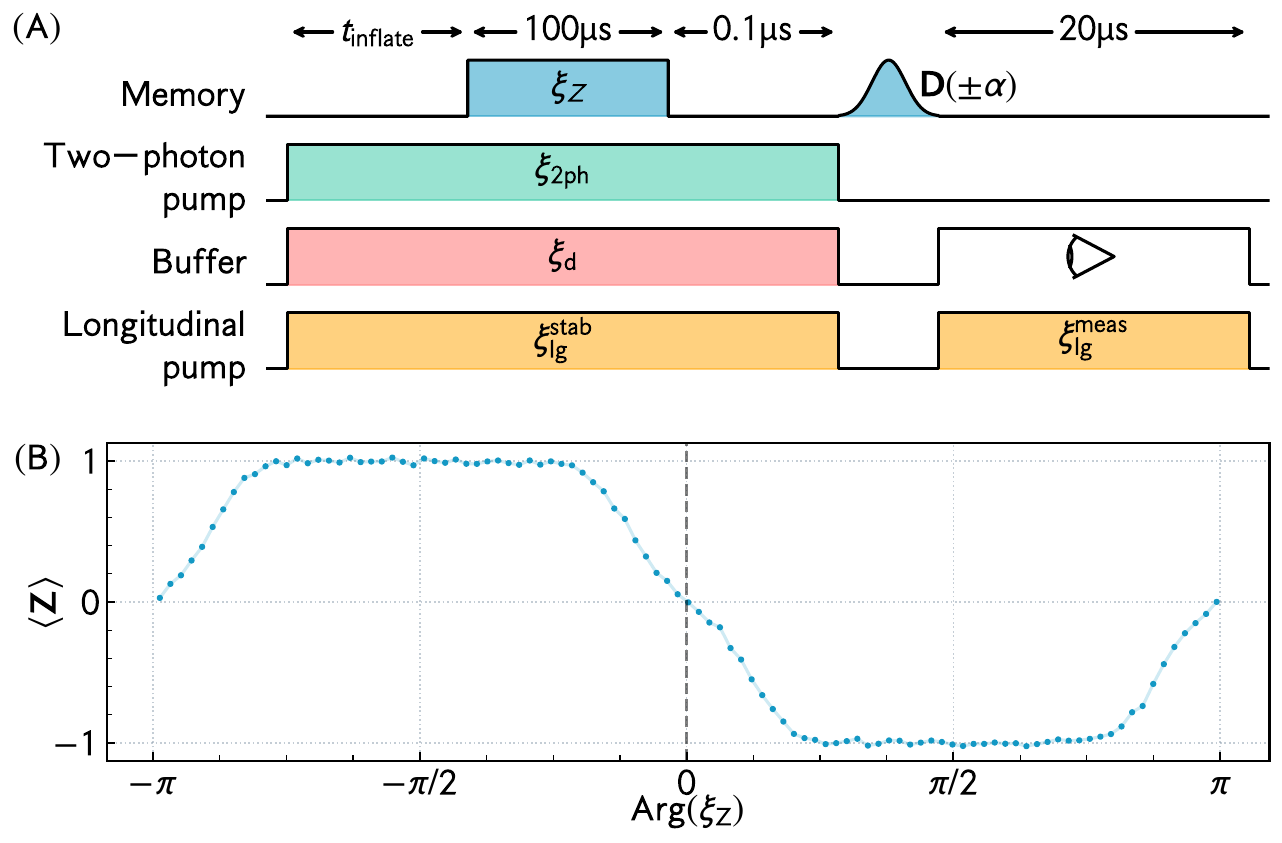}
    \caption{Zeno drive phase calibration.
        (A) Pulse sequence for the phase calibration.
        A cat state is prepared by stabilizing the cat subspace.
        A drive with a variable phase is applied to the stabilized cat state for \(100\;\rm \mu s\), followed by an additional \(100\;\rm ns\) of stabilization.
        Finally, the cat qubit \(\op{Z}\) operator is measured.
        (B) Measured cat qubit \(\op{Z}\) operator as a function of the drive phase for a Moon cat with deformation parameter \(\lambda = 0.38\) and size \(3.7\) photons.
        The black dashed line indicates the Zeno drive phase, chosen such that \(\langle\op{Z}\rangle = 0\).
    }\label{fig:zeno_phase_cal}
\end{figure}

\begin{figure}
    \centering
    \includegraphics[width=0.48\textwidth]{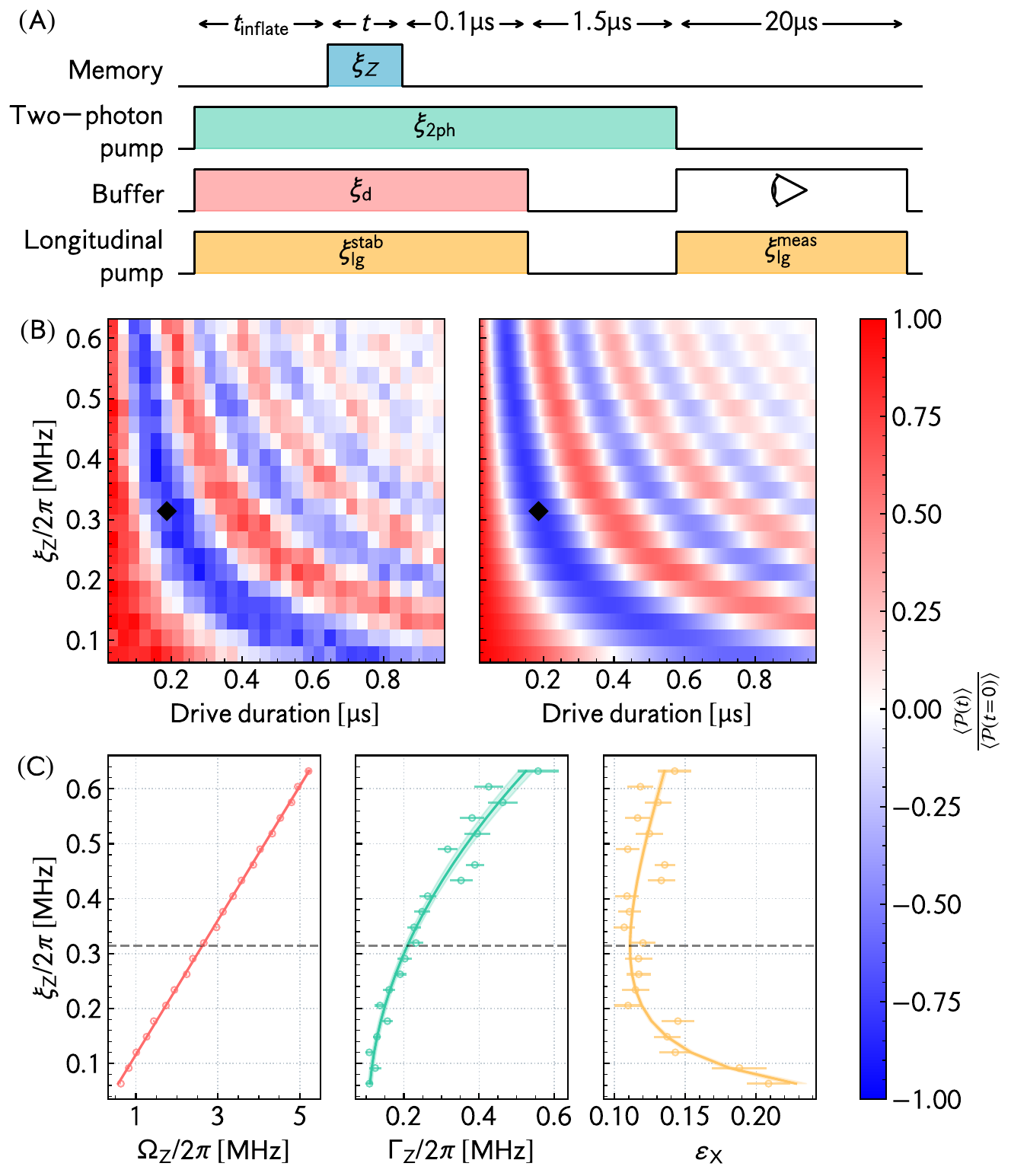}
    \caption{Zeno gate calibration.
        (A) Pulse sequence of the calibration.
        A cat state is prepared by stabilizing the cat subspace, followed by the application of a drive with variable amplitude \(\xi_Z\) to the stabilized cat state for a variable duration \(t\).
        This is followed by an additional \(100\;\rm ns\) of stabilization.
        Finally, the parity of the cat is measured by deflating the state to the 0/1 manifold and measuring the photon number in the memory.
        (B) Measured (left) and fitted (right) parity (color) as a function of the Zeno drive strength and duration.
        The black diamond indicates the optimal \(\pi \)-gate parameters.
        (C) Fitted gate parameters as a function of the Zeno drive strength (circles).
        The fitted parameters are modeled (lines) to extract the optimal Zeno gate parameters.
        The Rabi rate \(\Omega_Z\) increases linearly with the drive strength, while the phase-flip error rate \(\Gamma_Z\) increases quadratically.
        The fitted gate phase-flip errors \(\epsilon_X\) are deduced from the fit of the Rabi and loss rates.
    }\label{fig:zeno_calibration}
\end{figure}

In this section we detail the Zeno gate calibration.
The Zeno gate is executed by applying a drive to the memory mode while the cat is stabilized.
The drive phase is calibrated by sweeping the phase of the drive and measuring the cat qubit \(\op{Z}\) operator (Fig.~\ref{fig:zeno_phase_cal}).
When the drive aligns with the cat qubit axis, it transfers the population between the logical state.
Conversely, when the drive is orthogonal, no population transfer occurs.
The amplitude and duration of the Zeno \(pi\)-gate are calibrated by varying the duration and amplitude of the memory drive (Fig.~\ref{fig:zeno_calibration}), and measuring the cat qubit parity.
The parity time traces oscillate at the Rabi frequency \(\Omega_Z\) and decay at the phase-flip rate \(\Gamma_Z\).
By fitting these oscillations, we extract the gate parameters as a function of the drive strength.
Optimizing the parameters for the lowest phase-flip error rate, we find the optimal \(pi\)-gate parameters.
The bit-flip gate error is measured using the Bayesian adaptive bit-flip measurement, as described in Section~\ref{ap:bayesian_adaptive_measurement}.

%% file: 7_scalings.tex
\section{Error scalings}\label{ap:scalings}

\subsection{Photon number calibration}

\begin{figure}
    \centering
    \includegraphics[width=0.25\textwidth]{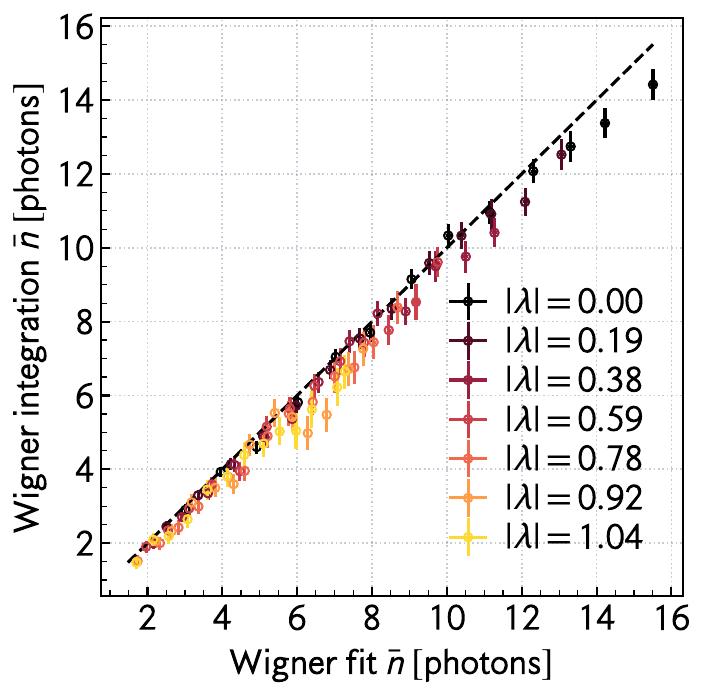}
    \caption{Measured mean photon number, extracted from Wigner function integration, plotted as a function of the mean photon number obtained from fitting the Wigner function for different deformation parameters strengths (color).
        The black dashed line represents the identity line.
    }\label{fig:nbar_calibration}
\end{figure}

To measure the mean photon number \(\bar{n}\) of cat states across the different scalings, we measured the Wigner function of the memory.
For a standard cat state, \(\bar{n}\) can be easily calculated as the square of the coherent state amplitude.
However, for Moon cats, the deformation of the Wigner function causes this method to underestimate \(\bar{n}\), as it neglects the squeezing-like photons in the cat state.
To address this, we performed a fit of the Wigner function using the analytical form of the Moon cat states.
To validate this approach, we compared it to an integration of the Wigner function, \(\bar{n} = \int \mathcal{W}(\beta, \beta^*)|\beta|^2 \mathrm{d}\beta^2 - \frac{1}{2}\).
Both methods show excellent agreement (Fig.~\ref{fig:nbar_calibration}).

\subsection{Buffer frequency fine-tunning}

\begin{figure}
    \centering
    \includegraphics[width=0.48\textwidth]{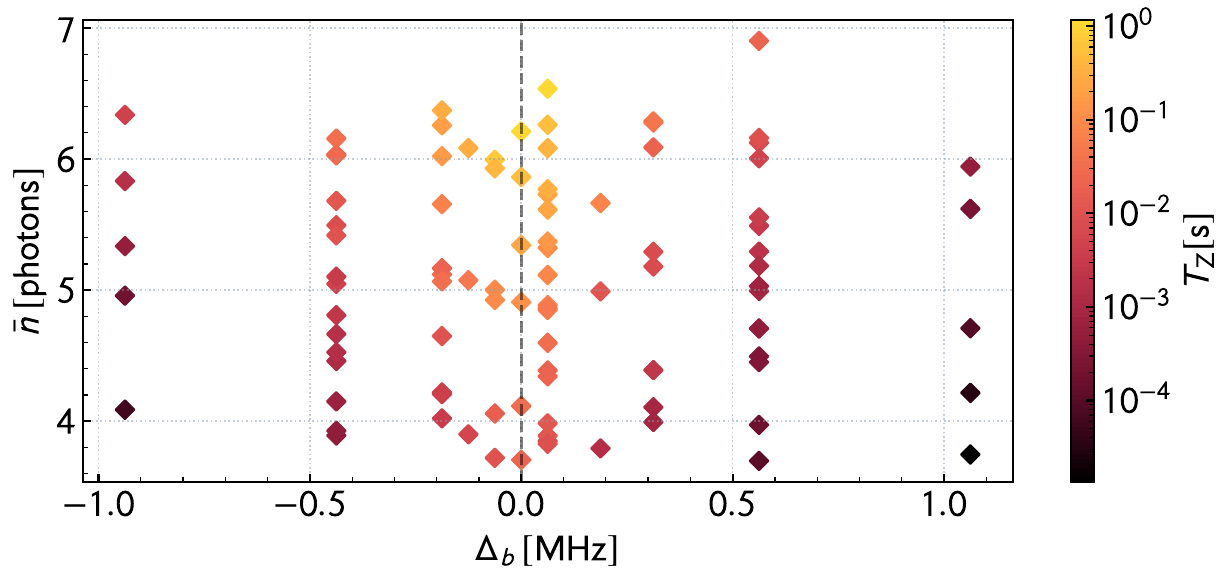}
    \caption{Bit-flip lifetime (color) as a function of the buffer frequency detuning \(\Delta_b\) and memory mean photon number \(\bar{n}\).
        The black dashed line marks the maximum bit-flip lifetime at a fixed \(\bar{n}\).
    }\label{fig:scalings_vs_delta_b}
\end{figure}

The buffer frequency detuning \(\Delta_b\) is a crucial parameter for optimizing the performances of the cat qubit.
To calibrate it precisely, we measured the bit-flip lifetime scaling across different \(\Delta_b\) values.
The buffer frequency used in all experiment in this paper is the that maximizes the bit-flip lifetime at a fixed \(\bar{n}\) (Fig.~\ref{fig:scalings_vs_delta_b}).

\subsection{Complete error scalings}

\begin{figure}
    \centering
    \includegraphics[width=0.48\textwidth]{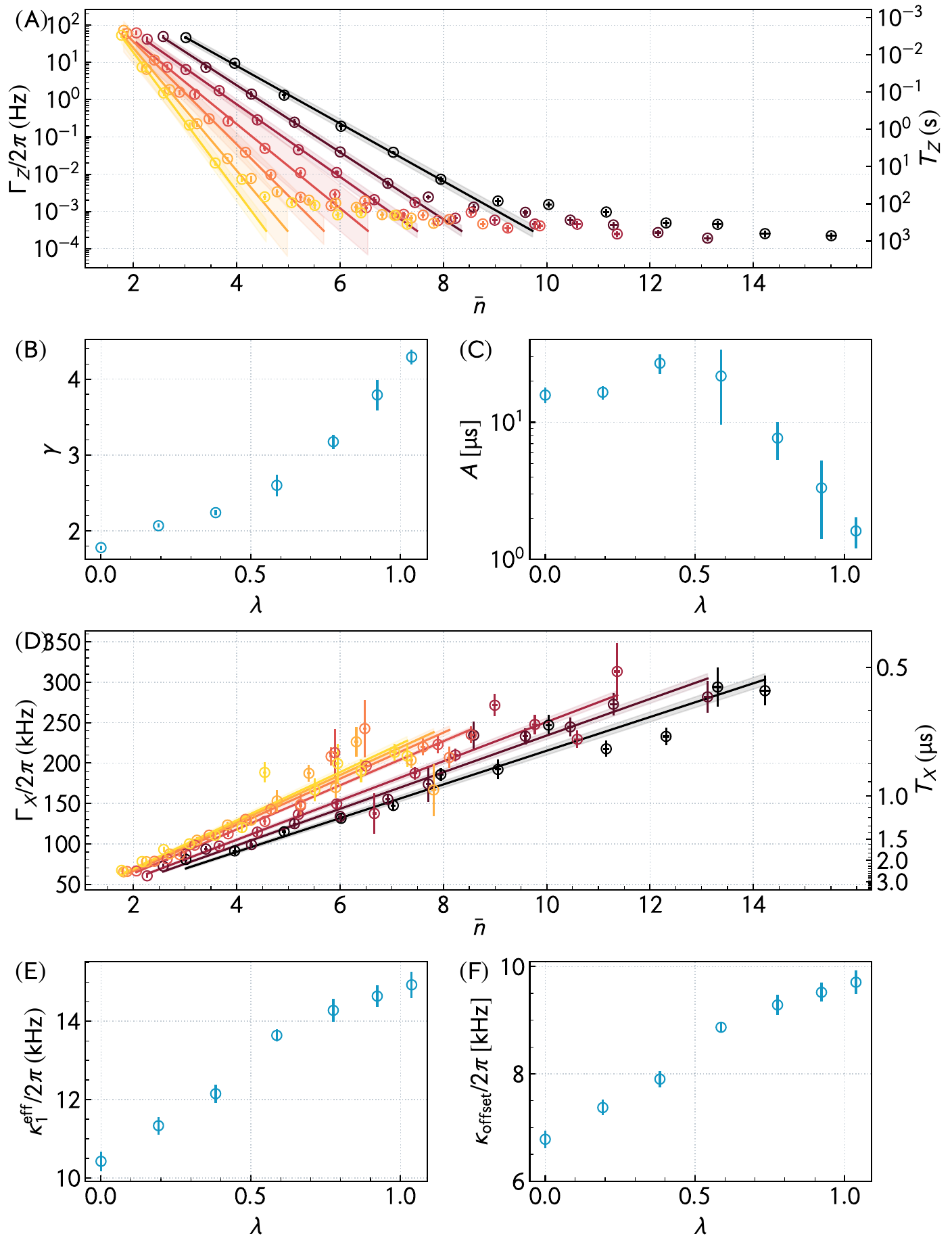}
    \caption{
        Complete idle error scaling.
    }\label{fig:idle_scaling_full}
\end{figure}

The scalings presented in the main text represent a subset of the complete set of measurements conducted.
In Fig.~\ref{fig:idle_scaling_full}, we show the full idle error scaling.
The entire measurement process was fully automated and took one week to complete.
The bit-flip lifetimes were fitted using an exponential dependence on the mean photon number, \(\Gamma_Z(\bar{n}) = A e^{-\gamma \bar{n}}\).
The dependency of the phase-flip lifetime on \(\bar{n}\) is expected to follow an affine relation~\ref{eq:Gamma_X}.
However, our measurements were not precise enough to independently extract both the thermal population and the loss rate.
The memory's thermal population, as determined from the Wigner function cut (Fig.~\ref{fig:photon_number_calibration}E) and from the evolution of memory state deflation (Fig.~\ref{fig:dephasing}B), shows good agreement.
Consequently, we fixed the thermal population in our fits of the phase-flip lifetimes, leaving the loss rate \(\kappa_1\) as the sole free parameter for each phase-flip scaling fit across different deformation strengths.

\begin{figure}
    \centering
    \includegraphics[width=0.384\textwidth]{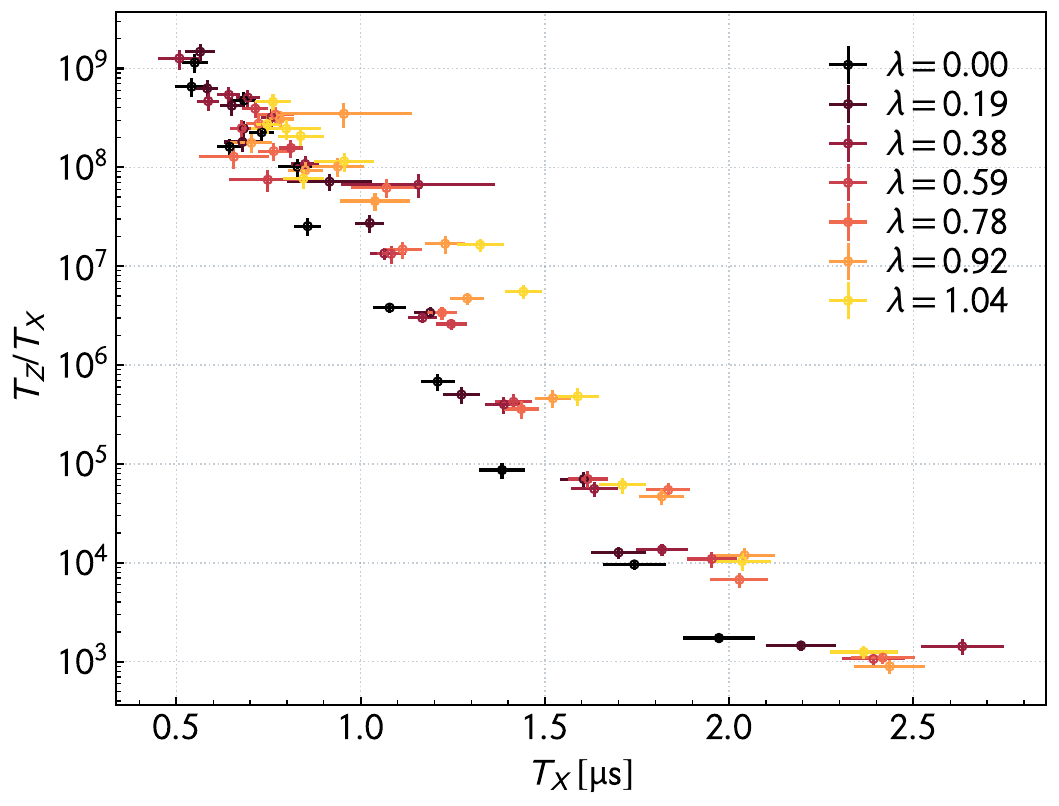}
    \caption{
        Lifetime bias as a function of the phase-flip lifetimes.
    }\label{fig:tx_vs_tz}
\end{figure}

Another representation of the data is presented in Fig.~\ref{fig:tx_vs_tz}, showing the lifetime bias as a function of the phase-flip lifetimes.

\begin{figure}
    \centering
    \includegraphics[width=0.48\textwidth]{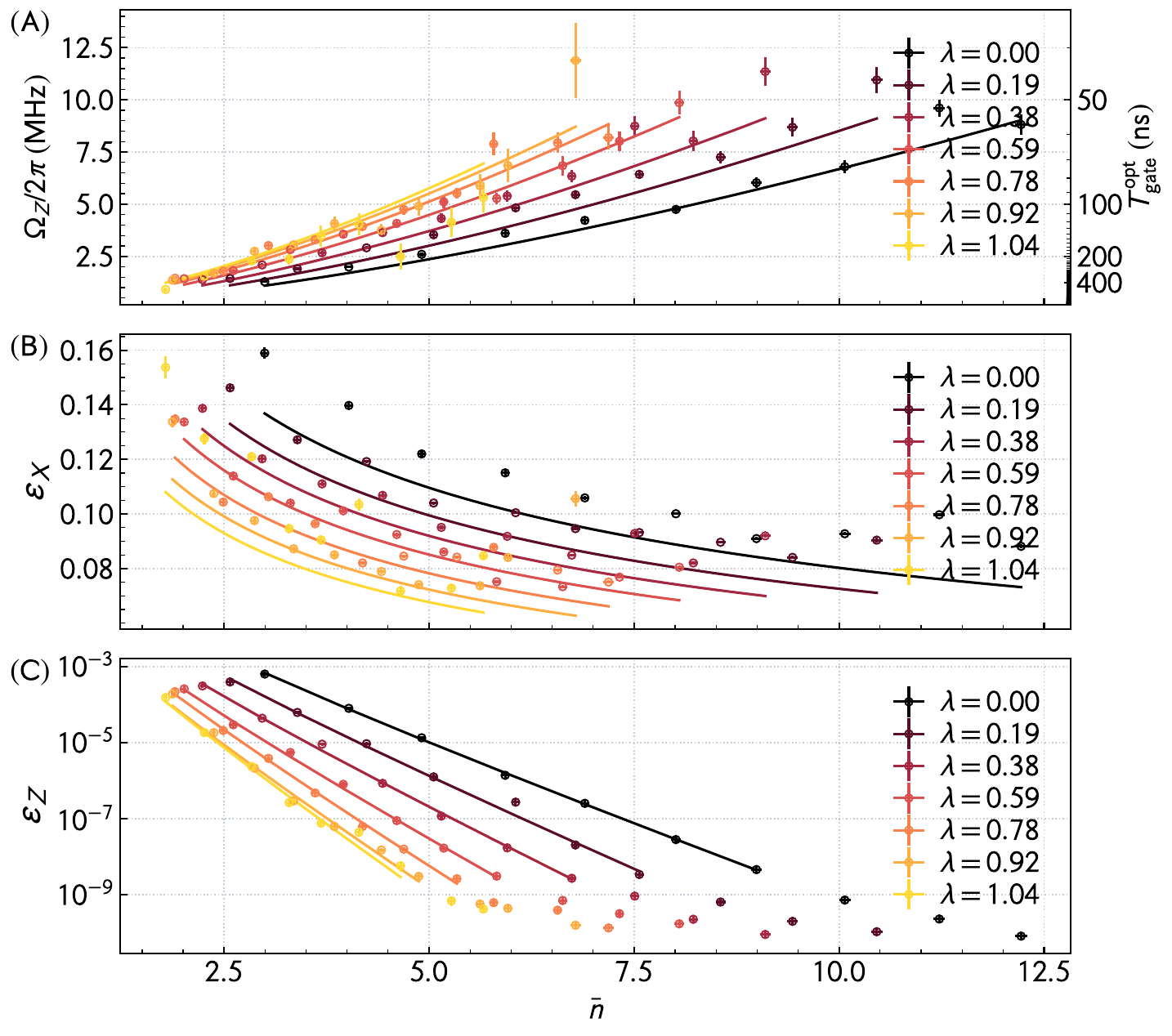}
    \caption{
        Complete Zeno gate error scaling.
    }\label{fig:zeno_scaling_full}
\end{figure}

In Fig.~\ref{fig:zeno_scaling_full}, we present the complete idle error scaling.
The full measurement was totally automated and took 4 days to complete.
The Zeno gate bit-flip error was extracted from bit-flip measurements when the Zeno pulse is played on the stabilized cat.

%% file: 8_theory.tex
\section{Theory}

\subsection{Lifetimes conventions}\label{ap:lifetimes_conventions}
The decay channels of the harmonic oscillator are captured by the Lindblad operators \(\sqrt{\kappa_1(1+n_{\rm th})}\,\op{a}\) for single-photon loss, \(\sqrt{\kappa_1 n_{\rm th}}\,\op{a}^\dag \) for single-photon gain, and \(\sqrt{2\kappa_\phi}\,\op{a}^\dag\op{a}\) for dephasing.
Under these definitions, the memory relaxation time is \(T_1 = 1/\kappa_1\), the pure dephasing time is \(T_\phi = 1/\kappa_\phi \), and the overall coherence time is \(T_2 = 1/(\kappa_1/2 + \kappa_\phi)\).

To characterize the idle lifetimes of the cat qubit, we introduce the decoherence rate \(\Gamma_{\rm O}\) of an observable \(\op{O}\), or equivalently its coherence time \(T_{\rm O} = 1/\Gamma_{\rm O}\).
These values are determined by measuring the exponential decay of the observable's expectation value, \(\langle \op{O}\rangle_t/\langle \op{O}\rangle_{t=0} = e^{-\Gamma_{\rm O} t}\).

In general, a quantum channel \(\mathcal{E}: \rho \mapsto \mathcal{E}(\rho)\) can be decomposed into an ideal target unitary \(\mathcal{U}\) and an error channel \(\mathcal{E}'\) via \(\mathcal{E} = \mathcal{E}' \circ \mathcal{U}\).
To quantify the error for cat qubit operations, we define the mean state-flip probability along the \(\op{O}\)-axis,
\begin{equation*}
    \epsilon_O = \frac{1}{2}\Bigl[\langle\psi_{\rm O}^-|\mathcal{E}'(|\psi_{\rm O}^+\rangle\langle\psi_{\rm O}^+|)|\psi_{\rm O}^-\rangle + \langle\psi_{\rm O}^+|\mathcal{E}'(|\psi_{\rm O}^-\rangle\langle\psi_{\rm O}^-|)|\psi_{\rm O}^+\rangle\Bigr].
\end{equation*}
This quantity is related to the observable's decay rate \(\Gamma_{\rm O}\) via \(\epsilon_O = [1 - e^{-\Gamma_{\rm O} t}]/2\).

\subsection{Moon cat theory}\label{ap:moon_cat_theory}

Under the parametric pumps, the system dynamics is described by the following master equation
\begin{equation}
    \begin{split}
         & \frac{d\op{\rho}}{dt} = -i[\op{H},\op{\rho}] + \mathcal{D}[  \sqrt{\kappa_\mathrm{b}}\op{b}](\op{\rho}),                          \\
         & \op{H} = \hbar g_2\op{a}^2\op{b}^\dag + \hbar g_l\op{a}^\dag\op{a}\op{b}^\dag + \hbar \xi_\mathrm{d} \op{b}^\dag + \mathrm{h.c.},
    \end{split}
\end{equation}

where we neglected single-photon losses and dephasing on the memory.
In the regime \(8\sqrt{|g_2 \xi_d|}  \ll \kappa_b\), we can  adiabatic eliminate the buffer mode~\cite{reglade_quantum_2024}.
The master equation on the reduced subsystem, composed of the memory only, reads,

\begin{equation}
    \begin{split}
         & \frac{d\op{\rho}}{dt} = \mathcal{D}[\op{L_2(\alpha,\lambda)} ](\op{\rho}),                                  \\
         & \op{L_2(\alpha,\lambda)} = \sqrt{\kappa_2}(\op{a}^2 - \alpha^2 + \lambda (\op{a}^\dag \op{a} -\alpha^2 ) ),
    \end{split}
\end{equation}

where \(\lambda = g_\mathrm{l}/g_2\) and \(\alpha^2 = -\frac{\xi_\mathrm{d}}{g_2(1+\lambda)}\).

This Lindblad dissipator stabilizes a 2D-manifold of cat-like states.
Below, we provide an analytical expression of the so-called moon cat states.
Secondly, we define a shifted Fock basis~\cite{chamberland_building_2022} tailored to the moon cat states.
In this basis, the memory can be conveniently decomposed in a two-level system encoding the logical information, and an oscillator, whose excitations describe leakage of the memory outside of the computational manifold.

\subsubsection{Moon cat states}\label{ap:moon_theory}

We seek for the two states \(\ket{\psi}\) that belongs to the kernel of \(\op{L_2}(\alpha,\lambda)\).
Expanding over the Fock basis \(\ket{\psi} = \sum_{n=0}^{\infty} \mu_n \ket{n}\), the coefficients follows the recurrence relation,
\begin{equation*}
    \mu_{n+2} = \frac{\alpha^2 +\lambda(\alpha^2 - n)}{\sqrt{(n+2)(n+1)}}\mu_n.
\end{equation*}
The moon cat states are the even and odd parity states \(\ket{\mathcal{C}_{\alpha, \lambda}^+} = N^+\sum_{p\geq0}\mu_{2p}\ket{2p}\) and \(\ket{\mathcal{C}_{\alpha, \lambda}^-} = N^-\sum_{p\geq0}\mu_{2p+1}\ket{2p+1}\).
Where, \(\mu_0=\mu_1 = 1\),
\begin{equation}
    \label{eq:moon_cat_analytical}
    \begin{split}
         & \mu_{2p} = \frac{1}{\sqrt{2p!}}\prod_{q=0}^{p-1}(\alpha^2 + \lambda(\alpha^2-2q))          \\
         & \mu_{2p+1} = \frac{1}{\sqrt{(2p+1)!}}\prod_{q=0}^{p-1}(\alpha^2 + \lambda(\alpha^2-2q-1)),
    \end{split}
\end{equation}
for \(p>0\) and \(N^\pm \) are normalization constants.
Since \(\alpha^2 + \lambda(\alpha^2-n)\) is minimal for \(n \approx \alpha^2 (1+\lambda)/\lambda \), the coefficients \(\mu_m\) are strongly diminished for \(m\geq \alpha^2 (1+\lambda)/\lambda \).
In particular, if the parameters are such that \( \alpha^2 (1+\lambda)/\lambda = 2q\) (resp. \(2q+1\)) is an integer, the coefficients \(\mu_{2p}\) (resp. \(\mu_{2p+1}\)) vanish exactly for \(p\geq q\).

\subsubsection{Shifted Fock basis for moon cat states}

Let us move in the frame defined by displaced Fock basis of displacement amplitude \(\alpha \) as introduced in~\cite{chamberland_building_2022}.
In this basis, the annihilation operator reads,
\begin{equation*}
    \op{a} \rightarrow \op{Z} \otimes \left(\op{\tilde{a}} + \alpha \right) + O(e^{-4|\alpha|^2}).
\end{equation*}
The logical state of the cat qubit is described by the Pauli operator \(\op{Z}\) acting on the first subsystem, while the gauge mode described by \(\op{\tilde{a}}\) corresponds to a harmonic oscillator centered on \(\ket{\op{Z} \alpha}\).
After this first transformation, the two-photon loss operator becomes,
\begin{equation*}
    \op{\tilde{L}_2^{(disp)}} =  \sqrt{\kappa_2}\;\op{1}\otimes \left( \op{\tilde{a}}^2 + \alpha(\lambda+2)(\op{\tilde{a}}+\frac{\lambda}{\lambda+2}\op{\tilde{a}}) \right).
\end{equation*}

Due to the presence of the term \(\lambda \op{a^\dagger} \op{a}\), the vacuum state of the gaude mode is not an eigenstate of \(\op{\tilde{L}_2^{(disp)}}\).
Below, we apply two more unitary transformations to bring this operator to a form where it annihilates the vacuum state.

We first apply a squeezing transformation with squeezing parameter \(r = \mathrm{th}^{-1}\frac{\lambda}{2+\lambda}\), such that

\begin{equation*}
    \op{\tilde{a}} \rightarrow \frac{1}{\sqrt{1+\lambda}}\left(\op{\tilde{a}} + \frac{\lambda}{2}(\op{\tilde{a}} - \op{\tilde{a}}^\dag)\right).
\end{equation*}

which leads to,

\begin{equation*}
    \op{\tilde{L}_2^{(disp,sq)}} =  2\alpha\sqrt{\kappa_2(1+\lambda)}\;\op{1}\otimes \left( \op{\tilde{a}} + \eta \big(\op{\tilde{a}}^2 +  \frac{\lambda^2}{4}(\op{\tilde{a}} + \op{\tilde{a}}^\dag)\big)^2 \right),
\end{equation*}
with \(\eta = (2\alpha\sqrt{\lambda+1})^{-1} \ll 1 \).
The first 2-level subsytem describes the state of the cat qubit, while the second mode \(\op{\tilde{a}}\) describes a harmonic oscillator, which is subject to a dissipation rate \( 4\alpha^2\kappa_2(1+\lambda)\).

The zero eigenstate of this operator is of the form \( \ket{0}+\eta\ket{\psi^{\perp}} + O(\eta^2) \), where we omitted the qubit state as \(\op{\tilde{L}_2}\) acts as the identity on it.
One can move to a basis where the pointer state is closer to the vacuum state, by applying the Schrieffer-Wolf transformation \(e^{\eta G}\), with \( G = \frac{\lambda^2}{4} ( a^\dagger +  {a^\dagger}^2 a + \frac{{a^\dagger}^3}{2} - h.c ) \).
In this new basis, using the Baker-Campbell-Hausdorff expansion, the two-photon loss operator reads,
\begin{equation}
    \label{eq:L2_tilde_eq}
    \op{\tilde{L}_2} =  2\alpha\sqrt{\kappa_2(1+\lambda)}\;\op{1}\otimes \left( \op{\tilde{a}} + \eta(1+\frac{\lambda^2}{2})\op{\tilde{a}}^2 + O(\eta^2) \right).
\end{equation}

Note that in this basis, the zero eigenstate of this operator is of the form \( \ket{0}+ O(\eta^2) \).
Additionally, the confinement rate of the gauge mode is increased upon deformation, as
\begin{equation}
    \label{eq:moon_confinement_rate}
    \kappa_{\rm conf} = 4\alpha^2\kappa_2(1+\lambda).
\end{equation}

\subsubsection{Comparison with dissipative squeezed cat qubits\label{ap:moon_cat_theory_sq_comparison}}

\begin{figure}[th!]
    \centering
    \includegraphics[width=0.48\textwidth]{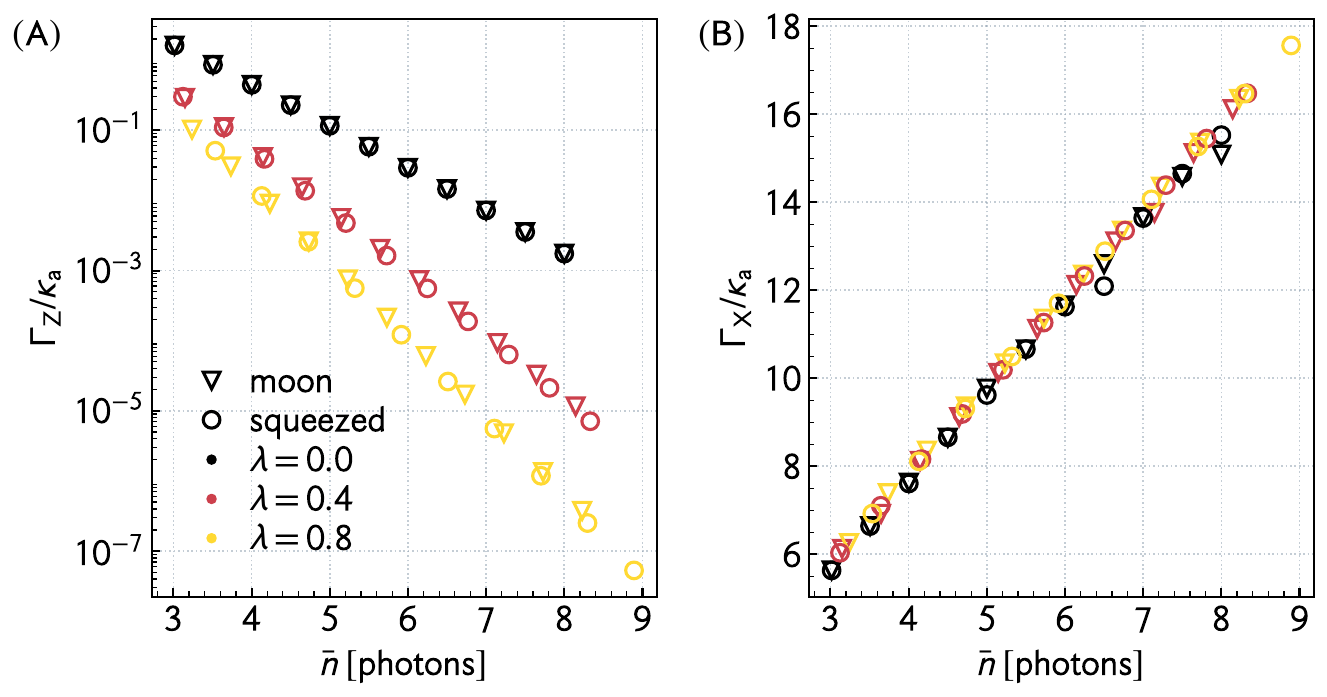}
    \caption{Simulated idle decay rates of moon cat qubits compared with equivalent squeezed cat qubits.
        The bit-flip decay rate \(\Gamma_Z\) (panel (A)) and phase-flip decay rate \(\Gamma_X\) (panel (B)), each normalized by the memory's single-photon loss rate \(\kappa_{\rm a}\), are plotted as functions of the mean photon number \(\bar{n}\) and deformation parameter \(\lambda \) (color-coded).
        Triangular markers represents moon cats and circular markers represents squeezed cats with \(r = \mathrm{th}^{-1}(\lambda / 2)\).
        Both qubits display steeper exponential suppression of the bit-flip decay rate with increasing deformation amplitude, with no associated increase in the phase-flip decay rate.
    }\label{fig:moon_vs_sq}
\end{figure}

The dissipative squeezed cat qubit is governed by a Lindblad operator~\cite{xu_autonomous_2023} of the form:
\begin{equation*}
    \op{L_2}(r) = \frac{\sqrt{\kappa_2}}{\mathrm{ch}^2 r}\op{S}(r)(\op{a}^2-\alpha'^2)\op{S}^\dag(r),
\end{equation*}
where \(\op{S}(r) = e^{\frac{r}{2}(\op{a}^2 + \op{a}^{\dag 2})}\) represents the squeezing operator.
Expanding \(\op{L_2}(r)\), we obtain:
\begin{equation*}
    \op{L_2}(r) = \sqrt{\kappa_2}( \op{a}^2 + \mathrm{th}^2 r \op{a}^{\dag 2} + \mathrm{th}\;r (\op{a}^\dag \op{a} + \op{a}\op{a}^\dag) - \alpha'^2).
\end{equation*}
The Lindblad operator for the moon cat qubit, given by Eq.~\ref{eq:Moon_dissipator}, is effectively equivalent to the above expression with the \(a^{\dag2}\) term omitted, and where we define \(\lambda = 2\mathrm{th}\;r\).
Notably, this term becomes negligible in the small-squeezing limit.
To analyze the robustness of the information encoded in either the moon cat or squeezed cat qubits, we consider the following operators:
\begin{equation*}
    \begin{aligned}
        \op{L_1}    & = \sqrt{\kappa_{\rm a}}\op{a},                 \\
        \op{L_\phi} & = \sqrt{2\kappa_{\rm \phi}}\op{a}^\dag\op{a},  \\
        \op{H}      & = - \hbar\frac{K_4}{2}\op{a}^{\dag 2}\op{a}^2.
    \end{aligned}
\end{equation*}
In Figure~\ref{fig:moon_vs_sq}, we present the simulated decay rates of the moon and squeezed cat qubits, using parameter values \(\kappa_2/\kappa_{\rm a} = 10^3\), \(\kappa_\phi/\kappa_{\rm a} = 20\) and \(K_4/\kappa_{\rm a} = 15\).
The results demonstrate that both squeezed and moon cat qubits achieve similar increase in noise bias, \(\Gamma_Z/\Gamma_X\), despite the fact that moon cat qubits are simpler to implement.

\subsection{Zeno gate theory}\label{ap:zeno_theory}
The zeno gate is implemented by applying a drive whose direction is orthogonal to the cat axis.
The master equation of the system reads,
\begin{equation}
    \label{eq:master_eq_Z}
    \frac{d\op{\rho}}{dt} = -i[\xi_Z (\op{a}+\op{a^\dagger}),\op{\rho} ] + \mathcal{D}[\sqrt{\kappa_\downarrow}\op{a}]\op{\rho} + \mathcal{D}[\sqrt{\kappa_\uparrow}\op{a^\dagger}]\op{\rho} + \mathcal{D}[\op{{L}_2}]\op{\rho},
\end{equation}
where \(\op{{L}_2}\) is defined in Eq.~\ref{eq:Moon_dissipator}.
Let us move in the basis introduced in Appendix~\ref{ap:moon_theory}.

Under these successive transformations, Eq.\ref{eq:master_eq_Z} reads,
\begin{equation*}
    \frac{d\op{\rho}}{dt} = -\frac{i}{\hbar}[\op{\tilde{H}_Z},\op{\rho} ] + \mathcal{D}[\op{\tilde{L}_1^{\downarrow}}]\op{\rho} + \mathcal{D}[\op{\tilde{L}_1^{\uparrow}}]\op{\rho} + \mathcal{D}[\op{\tilde{L}_2}]\op{\rho},
\end{equation*}
with
\begin{equation*}
    \begin{split}
        \op{\tilde{H}_Z} =            & \frac{\hbar \xi_Z}{\sqrt{1+\lambda}} \op{Z}\otimes(\op{\tilde{a}}+\op{\tilde{a}}^\dag -\eta \lambda^2\op{\tilde{a}}^\dagger \op{\tilde{a}})+2\hbar\xi_Z (\alpha - \frac{\eta\lambda^2}{2}) \op{Z}\otimes \op{1}, \\
        \op{\tilde{L}_1^\downarrow} = & \sqrt{\frac{\kappa_\downarrow}{1+\lambda}}\op{Z} \otimes \left(\op{\tilde{a}} + \frac{\lambda}{2}(\op{\tilde{a}} - \op{\tilde{a}}^\dag) + O(\eta)\right) + \sqrt{\kappa_\downarrow}\alpha\op{Z} \otimes \op{1} , \\
        \op{\tilde{L}_1^\uparrow} =   & \sqrt{\frac{\kappa_\uparrow}{1+\lambda}}\op{Z} \otimes \left(\op{\tilde{a}}^\dag - \frac{\lambda}{2}(\op{\tilde{a}} - \op{\tilde{a}}^\dag) + O(\eta)\right) + \sqrt{\kappa_\uparrow}\alpha\op{Z} \otimes \op{1},
    \end{split}
\end{equation*}
and \(\op{\tilde{L_2}}\) defined in Equation~\ref{eq:L2_tilde_eq}.

The superoperator \(\mathcal{D}[\op{\tilde{L}_2}]\) confines the gauge mode \(\op{\tilde{a}}\) to the vacuum state due to dissipation rate \(4\alpha^2\kappa_2(1+\lambda)\).
While the second term in the expression of \( \op{\tilde{H}_Z}\) acts solely on the qubit state and generates the desired rotation, the first term entangles the qubit state with the lossy gauge mode state, which results in an effective \(\op{\tilde{Z}}\) error channel on the qubit.
In the limit where \(\frac{\xi_Z}{\sqrt{1+\lambda}} \ll 4\alpha^2\kappa_2(1+\lambda)\), one can adiabatically eliminate the gauge mode~\cite{chamberland_building_2022}, while retaining second-order corrections to the qubit dynamics.
The master equation of the reduced qubit system reads,

\begin{equation*}
    \begin{split}
        \frac{d\op{\rho_g}(t)}{dt} = & \bigg((\kappa_\uparrow + \kappa_\downarrow) (\alpha^2 + \frac{\lambda^2}{4(1+\lambda)}) + \frac{\kappa_\uparrow}{1+\lambda} \\
                                     & + \frac{\xi_Z^2}{\alpha^2\kappa_2 (1+\lambda)^2}\bigg)\mathcal{D}[\op{Z}]\op{\rho_g}(t)                                     \\
                                     & - 2i(\alpha - \frac{\eta\lambda^2}{2})\xi_Z[\op{Z}, \op{\rho_g}(t)].
    \end{split}
\end{equation*}

The Rabi rate is given by \(\Omega = 4\xi_Z(\alpha - \frac{\lambda^2}{4\alpha\sqrt{\lambda+1}}) \), and the single-photon loss and gain contribution of the decay reads \(\Gamma_X^0 = 2(\alpha^2 + \frac{\lambda^2}{4(1+\lambda)} + \frac{n_\mathrm{th}}{(1+2 n_\mathrm{th})(1+\lambda)})\kappa_1^\mathrm{eff}\), where \(\kappa_1^\mathrm{eff} =\kappa_\uparrow + \kappa_\downarrow \) and \(n_\mathrm{th}\) is the temperature of the bath responsible for single-photon loss and gain.
The corrections with respect to the rates of the non-deformed cats (\(\lambda=0\)) are very small, while the non-adiabatic part of the error rate \(\Gamma_X^{\rm n.a.} = \frac{2\xi_Z^2}{\alpha^2\kappa_2{(1+\lambda)}^2}\) is divided by a factor \({(1+\lambda)}^2\).
This reduction can be attributed to two main contributions.
First, the confinement rate of the deformed cat is enhanced by a factor \(1+\lambda \) (Eq.~\ref{eq:moon_confinement_rate}).
Second, the leakage rate is further diminished by an additional factor of \(1+\lambda \), as the basis states of the moon cat qubit approach the eigenstates of the position operator \(\op{x} = \op{a} + \op{a}^\dag \) with increasing deformation (Fig.~\ref{fig:fig4}A-B).
An optimal gate time and amplitude can be computed by minimizing the phase-flip errors \(\epsilon_X = \frac{1}{2}(1-e^{-\Gamma_X T_\pi})\):
\begin{equation*}
    \begin{split}
        \Omega^* & = 4\alpha^3(1+\lambda) \sqrt{\kappa_2\kappa_1^\mathrm{eff}},                         \\
        p_Z^*    & \approx \frac{\pi}{2\alpha(1+\lambda)}\sqrt{\frac{\kappa_1^\mathrm{eff}}{\kappa_2}}.
    \end{split}
\end{equation*}

In this derivation of optimal rates, we neglected the contributions in \(\Gamma_X^0\) that do not depend on \(\alpha \).
The limit of adiabatic elimination of the gauge mode simplifies to \(\kappa_1^{\rm eff} \ll 16\kappa_2(1+\lambda)\) for the optimal rates.

\subsection{Moon cat concatenated with a repetition code}

Cat qubits, thanks to their large noise bias, are promising candidates for hardware efficient quantum correction.
As the dissipatively stabilized cat qubits possess a macroscopic bit-flip lifetime, it can be envisioned to build logical qubits by correcting only the phase-flip errors, while the bit-flip protection is handled by the autonomous stabilization of the cat qubits.
The repetition code, due to its minimal weight-2 connectivity requirement, is the simplest error correcting code for a concatenated cat qubit architecture~\cite{guillaud_repetition_2019}.
In this appendix, we compare the standard and moon cat qubits concatenated in a repetition code, and demonstrate that using moon cat qubits can lead to several orders of magnitude improvement in the logical error rate for a given physical error rate thanks to a higher threshold.

For a repetition code composed of \(d\) data cat qubits, the stabilizers are given by \( \{X_i X_{i+1}\}_{i \in \llbracket 1,d-1 \rrbracket}\) and \((d-1)/2\) phase-flip errors can be corrected.
The total logical error rate of a repetition code is given by \(\epsilon_L \approx p_{X_L} + p_{Z_L}\) (as \(p_{Y_L} = p_{X_L} p_{Z_L} \ll p_{X_L}, p_{Z_L}\)).
As the repetition is used as phase-flip error correcting code, a single bit-flip error in the quantum error correction circuit can create a logical bit-flip error.
Thus, \(p_{X_L}\) is given at the first order by the physical bit-flip error multiplied by the space-time volume of the quantum error correction circuit.
Hence, to compare the standard and moon cats on an equal footing, we must compare the two at a fixed bit-flip probability.
For the idle case or during a \(Z\) gate, we see that the bit-flip lifetime of a standard cat qubit at \(\bar n = 8\) approximately corresponds to the bit-flip lifetime of a moon cat qubit at \(\bar n = 4\) and \(\lambda = 1\) (see Figure~\ref{fig:fig3}F).

To measure the \(X_i X_{i+1}\) stabilizer without inducing bit-flip errors on the data cat qubits, the ancilla qubits need to also be cat qubits, and a bias-preserving CNOT gate has to be performed between them~\cite{guillaud_repetition_2019,puri_bias-preserving_2020}.
Promising results have been demonstrated towards a bias-preserving CNOT gate for standard cat qubits~\cite{jezouin_dissipative_2024}, and it is still to be demonstrated for moon cat qubits.
We make the choice for this appendix to assume that the bit-flip lifetime advantage of the moon cat is conserved during a bias-preserving CNOT gate, as it appears with master equation simulation.
Thus, we assume that the overall bit-flip error during quantum error correction for the standard cat qubit at \(\bar n = 8\) approximately corresponds to the one of the moon cat qubit at \(\bar n = 4\) and \(\lambda = 1\).

\begin{figure}[t]
    \centering
    \includegraphics[width=0.48\textwidth]{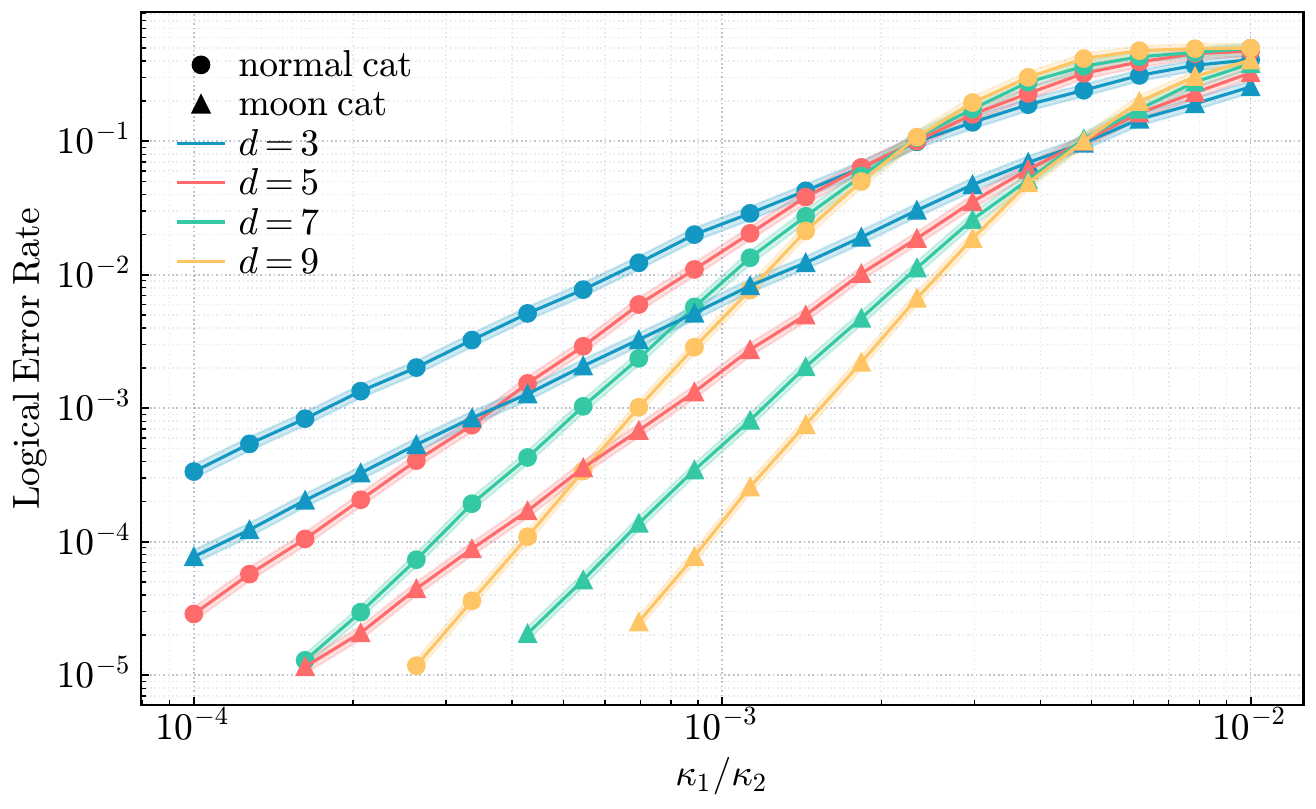}
    \caption{Logical phase-flip error rate of the repetition code as a function of the ratio of the single photon-loss rate and the two-photon dissipation rate rate \(\kappa_1/\kappa_2\).
        The circle markers indicate the standard cat qubits, while the triangle markers indicate the moon cat qubits at \(\lambda = 1\).
        The error rate is estimated with Monte Carlo simulation, with circuit-level noise, and the circuit is sampled until at least 1000 logical errors are observed.
        The transparent outlines indicate the error bar.
    }\label{fig:repcode}
\end{figure}

To evaluate \(p_{Z_L}\), we perform \(d\) rounds of error correction and evaluate the number of times the decoder incorrectly predicts the errors at the output of the circuit.
We use the library stim~\cite{gidney_stim_2021} and pymatching~\cite{higgott_sparse_2023}.
The precise error model used is detailed in Table~\ref{tab:error_model}.
The results are plotted in Figure~\ref{fig:repcode}.
We see that using moon cat qubits improves the phase-flip threshold of the repetition code.
This is due to the combination of two effects.
First, moon cat qubits have a lower photon number for the same bit-flip error, which results in a lower phase-flip error.
Second, moon cat qubits have lower non-adiabatic errors, resulting in lower phase-flip errors on ancilla cat qubits.

\begin{table}[bh!]
    \centering
    \begin{tabular}{|c|c|c|c|}
        \hline
                                 & \multirow{2}{*}{error type} & moon cat                                                                                 & standard cat                                                            \\
                                 &                             & (\(\bar n = 4\), \(\lambda = 1\))                                                        & (\(\bar n = 8\))                                                        \\

        \hline
        \hline
        \(\mathcal P_{\ket{+}}\) & \(Z\)                       & \(\bar n \kappa_1 T \)                                                                   & \(\bar n \kappa_1 T \)                                                  \\
        \hline
        \(\mathcal M_X\)         & \(Z\)                       & \(\bar n \kappa_1 T \)                                                                   & \(\bar n \kappa_1 T \)                                                  \\
        \hline
        \multirow{3}{*}{CNOT}    & \(Z_c \)                    & \(\bar n \kappa_1 T + \displaystyle\frac{\pi^2}{64 \bar n \kappa_2 (1 + \lambda)^2 T }\) & \(\bar n \kappa_1 T + \displaystyle\frac{\pi^2}{64 \bar n \kappa_2 T}\) \\
                                 & \(Z_t\)                     & \(0.5 \bar n \kappa_1 T \)                                                               & \(0.5 \bar n \kappa_1 T \)                                              \\
                                 & \(Z_c Z_t\)                 & \(0.5 \bar n \kappa_1 T \)                                                               & \(0.5 \bar n \kappa_1 T \)                                              \\
        \hline
    \end{tabular}
    \caption{
        \label{tab:error_model}Physical error probabilities of the operations used in the repetition code for the moon cat qubits and the standard cat qubit.
        The cat qubit errors (see~\cite{guillaud_repetition_2019,chamberland_building_2022} for a detailed analysis) depend on the time of the operations \(T\), the single-photon loss rate \(\kappa_1\), the two-photon dissipation rate \(\kappa_2\) and the average photon number \(\bar{n}\).
        In our simulation, we considered an identical time for all operations \(T = 1/\kappa_2\).
        In this case, the errors are parametrized by the parameter \(\kappa_1/\kappa_2\). The non-adiabatic errors of the CNOT gate for the moon cat can be derived analogously to those of the $Z$ gate by transitioning to the shifted Fock basis (see Appendix \ref{ap:zeno_theory}).
    }
\end{table}

To conclude, using moon cat qubits concatenated with repetition codes can lead to orders of magnitude improvement on the logical phase-flip, reducing even more the number of physical qubits necessary to build a logical qubit.